\documentclass[11pt]{article}
\usepackage{amsmath,amssymb,mathtools}
\usepackage{amsthm}
\usepackage{amsfonts}
\usepackage{graphicx}
\usepackage{subcaption}   
\usepackage{url}
\usepackage{natbib}
\usepackage{float}        
\usepackage{xcolor}
\usepackage{booktabs}    
\usepackage{multirow}    
\newtheorem{assumption}{Assumption}

\newtheorem{definition}{Definition}
\newtheorem{remark}{Remark}
\newtheorem{lemma}{Lemma}
\newtheorem{proposition}{Proposition}

\newtheorem{example}{Example}[section]

\newcommand{\mub}{{\pmb \mu}}
\newcommand{\Sigb}{{\pmb \Sigma}}

\newcommand{\Qb}{\mathbf{Q}}
\newcommand{\Fb}{\mathbf{F}}
\newcommand{\Xb}{\mathbf{X}}
\newcommand{\Yb}{\mathbf{Y}}
\newcommand{\xb}{\mathbf{x}}
\newcommand{\yb}{\mathbf{y}}

\begin{document}

\newcommand{\anon}{1}

\if1\anon
{
	\title{\bf Multidimensional Stochastic Dominance Test Based on Center-outward Quantiles}
	
	\author{
		Yiming Ma\thanks{Yiming Ma and Hang Liu are co-first authors.}, 
		Hang Liu\thanks{The author’s research was supported by the National Natural Science Foundation of China (NSFC) grant 12401372.}, 
		and 
		Weiwei Zhuang\thanks{\textbf{Corresponding author}. Email: \texttt{weizh@ustc.edu.cn}. The author's research was supported by NSFC grant 72571262.}\\ \\
		\textit{Department of Statistics and Finance, School of Management,}\\
		\textit{University of Science and Technology of China}
	}
	
	\maketitle
} \fi

\if0\anon
{
	\bigskip
	\bigskip
	\bigskip
	\begin{center}
		{\LARGE\bf Multidimensional Stochastic Dominance Test Based on Center-outward Quantiles}
	\end{center}
	\medskip
} \fi

\begin{abstract}
	
	Stochastic dominance (SD) provides a quantile-based partial ordering of random variables and has broad applications. Its extension to multivariate settings, however, is challenging due to the lack of a canonical ordering in 
	\(\mathbb{R}^d\) (\(d \geq 2\)) and the set-valued character of multivariate quantiles. Based on the multivariate center-outward quantile function in \cite{hallin2021distribution}, this paper proposes new first- and second-order multivariate stochastic dominance (MSD) concepts through comparing contribution functions defined over quantile contours and regions. To address computational and inferential challenges, we incorporate entropy-regularized optimal transport, which ensures faster convergence rate and tractable estimation. We further develop consistent Kolmogorov–Smirnov and Cramér–von Mises type test statistics for MSD, establish bootstrap validity, and demonstrate through extensive simulations good finite-sample performance of the tests. Our approach offers a theoretically rigorous, and computationally feasible solution for comparing multivariate distributions.

\end{abstract}

\noindent%
\vfill

\section{Introduction}

\subsection{Background and Motivation}

Stochastic dominance  provides a rigorous framework for comparing probability distributions under uncertainty and has been widely applied in finance, insurance, and environmental economics \citep{lizyayev2010stochastic, gilboa1989maxmin, pinar2022sensitivity}. In the univariate case, let \(X\) and \(Y\) be random variables with quantile functions \(Q_X(p)\) and \(Q_Y(p)\), \(p \in [0,1]\). Then \(X\) first-order stochastically dominates (FSD) \(Y\), denoted as \(X \succ_{FSD} Y\), if \(Q_X(p) \ge Q_Y(p)\) for all \(p\), with strict inequality for some \(p\). Second-order dominance (SSD), denoted as \(X \succ_{SSD} Y\), holds if $\int_0^p Q_X(u)\,du \ge \int_0^p Q_Y(u)\,du$ forall $p\in[0,1] $, with strict inequality for some \(p\).
FSD captures the preference of all decision-makers who always prefer greater outcomes, while SSD reflects the preferences of risk-averse agents who value not only greater means but also lower dispersion, accounting for both central tendency and risk.

While SD is effective in one dimension, it becomes inadequate in multivariate contexts where decisions depend on multiple correlated outcomes. In portfolio selection, investors must consider returns, risks, and liquidity simultaneously \citep{amihud1986asset}; in environmental policy, trade-offs arise among economic, social, and ecological objectives \citep{keeney1993decisions}. Extending SD to higher dimensions thus remains a fundamental yet challenging problem. 

\begin{example}[Residents' happiness analysis]\label{example.1}
	{\rm Residents' well-being depends on multiple factors such as income, health, education, and social relationships. To illustrate, suppose the happiness levels of Communities A and B are compared via the joint distributions of the income and health, which are modeled as follows. Community A follows a bivariate normal distribution with mean $\mub_{\rm A} = (50, 60)^\top$, where $^\top$ denotes the transpose operator, and covariance $\Sigb_A =\bigl[\begin{smallmatrix}100 & 50 \\ 50 & 100 \end{smallmatrix}\bigl]$, implying a correlation of $ 0.5$. Community B has normal marginals with mean $\mub_{\rm B} = (58, 58)\top$ and standard deviation 10, coupled via a Clayton copula (parameter $\theta = 2$) to induce lower-tail dependence, capturing joint occurrences of low income and poor health. As shown in Figure~\ref{fig:example1}, A has higher average health but lower income than B, while B exhibits greater density in the low-income–low-health region and complex cumulative distribution crossings. Most traditional MSD  methods cannot effectively cope with this complexity, especially if we consider more factors that affect happiness. In Appendix \ref{appendix.d}, it is shown that the MSD we defined has a more sensitive detection capability than the traditional way.
	}
	
	\begin{figure}[http]
		\centering
		\begin{subfigure}[b]{0.5\textwidth}
			\centering
			\includegraphics[width=\textwidth]{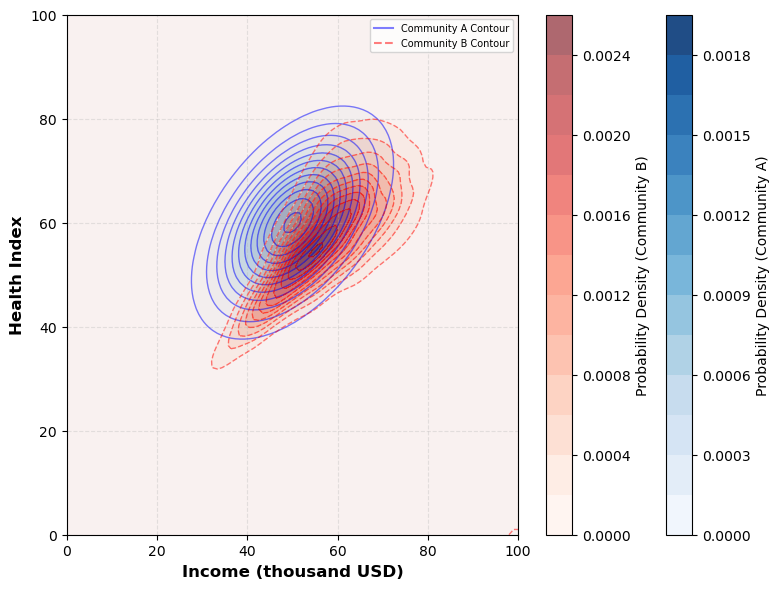}
		\end{subfigure}
		\begin{subfigure}[b]{0.4\textwidth}
			\centering
			\includegraphics[width=\textwidth]{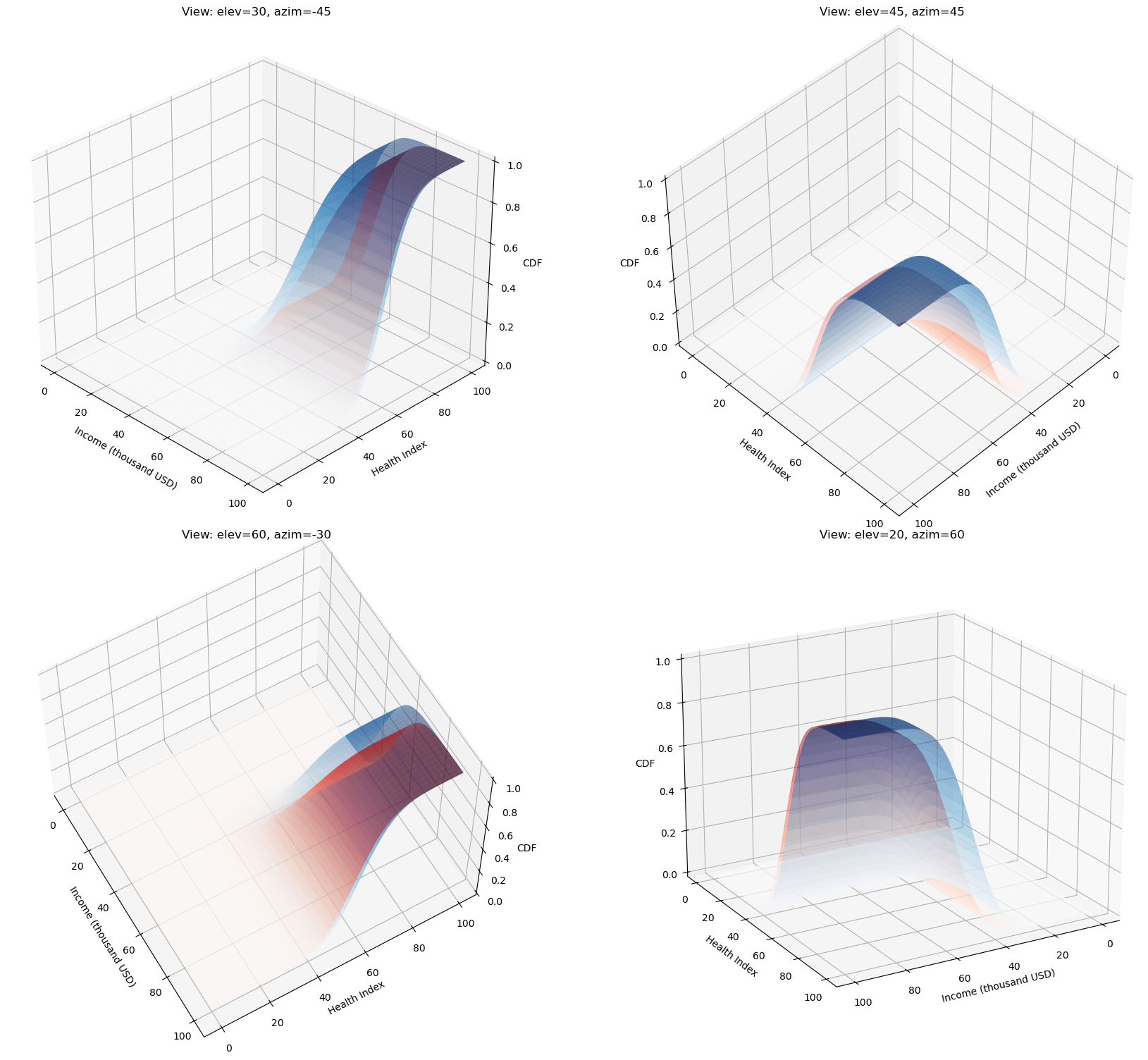}
		\end{subfigure}
		
		\caption{The left panel shows the distribution density plots and the corresponding contour plots for Communities A and B, while the right panel presents the cumulative distribution plots from four different perspectives.
		}
		\label{fig:example1}
	\end{figure}

\end{example}

\subsection{Existing Multivariate Approaches}

Three main approaches have been proposed to define MSD. 
(i) \textbf{Weighted Average Methods} convert multivariate distributions into univariate ones by aggregating dimensions through a weighted sum \citep{keeney1993decisions, marinoni2006benefits, mehdi2019stochastic}. Although simple, this reductionist strategy overlooks interdependencies and is sensitive to the choice of weights. 
(ii) \textbf{Marginal Stochastic Dominance} extends SD separately to each dimension, assuming dominance across all marginals \citep{lehmann1955ordered, post2005does}. Such methods often rely on additional assumptions, such as fixed dependence structures \cite{scarsini1988multivariate}, which may misrepresent joint behavior. 
(iii) \textbf{Copula-Based Methods} capture dependence structures explicitly \citep{joe1997multivariate, muller2001stochastic, decancq2010copula, siburg2024multivariate}. While flexible, they are computationally demanding and sensitive to copula specification and parameter estimation, especially in high dimensions.

In summary, existing MSD methods are limited by their neglect of dependence, restrictive assumptions, and poor scalability. These shortcomings necessitate a unified framework that efficiently captures complex dependence while providing theoretical robustness.

\subsection{Our Contributions}

Since classical SD is a quantile-based ordering in \(\mathbb{R}\), a natural path to MSD is via quantiles defined in \(\mathbb{R}^d\) (\(d \geq 2\)). This, however, presents two fundamental challenges:

First, the absence of a canonical ordering in \(\mathbb{R}^d\) complicates the definition of a quantile function. We adopt the solution of \cite{hallin2021distribution}, who leverage optimal transport (OT) theory to define a {\it center-outward quantile function} \(\Qb_{\pm}\) as the inverse of a {\it  center-outward distribution function} \(\Fb_{\pm}\), both of which inherit all the nice properties of their univariate counterparts.

Second, unlike the univariate case, a quantile in \(\mathbb{R}^d\) at a given level is usually a contour of dimension $d-1$. Consequently, even with a well-defined \(\Qb_{\pm}\), ordering these contours to define a practical MSD criterion remains a significant challenge.


This paper aims at tackling the second challenge by introducing novel MSD concepts that overcome the severe limitations of the existing ones. Specifically, we define first- and second-order MSD by comparing the average contribution of the mass over the $\Qb_{\pm}$-induced contours and regions, where the contribution is quantified via a function such as the distance from the origin. 

To conduct inference, we introduce two test statistics—Kolmogorov–Smirnov statistic and Cramér-von Mises statistic—and prove their consistency. Since the classical OT estimators suffer from slow convergence rates \citep{deb2021rates, gunsilius2022convergence, hutter2021minimax}, they are unsuitable for inference. We overcome this issue by employing entropy-regularized OT, which provides a computationally efficient approximation and an improved dimension-free \(N^{-1/2}\) converging rate \citep{goldfeld2024limit}. Furthermore, by using Hadamard differentiability \citep{fang2019inference, sun2021improved}, we establish bootstrap validity for the proposed tests.

Compared with previous methods such as \cite{rioux2024multivariate}, who use OT for hypothesis testing, our approach leverages OT to define multivariate quantiles themselves, yielding a unified notion of dominance that is both theoretically rigorous and computationally tractable. We also provide a data transformation method (Appendix~\ref{app.b}) applicable to general datasets.

The remainder of this paper is organized as follows. Section~2 introduces center-outward quantiles, defines multivariate FSD and SSD and establishes the theoretical properties. Section~3 develops the entropy-regularized OT approximation. Section~4 presents inference procedures and bootstrap validity results. Section~5 provides simulation results and real-data analysis. Proofs, additional numerical exercises and more details on OT are provided in the online supplementary material.

\section{Stochastic Dominance  in $\mathbb{R}^d$  ($d \geq 2$)}

In this section, we define multivariate stochastic dominance (MSD) in $\mathbb{R}^d$ and introduce corresponding  statistics with their asymptotic properties. To construct MSD, we first recall the center-outward quantile function proposed by \cite{hallin2021distribution}.

\subsection{Preliminary: center-outward quantile function}


Throughout, we let $\mathcal{P}_d(\mathcal{X})$ denote the family of all absolutely continuous probability measures $\mathrm P$ supported on a convex subset of $\mathcal{X}\subset\mathbb{R}^d$, with probability density function $f$. 
Let $\mathbb{S}^d$ and $\mathcal{S}^{d-1}$ denote the $d$-dimensional unit ball and the corresponding $d-1$-dimensional hypersphere, respectively. Denote by~${\rm U}_d$ the spherical uniform distribution on~${\mathbb S}^d$, that is, the product of a uniform measure on the hypersphere ${\mathcal S}^{d-1}$ and a uniform on the unit interval of distances to the origin. 

In the univariate case, the left-to-right canonical ordering on $\mathbb{R}$ allows for straightforward definitinos of the quantile and distribution functions. 
This direct extension to the multivariate setting is infeasible, since no canonical order exists in $\mathbb{R}^d$. As a generalization, the center-outward quantile function is defined through the measure transport between ${\rm P}$ on $\mathbb{R}^d$ and ${\rm U}_d$ on ${\mathbb S}^d$, with which an ordering according to the distance to the origin can be exploited. 
The formal definition is given below.

\begin{definition}
	{\rm For $\mathrm{P}\in\mathcal{P}_d$, the {\it center-outward distribution function}~$\mathbf{F}_{\pm}$ of $\rm P$ is the a.e. unique gradient of a convex potential $\phi: \mathbb{R}^d \to \mathbb{R}$ and $\mathbf{F}_{\pm} = \nabla \phi$ pushes $\mathrm{P}$ forward to ${\rm U}_d$. 
		The \emph{center-outward quantile function} \( \mathbf{Q}_{\pm} \) is defined as $\mathbf{Q}_{\pm} = \mathbf{F}_{\pm}^{-1}$.
	}
\end{definition}\label{def.QF}

The following result establishes the validity of $\mathbf{Q}_{\pm}$ as a quantile function. Part (i) is from \cite{mccann1995existence}, parts (ii) and (iii) are from \cite{hallin2021distribution}, 	and part (iv) is from \citet[Theorem 9.4]{villani2009optimal}.
\begin{proposition}\label{pro.0}
	Let $\mathrm{P} \in \mathcal{P}_d$, with center-outward quantile function $\mathbf{Q}_{\pm}$. Then,
	
	\begin{itemize}
		\item[(i)] \(\mathbf{Q}_{\pm}\) is the gradient of a convex function \(\psi\) and is unique \({\rm P}\)-almost everywhere;
		\item[(ii)] \(\mathbf{Q}_{\pm}\) serves as a probability integral transformation on \(\mathbb{R}^d\), meaning that \(\mathbf{S} \sim \mathrm{U}_d\) if and only if \(\mathbf{Q}_{\pm}(\mathbf{S}) \sim \mathrm{P}\);
		\item[(iii)] the set \(\mathbf{Q}_{\pm}(\mathbf{0})\) is compact and of Lebesgue measure zero. The restrictions of \(\mathbf{Q}_{\pm}\) to \(\mathbb{S}^d \setminus \{\mathbf{0}\}\) are homeomorphisms between \(\mathbb{R}^d \setminus \mathbf{Q}_{\pm}(\mathbf{0})\) and \(\mathbb{S}^d \setminus \{\mathbf{0}\}\). When \(d = 1, 2\), the set \(\mathbf{Q}_{\pm}(\mathbf{0})\) is a single point, and \(\mathbf{Q}_{\pm}\) becomes a homeomorphism between \(\mathbb{R}^d\) and \(\mathbb{S}^d\);
		\item[(iv)] if \(\mathrm{P}\) has finite second moment, then \(\mathbf{Q}_{\pm}\) of \(\mathrm{P}\) coincides with the \(L_2\)-optimal transport map from \(\mathrm{U}_d\) to \(\mathrm{P}\). Specifically, \(\mathbf{Q}_{\pm}\) is the a.e. unique solution to the optimization problem
		\begin{equation}\label{equ.otmap}
			\inf_{T} \int_{\mathbb{S}^d} \left\| T(\mathbf{s}) - \mathbf{s} \right\|^2 \, \mathrm{d}\mathrm{U}_d \quad \text{subject to} \quad T_{\#} \mathrm{U}_d = \mathrm{P},
		\end{equation}
		where $T\# \mathrm{U}_d = \mathrm{P}$ indicates that $T$ pushes $\mathrm{U}_d$ forward to $\mathrm{P}$, i.e., $(T\#\mathrm{U}_d)(B) = \mathrm{U}_d(T^{-1}(B))$ for any $B \in \mathcal{B}(\mathcal{Y})$, with $\mathcal{B}(\mathcal{Y})$ denoting the Borel $\sigma$-algebra on $\mathcal{Y}$.
	\end{itemize}
\end{proposition}


Since by \citet[Proposition 2.1]{hallin2021distribution}, 
$\|\mathbf{F}_{\pm}(\mathbf{Z})\|$ is uniform over $[0, 1)$, we can define, for \( p \in [0,1) \), the \textit{center-outward quantile region} and the \textit{center-outward quantile contour} of order \( p \) as $ \mathbb{C}(p) := \mathbf{Q}_{\pm} (p\mathbb{S}^d)$ and $\mathtt{C}(p) := \mathbf{Q}_{\pm} (p \mathcal{S}^{d-1}),$ respectively. It follows from Proposition~\ref{pro.0} that \( \mathbb{C}(p) \) satisfies \( \mathrm{P}(\mathbb{C}(p)) = p \), and that \( \mathtt{C}(p) \) forms the boundary of \( \mathbb{C}(p) \).
\begin{remark}
	{\rm Note that in Definition~\ref{def.QF}, no moment condition is imposed on \( {\rm P} \). When
		the second moment exists,  Proposition \ref{pro.0} (iv)  provides a feasible approach for computing the center-outward quantile function by solving an optimal coupling problem. More details on optimal transport is provided in Appendix~\ref{sec.OT}.}
\end{remark}


\subsection{MSD}\label{sec.3.1}

Univariate first-order stochastic dominance occurs when a distribution has more mass concentrated on the right side of the real line. We extend this principle to the center-outward framework by proposing a multivariate analogue: a dominating distribution is the one with more mass concentrated on quantile contours further from the origin.

This definition relies on a fundamental assumption: in any given direction, a greater distance from the origin implies a stronger contribution to dominance. This assumption reflects the natural relationship between distance and intensity and has been widely observed across disciplines. In economics, a larger deviation of quantity and price from equilibrium represents stronger market disequilibrium \citep{hendricks2014effect, blanchard1989dynamic}. In game theory, higher joint payoffs indicate more optimal strategic outcomes \citep{ichiishi1983game}. In physics, the electric field strength decays with distance \citep{griffiths2023introduction}. Similar relationships appear in machine learning, optimization, and engineering, where greater distance often corresponds to higher performance or stronger effects. This “distance–dominance” assumption therefore provides an intuitive and broadly applicable foundation for our subsequent definitions. For general data applications, a transformation procedure ensuring this assumption’s validity is presented in Appendix~\ref{app.b}.

In order to define MSD that also takes into account of uneven contributions of different spatial positions, we further introduce a context-dependent weight function \( \rho \).  In many contexts, such as income and health, once an indicator exceeds a threshold, further increases have limited impact, akin to poverty line truncation in univariate stochastic dominance \citep{barrett2014consistent}. This idea of diminishing returns also extends to other applications, where \(\rho\) can represent weighting in portfolio optimization, feature sensitivity in machine learning, or risk management in finance. 
Throughout, we assume \(\rho\) is non-negative and radially monotonic, i.e.,  for all $\mathbf{x} \in \mathbb{R}^d$ and $t_1 \geq t_2 \geq 0$, $\rho(\mathbf{x}) \geq 0$ and 
$\rho(t_1 \mathbf{x}) \geq \rho(t_2 \mathbf{x})$.  Throughout the paper, $\rho$ is chosen as the Euclidean norm, which treats all directions uniformly.

Now we are ready to define multivariate stochastic dominance. Let $\mathbf{X}$ and $\mathbf{Y}$ be two $d$-dimensional random variables, with distributions ${\rm P}_{\Xb}, {\rm P}_{\Yb} \in \mathcal{P}_d$, density \( f_{\mathbf{X}} \) and \( f_{\mathbf{Y}} \), and quantile contours $\mathtt{C}_{\mathbf{X}}(p)$ and $\mathtt{C}_{\mathbf{Y}}(p)$. The first- and second-order multivariate stochastic dominance are defined below.

\begin{definition}\label{def.msd1}
	{\rm	Define the {\it first-order contribution function} of $\Xb$ as
		\begin{equation} 
			\mathrm{M}_{\mathbf{X}}(p) = \dfrac{\int_{\mathtt{C}_{\mathbf{X}}(p)} \rho(\mathbf{x}) f_{\mathbf{X}}(\mathbf{x}) \, d\mathcal{H}^{d-1}(\mathbf{x})}{\int_{\mathtt{C}_{\mathbf{X}}(p)} f_{\mathbf{X}}(\mathbf{x}) \, d\mathcal{H}^{d-1}(\mathbf{x})},
		\end{equation}
		where \( \mathcal{H}^{d-1} \) denotes the \((d-1)\)-dimensional Hausdorff measure. 
		We say that \( \mathbf{X} \) {\it first-order stochastically dominates} \( \mathbf{Y} \), denoted as \( \mathbf{X} \succeq^1 \mathbf{Y} \), if \( \mathrm{M}_{\mathbf{X}}(p) \geq \mathrm{M}_{\mathbf{Y}}(p) \) for all \( p \in (0,1) \).}
\end{definition}

While quantile contours tend to lie further from the origin for distributions with greater dispersion, one might conjecture that first-order stochastic dominance is determined solely by variance. The following example demonstrates that this is not the case.



\begin{example}\label{example.2}
	\rm Let $\mathbf{X}_t$ be a random vector defined by 
	$\mathbf{X}_t = (tZ_1,Z_2)^\top$ if $Z_1>0$ and $\mathbf{X}_t=(Z_1,Z_2)^\top$ otherwise, 
	where $(Z_1,Z_2)\sim\mathcal{N}(\mathbf{0},\boldsymbol{I}_2)$. 
	Let $\mathbf{Y}\sim\mathcal{N}(\mathbf{0},\operatorname{diag}(4,1))$. 
	Then $\text{Cov}(\mathbf{X}_t)=\operatorname{diag}\!\left(\frac{t^2+1}{2}-\frac{(t-1)^2}{2\pi},1\right)$ and 
	$\text{Cov}(\mathbf{Y})=\operatorname{diag}(4,1)$. 
	When $t=2.9$, the first diagonal element of $\text{Cov}(\mathbf{X}_{2.9})$ is about $4.13>4$, so $\mathbf{X}_{2.9}$ has greater variance than $\mathbf{Y}$. 
	However, as illustrated in Appendix~\ref{appendix.d}, $\mathbf{Y} \succeq^1 \mathbf{X}$ holds.
\end{example}

Let $\mathbb{C}_{\mathbf{X}}(p)$ and $\mathbb{C}_{\mathbf{Y}}(p)$ be the center-outward $p$-th quantile regions  of $\mathbf{X}$ and $\mathbf{Y}$, respectively. Analogous to univariate second-order stochastic dominance, we define its multivariate counterpart by comparing the integrals of $\rho$ over these quantile regions.


\begin{definition}\label{def.msd2}
	{\rm Define the {\it second-order contribution function} of $\Xb$ as $\mathbb{M}_{\mathbf{X}}(p) = \int_{\mathbb{C}_{\mathbf{X}}(p)} \rho(\mathbf{x}) \, f_{\mathbf{X}}(\mathbf{x}) \, d\mathcal{H}^{d}(\mathbf{x}), $
		where \( \mathcal{H}^{d} \) denotes the \(d\)-dimensional Hausdorff measure.\footnote{Since the \(d\)-dimensional Hausdorff and Lebesgue measures only differ by a scaling factor, the Lebesgue measure can be used instead.} 
		We say that \( \mathbf{X} \) {\it second-order stochastically dominates} \( \mathbf{Y} \), denoted as \( \mathbf{X} \succeq^2 \mathbf{Y} \), if
		$\mathbb{M}_{\mathbf{X}}(p) \;\ge\; \mathbb{M}_{\mathbf{Y}}(p)$ for all $p \in (0,1)$.}
\end{definition}

\begin{remark}
	{\rm Because our construction relies on center‑outward quantiles, 
		both the average radial distances over quantile regions and their contours are invariant under such rotations. Consequently, the  multivariate stochastic dominance ordering is rotation‑invariant.}
	
	%
	\end{remark}

	By definition, first-order SD is contour-based, emphasizing localized dominance on specific quantile contours. Second-order SD, on the other hand, is region-based, assessing cumulative advantage by aggregating contributions over entire regions. Clearly, \( \mathbf{X} \succeq^1 \mathbf{Y} \) implies \( \mathbf{X} \succeq^2 \mathbf{Y} \).

	Our multivariate SD framework overcomes the challenge of multivariate ordering by introducing the quantile-contour/region-based contribution functions $\mathrm{M}_{\Xb}$ and $\mathbb{M}_{\Xb}$. Since the underlying center-outward quantile approach is nonparametric, the framework is well-suited for data with complex dependence structures.
	
	\subsection{Estimation of the contribution functions}
	
	
	Let $\Xb_1, \ldots, \Xb_N$ be i.i.d. copies of $\Xb$. In this section, we propose the empirical versions of $\mathrm{M}_\mathbf{X}$ and $\mathbb{M}_\mathbf{X}$, and we establish the corresponding asymptotics. For simplicity, we will omit subscripts (e.g., \( \mathbf{X} \) in \( \mathbb{M}_\mathbf{X} \)) when no confusion arises. 
	
	Let \( \widehat{\mathbf{Q}}_{\pm}^{(N)} \) denote the empirical quantile function defined on a regular grid \(\mathfrak{G}_N := \{\mathfrak{g}_i\}_{i=1}^N \); computation details for both are given in Appendix \ref{appendix.c}.  Estimation of $\mathrm{M}_\mathbf{X}$ relies on interpolation of $\widehat{\mathbf{Q}}_{\pm}^{(N)}$.  A smooth interpolation is provided by \cite{hallin2021distribution}, yielding continuous empirical center-outward quantile contours and regions. This interpolation procedure, however, is of high computational cost when $N$ is large. We offer an alternative perspective by noting that 
	\[
	\begin{aligned}
\mathrm{M}(p) &= \frac{\int_{\mathtt{C}(p)} \rho(\mathbf{x}) f(\mathbf{x}) \, d\mathcal{H}^{d-1}(\mathbf{x})}{\int_{\mathtt{C}(p)} f(\mathbf{x}) \, d\mathcal{H}^{d-1}(\mathbf{x})} \\
&= \frac{\int_{p\mathcal{S}^{d-1}} \rho(\mathbf{Q}_{\pm}(\mathbf{s})) f(\mathbf{Q}(\mathbf{s})) \det (J)  d\mathcal{H}^{d-1}(\mathbf{s})}{\int_{p\mathcal{S}^{d-1}} f(\mathbf{Q}_{\pm}(\mathbf{s})) \det(J) d\mathcal{H}^{d-1}(\mathbf{s})} \\
&= \frac{\int_{p \mathcal{S}^{d-1}} \rho(\mathbf{Q}(\mathbf{s})) g(\mathbf{s}) \, d\mathcal{H}^{d-1}(\mathbf{s})}{\int_{p \mathcal{S}^{d-1}} g(\mathbf{s}) \, d\mathcal{H}^{d-1}(\mathbf{s})}, 
\end{aligned}
\]
where $g$ is the density function of $\mathrm{U}_d$, and $J$ is the Jacobi matrix of $\mathbf{Q_{\pm}}$. Thus, the empirical version of \(\mathrm{M}(p) \) can be computed on the set $\left\{ \mathfrak{g}_{i} : \| \mathfrak{g}_{i}\| = p \right\}_{i=1}^N$. 
To facilitate both estimation and theoretical analysis, we employ the kernel regression method for estimating \(\mathrm{M}\). Given a bandwidth {\( b \in (0, 1)\)}, a uniform-kernel-based estimator\footnote{Although other widely-employed kernels are possible, here we consider the uniform kernel only for simplicity.} of \( \mathrm{M}(p) \) is 
\begin{equation}\label{equ.2}
\widehat{\mathrm{M}}_{b,N}(p) = \frac{1}{ \# (\mathtt{I}_p(b))} \sum_{i \in \mathtt{I}_p(b)} \rho(\widehat{\mathbf{Q}}_{\pm}^{(N)}(\mathfrak{g}_i)) ,
\end{equation}
where \( \mathtt{I}_p(b) = \left\lbrace i \in \{1,\ldots,N\} : p - b< \| \mathfrak{g}_i \| \leq p + b\right\rbrace \). 


Similarly, the empirical version of $\mathbb{M}(p)$ can be written as
\begin{equation}\label{equ.3}
\widehat{\mathbb{M}}_{N}(p) = \frac{1}{N} \sum_{i=1}^N \rho(\widehat{\mathbf{Q}}^{(N)}_{\pm}(\mathfrak{g_i}))\mathbf{I}
\left(  {i \in \mathtt{J}_p}\right),
\end{equation}
where $\mathtt{J}_p = \left\lbrace j \in \{1,\ldots,N\}: \| \mathfrak{g}_j \| \leq p \right\rbrace \).

The following mild assumptions are needed to establish the consistency of $\widehat{\mathrm{M}}_{h,N}$ and $\widehat{\mathbb{M}}_{N}$. 

\begin{assumption}\label{assump.1}
The probability density function \( f \) is continuous.
\end{assumption}

\begin{assumption}\label{assump.2}
The function \( \rho \) is  Lipschitz continuous and continuously differentiable.
\end{assumption}

\begin{assumption}\label{assump.3}
\( b \to 0 \) as $N \to \infty$. 
\end{assumption}

We can naturally define \( \mathrm{M}(1) = \lim_{q \to 1^{-}} \mathrm{M}(q) \) and \( \mathbb{M}(1) = \lim_{p \to 1^{-}} \mathbb{M}(p) \) if their limits exist. Proposition~\ref{pro.1} states that both $\widehat{\mathrm{M}}_{h,N}$ and $\widehat{\mathbb{M}}_{N}$ are uniform consistent.

\begin{proposition}\label{pro.1}
Under  Assumptions \ref{assump.1}, \ref{assump.2}, \ref{assump.3},  we have  $\underset{N \rightarrow \infty}{\lim} \underset{0<p \leq 1}{\sup}   \vert     \widehat{\mathrm{M}}_{b,N}(p)  -  \mathrm{M}(p)  \vert  \overset{a.s.}{ \rightarrow} 0. $
Under  Assumptions \ref{assump.1}, \ref{assump.2}, we have
$  \underset{N \rightarrow \infty}{\lim} \underset{0<p \leq 1}{\sup}   \vert     \widehat{\mathbb{M}}_{N}(p)  -  \mathbb{M}(p)  \vert  \overset{a.s.}{ \rightarrow} 0. $
\end{proposition}

The empirical quantile function is derived via barycentric projection of discrete optimal transport plans (see Appendix~\ref{appendix.c} for details).
The convergence rate of empirical optimal transport problems has been extensively studied in the literature. For instance, \cite{talagrand1994transportation} and \cite{barthe2013combinatorial} show that the Wasserstein distance \( W_2(\widehat{\mu}_N, \mu) \), where \( \hat{\mu}_N \) is the empirical version of \( \mu \), converges at rate \( N^{-1/d} \). Under milder assumptions, \cite{deb2021rates} established faster convergence rates of \( N^{-2/d} \). Building on this foundation, we provide the average convergence rate for $\widehat{\mathbb{M}}_{N}(p)$.

\begin{proposition}\label{pro.2}
Let Assumptions \ref{assump.1} and \ref{assump.2} hold. Additionally, assume that $\mathbf{Q}_{\pm}$ is $L$-Lipschitz continuous, and that the moment condition $\mathbb{E} \exp(t\|\mathbf{X}\|^\alpha) < \infty$ holds for some constants $t > 0$ and $\alpha > 0$.
Then, for any \(p \in (0,1)\), we have $\mathbb{E} \left| \widehat{\mathbb{M}}_{N}(p) - \mathbb{M}(p) \right| \leq C \sqrt{r_d^{(N)}} $
for some constant \(C > 0\), where
\[
r_d^{(N)} := \begin{cases}
	N^{-1/2} & \text{for } d = 2, 3, \\
	N^{-1/2} \log{(1+N)} & \text{for } d = 4, \\
	N^{-2/d} & \text{for } d \geq 5.
\end{cases}
\]
\end{proposition}

\begin{remark}
{\rm	Lipschitz  continuity of  $\mathbf{Q}_{\pm}$ is a standard assumption for deriving convergence rates of optimal transport maps \citep{gigli2011holder, hutter2021minimax}. Indeed, $\mathbf{Q}_{\pm}$ is Lipschitz  continuous when the underlying potential function satisfies some smoothness conditions.  Recall that $\mathbf{Q} = \nabla \varphi$ almost surely  for a convex function $\varphi$.  If  $\varphi$ is strongly convex, then there exists $C > 0$ such that $ \varphi(\mathbf{y}) \geq \varphi(\mathbf{x}) + \nabla \varphi(x)^\top (\mathbf{y - x}) + \frac{C}{2} \| \mathbf{y - x} \|^2.$ This implies that the gradient $\nabla \varphi$ is Lipschitz continuous with constant $L = C$ such that $ \| \nabla \varphi(\mathbf{x}) - \nabla \varphi(\mathbf{y}) \| \leq L \| \mathbf{x - y }\|.$ In our  setting in Section \ref{sec.emsd}, since distributions are supported on compact sets, twice differentiability of \(\varphi\) is sufficient for Lipschitz continuity of  \( Q \), as guaranteed by the Taylor expansion. A similar strong convexity condition can be found in \cite{ghosal2022multivariate}.
}	
\end{remark}

\begin{remark}
{\rm	Since \( \mathtt{C}(p) \) is a zero-measure set in $\mathbb{R}^d$, we cannot directly obtain similar conclusions for \( \widehat{\mathrm{M}}_{N} \).  However, letting   $	\widetilde{\mathrm{M}}_{b,N}(p) = \frac{1}{\#(\mathtt{I}_p(b))} \sum_{i=1}^N \rho(\mathbf{Q}_{\pm}(\mathfrak{g}_i)) \mathbf{I} \left( i \in \mathtt{I}_p(b) \right)$,
	then it can be shown that $\mathbb{E} \left| \widehat{\mathrm{M}}_{b,N}(p) - \widetilde{\mathrm{M}}_{b,N}(p) \right|$  has the same bounds as in Proposition \ref{pro.2}.
}
\end{remark}

\section{Entropic MSD}\label{sec.emsd}

While we have introduced first‑ and second‑order MSD and their empirical counterparts, and established uniform consistency and convergence rates for \( \widetilde{\mathrm{M}}_{h,N} \) and \( \widehat{\mathbb{M}}_{N} \), applying these results in practice requires an empirical OT map that converges at rate no less than $N^{-1/2}$. In high dimensions, empirical OT maps often fail to achieve this rate, since the rate deteriorates as \( d \) increases and depends on the smoothness of the underlying densities \citep{dowson1982frechet, santambrogio2015optimal, deb2021rates, gunsilius2022convergence}. Moreover, OT computation is prohibitively expensive, with classical algorithms exhibiting \( O(N^3 \log N) \) complexity \citep{peyre2019computational}, making them impractical for large-scale or high-dimensional applications.

This section addresses these challenges by proposing entropic first- and second-order contribution functions via entropy-regularized optimal transport, and establishes their asymptotic properties. We start from defining the entropic center-outward quantile function.

\subsection{Entropic center-outward quantile function}\label{sec.4}

For the entropic regularization framework, we assume the distributions have compact support. More general measures can be approximated by conditioning on sufficiently large compact sets. Let $\mathbf{X}$ and $\mathbf{Y}$ be random vectors with probability measures $\mu$ and $\nu$, and let $\Pi(\mu, \nu)$ be the set of all joint probability measures with marginals $\mu$ and $\nu$. The entropic Kantorovich problem aims to find $\pi \in \Pi(\mu, \nu)$ that minimizes the entropy‑regularized transport cost:
\begin{equation}\label{equ.6}
\inf_{\pi \in \Pi(\mu, \nu)} \left\{ \int_{\mathbb{R}^d \times \mathbb{R}^d} \frac{1}{2}\|\mathbf{x} - \mathbf{y}\|^2 d\pi(\mathbf{x}, \mathbf{y}) + \varepsilon D_{\mathrm{KL}}(\pi \| \mu \otimes \nu) \right\},
\end{equation}
where $D_{\mathrm{KL}}(\pi \| \mu \otimes \nu) := \int_{\mathbb{R}^d \times \mathbb{R}^d} \log\left(\frac{d\pi}{d(\mu \otimes \nu)}\right) d\pi(\mathbf{x}, \mathbf{y})$ is the Kullback-Leibler divergence, and the regularization parameter $\varepsilon > 0$ governs the trade-off between the original transport cost and entropic smoothing. Given a solution $\pi_\varepsilon$ to \eqref{equ.6}, following \cite{deb2021rates} and \cite{pooladian2021entropic}, the entropic optimal transport map is estimated via barycentric projection as
\begin{equation}\label{equ.11}
\mathbf{T}^\varepsilon(\mathbf{x}) := \mathbb{E}_{\pi_\varepsilon}[\mathbf{Y} \mid \mathbf{X} = \mathbf{x}] = \int \mathbf{y} \, \mathrm{d}\pi_\varepsilon^\mathbf{x}(\mathbf{y});
\end{equation}
see Appendix~\ref{sec.EOT} for an explicit form of $\mathbf{T}^\varepsilon(\mathbf{x})$.

We can define a similar concept of center-outward quantile based on the entropic optimal transport map. Specifically, we substitute \( \mathrm{P }\) and \(\mathrm{U}_d \) into \eqref{equ.6}, and then, based on \eqref{equ.11}, we obtain an entropic optimal transport map from  \(\mathrm{U}_d \)  to  \( \mathrm{P} \), denoted as \(\mathbf{Q}_{\pm}^\varepsilon\). We refer to this as the entropic quantile function.  Similarly, the entropic  version of  {center-outward quantile region} and {contour} of order $p$ can be defined as
$\mathbb{C}^{\varepsilon}(p) := \mathbf{Q}_{\pm}^{\varepsilon}(p \mathbb{S}^d)$ and $ \mathtt{C}^{\varepsilon}(p) := \mathbf{Q}_{\pm}^{\varepsilon}(p \mathcal{S}^{d-1}),$
respectively. We will demonstrate theoretically that \(\mathbf{Q}_{\pm}^\varepsilon\) is a reliable approximation to \(\mathbf{Q}_{\pm}\) (see Proposition~\ref{pro.4}).


Proposition \ref{pro.3} states that the entropic quantile regions are nested and simply connected, ensuring topological consistency within the regularization framework, as discussed in \cite{hallin2021distribution}. The proof further establishes the continuity of the regularized quantile map \(\mathbf{Q}_{\pm}^{\varepsilon}\). When restricted to a compact domain \(p\overline{\mathbb{S}}_d\), this map becomes a homeomorphism onto its image.

\begin{proposition}\label{pro.3} 
Given \( 0 < p_1 < p_2 < 1 \), the following strict inclusions hold: \(\mathbb{C}^{\varepsilon}(p_1)  \subsetneq \mathbb{C}^{\varepsilon}(p_2)\). Moreover, \(\mathbb{C}^{\varepsilon}(p)\) is simply connected. 
\end{proposition}



It is well-known that the entropic optimal transport cost converges to the unregularized cost as $\varepsilon \to 0$. In our setting, Proposition~\ref{pro.4} establishes both uniform convergence and probabilistic consistency of the entropic center-outward quantile and related quantities as $\varepsilon \to 0$, providing a solid theoretical justification for using entropic regularization.

\begin{proposition}\label{pro.4}
Given a sequence $\varepsilon_1, \varepsilon_2, \ldots, \varepsilon_k, \ldots$ that converges to $0$, the following results hold:

\begin{enumerate}
	\item[(i)]  $	\sup_{\mathbf{s} \in \mathbb{S}^d} \left| \mathbf{Q}_{\pm}^{\varepsilon_k}(\mathbf{s}) - \mathbf{Q}_{\pm}(\mathbf{s}) \right| \to 0 \quad \text{as } k \to \infty. $		
	\item[(ii)] For every $\mathbf{u} \in \mathbb{S}^d$ and $\epsilon > 0$, $\mathbb{P} \left( \mathbf{Q}_{\pm}^{\varepsilon_k}(\mathbf{u}) \notin \mathbf{Q}_{\pm}(\mathbf{u}) + \epsilon \mathbb{S}^d \right) \to 0 \quad \text{as } k \to \infty.$
	
	\item[(iii)] For every $p \in (0, 1)$ and $\epsilon > 0$, $ 	\mathbb{P} \left( \mathbb{C}^{\varepsilon_k}(p) \not\subset \mathbb{C}(p) + \varepsilon \mathbb{S}^d \right) \to 0 $ and $\mathbb{P} \left( \mathtt{C}^{\varepsilon_k}(p) \not\subset \mathtt{C}(p) + \epsilon \mathbb{S}^d \right) \to 0 $ as $ k \to \infty.$
\end{enumerate}
\end{proposition}


\subsection{Entropic dominance criteria}

Substituting \(\mathbf{Q}_{\pm}\) with \(\mathbf{Q}_{\pm}^{\varepsilon}\) in Definitions \ref{def.msd1} and \ref{def.msd2} yields entropic contribution functions and the corresponding definitions of stochastic dominance.

\begin{definition}\label{def.emsd}
The entropic versions of Definitions \ref{def.msd1} and \ref{def.msd2} are given as follows.
\begin{enumerate}
	\item[(i)] Define the \emph{first-order entropic contribution function} of $\Xb$ as ${\mathrm{M}}_{\mathbf{X}}^{\varepsilon} (p) = \dfrac{\int_{\mathtt{C}_{\mathbf{X}}^{\varepsilon}(p)} \rho(\mathbf{x})f(\mathbf{x}) \, d\mathcal{H}^{d}(\mathbf{x})}{\int_{\mathtt{C}_{\mathbf{X}}^{\varepsilon}(p)} f(\mathbf{x}) \, d\mathcal{H}^{d}(\mathbf{x})}.$ Similarly, define ${\mathrm{M}}_{\mathbf{Y}}^{\varepsilon}(p)$ for $\mathbf{Y}$. We say that $\mathbf{X}$ \emph{entropic first-order stochastically dominates} $\mathbf{Y}$, denoted $\mathbf{X} \succeq^{1,\varepsilon} \mathbf{Y}$, if for all $p \in (0,1)$, $ 	{\mathrm{M}}_{\mathbf{X}}^{\varepsilon}(p) \geq {\mathrm{M}}_{\mathbf{Y}}^{\varepsilon}(p).$
	\item[(ii)] Define the \emph{second-order entropic contribution function} of $\Xb$ as ${\mathbb{M}}^{\varepsilon}_\mathbf{X} (p) = \int_{\mathbb{C}^{\varepsilon}_\mathbf{X}(p)} \rho(\mathbf{x})f(\mathbf{x}) \, d\mathcal{H}^{d}(\mathbf{x}).$
	Similarly, define ${\mathbb{M}}^{\varepsilon}_\mathbf{Y}(p)$ for $\mathbf{Y}$. We say that $\mathbf{X}$ \emph{entropic second-order stochastically dominates} $\mathbf{Y}$, denoted $\mathbf{X} \succeq^{2,\varepsilon} \mathbf{Y}$, if for all $p \in (0,1)$, ${\mathbb{M}}^{\varepsilon}_\mathbf{X}(p) \geq {\mathbb{M}}^{\varepsilon}_\mathbf{Y}(p).$
\end{enumerate}
\end{definition}

For simplicity, we will omit subscripts (e.g., \( \mathbf{X} \) in \( \mathbb{M}^{\varepsilon}_\mathbf{X} \)) when no confusion arises. The following result establishes the uniform convergence of the entropic contribution functions to their unregularized counterparts as \(\varepsilon \to 0\), which justifies the use of the entropic ones in applications.


\begin{proposition}\label{pro.7}
For a sequence \(\varepsilon_1, \varepsilon_2, \ldots, \varepsilon_k, \ldots\) that converges to \(0\), we have 
$$ \underset{k \rightarrow \infty}{\lim} \underset{0<p\leq1}{\sup}   \vert     {\mathrm{M}}^{\varepsilon_k}(p)  -  \mathrm{M}(p)  \vert  =  0 \quad \text{and} \quad \underset{k \rightarrow \infty}{\lim} \underset{0<p\leq1}{\sup}   \vert     {\mathbb{M}}^{\varepsilon_k}(p)  -  \mathbb{M}(p)  \vert = 0 .$$ 
\end{proposition}

The empirical \(\widehat{\mathbf{Q}}^{ (\varepsilon, N)}_{\pm} \) is constructed by solving the discrete entropic Kantorovich problem  and  using the barycenter projection method. Details can be found in  Appendix \ref{appendix.c}. Replacing  \(\widehat{\mathbf{Q}}^{ ( N)}_{\pm} \)  with \(\widehat{\mathbf{Q}}^{ (\varepsilon, N)}_{\pm} \) in \eqref{equ.2} and \eqref{equ.3}, we obtain the empirical estimators of ${\mathrm{M}}^{\varepsilon} (p)$ and ${\mathbb{M}}^{\varepsilon} (p)$, denoted as  $ \widehat{\mathrm{M}}^{\varepsilon}_{b,N}(p) = \frac{1}{ \# (\mathtt{I}_p(b))} \sum_{i \in \mathtt{I}_p(b)} \rho(\widehat{\mathbf{Q}}^{ (\varepsilon, N)}_{\pm}(\mathfrak{g_i})) $ and $	\widehat{\mathbb{M}}^{\varepsilon}_{N}(p) = \frac{1}{N} \sum_{i=1}^N \rho(\widehat{\mathbf{Q}}^{ (\varepsilon, N)}_{\pm}(\mathfrak{g_i}))\mathbf{I}
\left(  {i \in \mathtt{J}_q}\right).$


Analogous to Proposition \ref{pro.1}, Proposition~\ref{pro.entrouni} establishes the uniform convergence of $\widehat{\mathrm{M}}^{\varepsilon}_{h,N}$ and $\widehat{\mathbb{M}}^{\varepsilon}_{N}$. The proof relies on the uniform convergence of the empirical entropic OT map, a result which is of independent interest and is stated as follows.


\begin{lemma}\label{lem.entromap}
Let $\mathcal{X}$ and $\mathcal{Y}$ be compact metric spaces, and let $\mu \in \mathcal{P}_d(\mathcal{X})$ and $\nu \in \mathcal{P}_d(\mathcal{Y})$. Suppose the cost function \(c: \mathcal{X} \times \mathcal{Y} \to (-\infty,\infty)\) is continuous and the regularization parameter \(\varepsilon > 0\) is fixed.
Denote by $\mathbf{T}^{\varepsilon}$ the entropic optimal transport map from $\mu$ to $\nu$, and by $\widehat{\mathbf{T}}^{\varepsilon}$ the empirical entropic map from $\widehat{\mu}_{N_1}$ to $\widehat{\nu}_{N_2}$.  
Then, 
\(
\sup_{\mathbf{x} \in \mathcal{X}} \|\widehat{\mathbf{T}}^{\varepsilon}(\mathbf{x}) - \mathbf{T}^{\varepsilon}(\mathbf{x})\| \to 0 
\) as $ N_1, N_2 \to \infty.$
\end{lemma}


\begin{proposition}\label{pro.entrouni}
Under  Assumptions \ref{assump.1}, \ref{assump.2}, \ref{assump.3},  we have $\underset{N \rightarrow \infty}{\lim} \underset{0<p\leq1}{\sup}   \vert     \widehat{\mathrm{M}}^{\varepsilon}_{ b,N}(p)  -  {\mathrm{M}}^{\varepsilon} (p) \vert  \overset{a.s.}{ \rightarrow} 0.  $
Under  Assumptions \ref{assump.1}, \ref{assump.2}, we have $ \underset{N \rightarrow \infty}{\lim} \underset{0<p\leq1}{\sup}   \vert     \widehat{\mathbb{M}}^{\varepsilon}_{N}(p)  -  {\mathbb{M}}^{\varepsilon} (p) \vert  \overset{a.s.}{ \rightarrow} 0. $ 
\end{proposition}

Based on the results from \cite{pooladian2021entropic}, we can establish the convergence rate of \( \widehat{\mathbb{M}}_{\mathbf{X},N}^{\varepsilon} \) to $ \mathbb{M}(p)$.

\begin{proposition}\label{pro.8}
Let $\psi$ and $\phi$ denote the optimal transport potentials (defined in \eqref{equb3} of Appendix \ref{sec.OT}) for $\mu = \mathrm{U}_d$ and $\nu = \mathrm{P}_{\mathbf{X}}$. Assume that $\psi \in \mathcal{C}^2(\mathbb{S}^d)$ and $\phi \in \mathcal{C}^{\alpha+1}(\mathcal{X})$ for some $\alpha > 1$. Define \(\varepsilon_{N} \asymp N^{-\frac{1}{d+\bar{\alpha}+1}}\), where \( \bar{\alpha} = \alpha \wedge 3 \). Then, for $p\in (0,1]$,  
$$
\mathbb{E} \left\vert \widehat{\mathbb{M}}_{ N}^{\varepsilon_N}(p) - \mathbb{M}(p) \right\vert \leq C \left( N^{-\frac{(\bar{\alpha}+1)}{2(d+\bar{\alpha}+1)}}\log N \right)^2,
$$
where \( C \) is a constant depending on $p$ .
\end{proposition}

\begin{remark}
{\rm	This proposition ensures that the gap between \( \widehat{\mathbb{M}}_{\mathbf{X},N}^{\varepsilon} \) and $ \mathbb{M}(p)$ is sufficiently small. In fact, \cite{goldfeld2024limit} provides the central limit theorem for entropic optimal potentials, based on which we will construct a testing procedure in the next section.}
\end{remark}


Proposition~\ref{pro.8} provides a convergence rate under the regularity conditions $\psi \in C^{2}$ and $\varphi \in C^{\alpha+1}$. In practice, such smoothness can be assessed via empirical diagnostics, such as numerically estimating derivatives and Hölder exponents from the data. More broadly, the assumptions can be relaxed to weaker function classes—for example, $\psi \in W^{2,p}$ and $\varphi \in W^{\alpha+1,p}$ for some $p>2$, or piecewise $C^{2}$ functions with bounded variation. Under these relaxed conditions, the qualitative consistency of the estimator remains valid, albeit with a potentially slower convergence rate. Quantifying the precise relationship between regularity strength and convergence speed offers a valuable direction for future research.

\section{Hypothesis test of MSD} \label{sec:hypothesis}


%
%

This section develops a hypothesis test for entropic dominance. At each probability level \( p \), we compare estimators of \( \mathrm{M}^{\varepsilon} \) and \( \mathbb{M}^{\varepsilon} \) to identify the dominance structure. A detailed implementation procedure is provided, with supporting theory.

For consistency with the non-entropic case, we fix \(\varepsilon\) sufficiently small. Our theory shows this ensures the entropic OT-based test approximates the non-entropic one. We then construct the test statistic and analyze its asymptotic properties.
We propose Kolmogorov-Smirnov and Cramér-von Mises type statistics. As the asymptotic distribution is intractable, we use a bootstrap method for inference and prove its validity. The testing problem is formally introduced below.

Consider two distributions \(\mathbf{X}\) and \(\mathbf{Y}\), and test whether \(\mathbf{X} \succeq^{1,\varepsilon} \mathbf{Y}\). The hypotheses can be stated as: 
$
H^1_0: \mathbf{X} \succeq^{1,\varepsilon} \mathbf{Y}$ and $H^1_1: \mathbf{X} \not\succeq^{1,\varepsilon} \mathbf{Y},
$
which correspond to $H^1_0: \mathrm{M}_1^{\varepsilon}(p) \geq \mathrm{M}_2^{\varepsilon}(p) \text{ for all } p \in (0,1) $ and $ H^1_1: \mathrm{M}_1^{\varepsilon}(p) < \mathrm{M}_2^{\varepsilon}(p) \text{ for some } p \in (0,1).$
Similarly, to test whether \(\mathbf{X} \succeq^{2,\varepsilon} \mathbf{Y}\), we state $H^2_0: \mathbf{X} \succeq^{2,\varepsilon} \mathbf{Y}$ and  $H^2_1: \mathbf{X} \not\succeq^{2,\varepsilon} \mathbf{Y}$, 
which correspond to $H^2_0: \mathbb{M}_1^{\varepsilon}(p) \geq \mathbb{M}_2^{\varepsilon}(p) \text{ for all } p \in (0,1) $ and $ H^2_1: \mathbb{M}_1^{\varepsilon}(p) < \mathbb{M}_2^{\varepsilon}(p) \text{ for some } p \in (0,1).$
Note that the alternatives \(H^1_1\) and \(H^2_1\) also include cases where no dominance relation exists between \(\mathbf{X}\) and \(\mathbf{Y}\).

\subsection{Construction of test statistics}


Let $\mathrm{T}^{\varepsilon}(p) = \mathrm{M}_2^{\varepsilon}(p) - \mathrm{M}_1^{\varepsilon}(p)$ and $\mathbb{T}^{\varepsilon} = \mathbb{M}_2^{\varepsilon}(p) - \mathbb{M}_1^{\varepsilon}(p)$. Given $N_1$ observations from $\mathbf{X}$ and $N_2$ observations  from $\mathbf{Y}$, their empirical counterparts are $\widehat{\mathrm{T}}^{\varepsilon}_{N_1,N_2}(p) = \widehat{\mathrm{M}}^{\varepsilon}_{N_2}(p) - \widehat{\mathrm{M}}^{\varepsilon}_{N_1}(p)$ and $\widehat{\mathbb{T}}^{\varepsilon}_{N_1,N_2}(p) = \widehat{\mathbb{M}}^{\varepsilon}_{N_2}(p) - \widehat{\mathbb{M}}^{\varepsilon}_{N_1}(p)$. When no confusion arises, we denote them simply as $\widehat{ \mathrm{T} }$ and $\widehat{ \mathbb{T} }$, respectively. Setting $N = N_1 + N_2$ and $r_N = N_1 N_2 / N$, we obtain the following result.

\begin{lemma}\label{lem.1}
Assuming $N_1,N_2 \rightarrow \infty$ and $\frac{N_1}{N_1+ N_2} \rightarrow \lambda \in (0, 1)$, then  
$$r_N^{1/2}(\widehat{ \mathrm{T} } -\mathrm{T})\rightsquigarrow\bar{\mathcal{J}}:=\lambda^{1/2}\mathcal{J}_2-(1-\lambda)^{1/2}\mathcal{J}_1 \quad \text{and} \quad  r_N^{1/2}(\widehat{ \mathbb{T}} -\mathbb{T})\rightsquigarrow\bar{\mathcal{K}}:=\lambda^{1/2}\mathcal{K}_2-(1-\lambda)^{1/2}\mathcal{K}_1,$$  
where $\mathcal{J}_2, \mathcal{J}_1,\mathcal{K}_1,\mathcal{K}_2 \in \mathcal{C}[0, 1]$.
\end{lemma}

Let  $\mathcal{C}[0,1]$ denote the space of continuous functions on $\left[0,1\right]$. We consider the test statistics introduced by \cite{barrett2014consistent}, which take the form  $ r_N^{1/2} \mathcal{F}(\widehat{\mathrm{T}})$ and $ r_N^{1/2} \mathcal{F}(\widehat{\mathbb{T}})$,
where $\mathcal{F}: \mathcal{C}[0,1] \rightarrow \mathbb{R}$ is a functional that, loosely speaking, measures the size of the positive part of $\widehat{\mathrm{T}}$ and $\widehat{\mathbb{T}}$. We assume the following property of $\mathcal{F}$.
\begin{assumption}\label{assump.4}
The functional $\mathcal{F}: \mathcal{C}[0,1] \rightarrow \mathbb{R}$ satisfies that for
any $h \in \mathcal{C}[0,1]$,
\begin{enumerate}
	\item[(i)] if $h(p) \leq 0$ for all $p \in [0,1]$ and $h(p) = 0$ for some $p \in [0,1]$, then $\mathcal{F}(h) = 0$;
	\item[(ii)] if $h(p) > 0$ for some $p \in (0,1)$, then $\mathcal{F}(h) > 0$.
\end{enumerate}
\end{assumption}

Under Assumption \ref{assump.4}, the null hypotheses $H^1_0$ and $H^2_0$ hold if and only if (iff) $\mathcal{F}(\widehat{\mathrm{T}}) = 0$ and $\mathcal{F}(\widehat{\mathbb{T}}) = 0$, respectively, while the alternatives  $H^1_1$ and $H^2_1$ hold iff $\mathcal{F}(\widehat{\mathrm{T}}) > 0$ and $\mathcal{F}(\widehat{\mathbb{T}})  > 0$.

As in \cite{sun2021improved} and \cite{barrett2014consistent}, we mainly focus on two specific choices of \(\mathcal{F}\), denoted by \(\mathcal{S}\) and \(\mathcal{I}\). For \( h \in \mathcal{C}[0,1] \), these functionals are defined as  
\[
\mathcal{S}(h) = \sup_{p \in (0,1]} h(p), \quad \text{and} \quad \mathcal{I}(h) = \int_0^1 \max\{h(p), 0\} \, \mathrm{d}p,
\]
which are commonly referred to as the Kolmogorov-Smirnov and Cramér-von Mises statistics, respectively. Clearly, both \(\mathcal{S}\) and \(\mathcal{I}\) satisfy Assumption \ref{assump.4}.


\cite{barrett2014consistent} observe that if $\mathcal{F}$ is monotone and positive homogeneous of degree one, then under the null hypothesis, the inequality
\(r_N^{1/2} \mathcal{F}(\hat{h}) \le \mathcal{F}\big(r_N^{1/2} (\hat{h} - h)\big)\)
holds uniformly, with equality if and only if $h = 0$. Lemma~\ref{lem.1} provides the weak convergence of $r_N^{1/2} (\hat{h} - h)$; applying the continuous mapping theorem then yields the asymptotic distribution of the right‑hand side, provided $\mathcal{F}$ is continuous. A bootstrap approximation of this limit distribution gives a critical value, ensuring the test's asymptotic size equals the nominal level when $h=0$, and never exceeds it elsewhere under the null. Although the distribution of $r_N^{1/2} (\hat{h} - h)$ can be consistently bootstrapped, the approach in Section~\ref{sec.asymptoticresults} directly approximates the upper quantiles of the limiting distribution of $r_N^{1/2} (\mathcal{F}(\hat{h}) - \mathcal{F}(h))$ rather than those of a conservative upper bound, with the aim of improving power.


\subsection{Asymptotic results for MSD test} \label{sec.asymptoticresults}

%
%
%

Building on recent advances in entropic optimal transport, dual potentials, and Sinkhorn divergence \citep{rioux2024multivariate}, together with the functional delta method for  differentiable maps \citep{kosorok2008introduction, fang2019inference}, we derive the asymptotic properties of the test statistics $\mathcal{S}$ and $\mathcal{I}$.
Following \cite{sun2021improved}, we use a bootstrap scheme based on the directional functional delta method to approximate the upper quantiles of the limiting distributions $\mathcal{F}_{\mathrm{T}}^{\prime}(\bar{\mathcal{J}})$ and $\mathcal{F}_{\mathbb{T}}^{\prime}(\bar{\mathcal{K}})$ under the following Assumption.

\begin{assumption}\label{assump.5}
The functional $\mathcal{F}: \mathcal{C}[0,1] \to \mathbb{R}$ is Hadamard directionally differentiable at $g \in \mathcal{C}[0,1]$, with directional derivative $\mathcal{F}_g^{\prime}: \mathcal{C}[0,1] \to \mathbb{R}$.
\end{assumption}
Although $\mathcal{S}$ and $\mathcal{I}$ are not fully Hadamard differentiable, they are Hadamard directionally differentiable, allowing the use of a directional functional delta method (The definitions of the two functional differentials can be found in Appendix~\ref{sec.appendixf}).

\begin{remark}\label{re.11}
{\rm 	Define the set  $ \Psi(g) = \arg\max_{p \in [0,1]} g(p).$
	Lemma S.4.9 of \cite{fang2019inference} establishes that the functional \(\mathcal{S}\) satisfies Assumption \ref{assump.5}, with the directional derivative given by $ 	\mathcal{S}_{g}^{\prime}(h) = \sup_{p \in \Psi(g)} h(p), $ where $ h \in \mathcal{C}[0,1].$}
	\end{remark}
	
	\begin{remark}\label{re.12}
{\rm 	Define the sets  $B_{0}(g) = \{p \in [0,1]: g(p) = 0\}  $ and $B_{+}(g) = \{p \in [0,1]: g(p) > 0\}.$
	Lemma S.4.5 of \cite{fang2019inference} establishes that the functional \(\mathcal{I}\) satisfies Assumption \ref{assump.5}, with the directional derivative given by $ 	\mathcal{I}_{g}^{\prime}(h) = \int_{B_{+}(g)} h(p) \mathrm{d}q + \int_{B_{0}(g)} \max\{h(p), 0\} \mathrm{d}q, $ where $ h \in \mathcal{C}[0,1].$}
	\end{remark}

	%
	%
	%
	%
	
	Remarks \ref{re.11} and \ref{re.12} indicate that directional derivatives exist for both statistics, implying the existence of their asymptotic forms. In fact, we can derive the following conclusion using the method of the delta function. By Theorem 2.1 in \cite{fang2019inference} and the continuous mapping theorem (Theorem 7.7 \cite{kosorok2008introduction}), we have the following. 
	\begin{proposition}\label{pro.9}
Under Assumptions \ref{assump.1}, \ref{assump.2}, \ref{assump.3}, \ref{assump.4} and \ref{assump.5},  as $N \rightarrow \infty$, we have $ 	r_N^{1/2} (\mathcal{F}(\widehat{\mathrm{T}}) - \mathcal{F}(\mathrm{T})) \rightsquigarrow \mathcal{F}_{\mathrm{T}}^{\prime}(\bar{\mathcal{J}}) \quad \mathrm{in~} \mathbb{R}$, and $ r_N^{1/2} (\mathcal{F}(\widehat{\mathbb{T}}) - \mathcal{F}(\mathbb{T})) \rightsquigarrow \mathcal{F}_{\mathbb{T}}^{\prime}(\bar{\mathcal{K}}) \quad \mathrm{in~} \mathbb{R}.$
\end{proposition}

Let $\mathbf{W}_{N_1} = (W_1^{\mathbf{X}}, \dots, W_{N_1}^{\mathbf{X}})$ and $\mathbf{W}_{N_2} = (W_1^{\mathbf{Y}}, \dots, W_{N_2}^{\mathbf{Y}})$ follow a multinomial distribution with parameters $N_1$ (resp. $N_2$) and uniform probabilities.
Each \( W_i^{\mathbf{X}} \) (resp. \( W_j^{\mathbf{Y}} \)) represents the number of times \( \mathbf{X}_i \) (resp. \( \mathbf{Y}_j \)) is selected in the bootstrap, with total counts $\sum_{i=1}^{N_1} W_i^{\mathbf{X}} = N_1$ and $\sum_{j=1}^{N_2} W_j^{\mathbf{Y}} = N_2. $
The bootstrap empirical distributions are defined as $ \mathrm{P}_{N_1}^{B} = \sum_{i=1}^{N_1} \frac{W_i^{\mathbf{X}}}{N_1} \delta_{\mathbf{X}_i}$ and $\mathrm{P}_{N_2}^{B} = \sum_{j=1}^{N_2} \frac{W_j^{\mathbf{Y}}}{N_2} \delta_{\mathbf{Y}_j}.$

To analyze the asymptotic behavior, we employ weak convergence conditional on the data, with bootstrap weights \( (\mathbf{W}_{N_1}, \mathbf{W}_{N_2}) \). { We use \( \overset{\rm P}{\underset{W}{\rightsquigarrow}} \)} to denote weak convergence conditional on the data in probability, as defined in \cite{kosorok2008introduction}.
Let \(\hat{\mathrm{T}}^B\) and \(\hat{\mathbb{T}}^B\) denote the statistics computed from the bootstrap empirical distributions \(\mathrm{P}_{N_1}^B\) and \(\mathrm{P}_{N_2}^B\), corresponding to \(\hat{\mathrm{T}}\) and \(\hat{\mathbb{T}}\), respectively.
The following result shows that the bootstrap processes $ r_N^{1/2}(\hat{\mathrm{T}}^B - \hat{\mathrm{T}})$ and  $r_N^{1/2}(\hat{\mathbb{T}}^B - \hat{\mathbb{T}})$
consistently approximate the weak limits of $ r_N^{1/2}(\hat{\mathrm{T}} - \mathrm{T})$ and $r_N^{1/2}(\hat{\mathbb{T}} - \mathbb{T}),$
respectively. This follows from the functional delta method for the bootstrap \citep{van1996weak}.

\begin{proposition}\label{pro.10} 
Under Assumptions \ref{assump.1}, \ref{assump.2}, \ref{assump.3}, and \ref{assump.4},  as \( N \to \infty \), $ 	r_N^{1/2}(\widehat{\mathrm{T}}^{B}-\widehat{\mathrm{T}}) \overset{\rm P}{\underset{W}{\rightsquigarrow}} \bar{\mathcal{J}} $ and  $r_N^{1/2}(\widehat{ \mathbb{T}}^{B}-\widehat{ \mathbb{T}}) \overset{\rm P}{\underset{W}{\rightsquigarrow}}\bar{\mathcal{K}} $ in $ \mathcal{C}[0,1].$
Moreover, the convergence holds with probability one if the empirical processes involved satisfy uniform tightness conditions.
\end{proposition}

The estimated functional \( \widehat{\mathcal{F}}'_{g} \) should consistently approximate the directional derivative \( \mathcal{F}'_{g} \). Naturally, the choice of \( \widehat{\mathcal{F}}'_{g} \) depends on the selection of \( \mathcal{F} \) used to construct the test statistic. 
We propose specific functionals \( \widehat{\mathcal{S}}'_{g} \) and \( \widehat{\mathcal{I}}'_{g} \) when \( \mathcal{F} \) is chosen as \( \mathcal{S} \) or \( \mathcal{I} \), respectively. These functionals depend on a positive tuning parameter \( \tau_N \) and a regularized estimator \( \widehat{V}(p) \) of the variance of \( r_N^{1/2} \widehat{g}(p) \), which will be discussed below.

We estimate \( \mathcal{S}'_{g} \) and \( \mathcal{I}'_{g} \) using the functionals $ \widehat{\mathcal{S}}^{\prime}_{g}(h) = \sup_{p \in \widehat{B_0}(g)} h(p)$ and  $\widehat{\mathcal{I}}^{\prime}_{g}(h) = \int_{\widehat{B_0}(g)} \max\{h(p), 0\} dp,$ where $ h \in \mathcal{C}[0,1].$
Here, the estimated contact set is defined as
\begin{equation}\label{equ.contact}
\widehat{B_0}(g) = \left\{ p \in [0,1] : |r_N^{1/2} \hat{g}(p)| \leq \tau_N\widehat{V}(p)^{1/2} \right\},
\end{equation}
where \( \widehat{V}(p) \) is an estimator of the variance of \( r_N^{1/2} \hat{g}(p) \).

\begin{remark}
{\rm	The estimated set \( \widehat{B_0}(g) \) is used to approximate both \( B_0(g) \) and \( \Psi(g) \) when \( g = \mathrm{T} \) or \( g = \mathbb{T} \). Under the null hypothesis, both \( \mathrm{T} \) and \( \mathbb{T} \) are zero, implying \( \Psi(g) = B_0(g) \), and thus a single estimator suffices. Furthermore, the set \( B_+(g) \), which appears in the Hadamard directional derivative of \( \mathcal{I} \), is always empty under the null hypothesis, eliminating the need for its estimation.}
\end{remark}	

To ensure numerical stability, we introduce a regularized variance estimator $ \widehat{V}(p) = \widetilde{V}(p) \lor \nu,$
where \( \widetilde{V}(p) \) is an estimate of the variance of \( r_N^{1/2} \hat{g}(p) \) and \( \nu \) is a small positive constant (e.g., \( \nu = 0.001 \)). Regularization prevents the threshold in \eqref{equ.contact} from approaching zero.
Although directly computing \( \widetilde{V}(p) \) can be complex, a bootstrap-based estimator \( V^B(p) \) may be used instead. The primary purpose of the variance estimate is to provide an appropriate scaling factor for the test statistic, rather than modifying the asymptotic distribution or affecting inferential conclusions. Therefore, a highly precise estimate is not required.

We present Assumption \ref{assump.6} along with its sufficient conditions:

\begin{assumption}\label{assump.6}
The estimated functional \( \widehat{\mathcal{F}}_{g}^{\prime}:\mathcal{C}[0,1] \to \mathbb{R} \) associated with \( \mathcal{F}_{g}^{\prime} \) satisfies the following property: for every compact set \( K \subseteq \mathcal{C}[0,1] \) and for any \( \varepsilon > 0 \), $ \mathbb{P}\left(\sup_{h\in K}\left|\widehat{\mathcal{F}}_{g}^{\prime}(h)-\mathcal{F}_{g}^{\prime}(h)\right|>\varepsilon\right) \to 0 \quad \text{as } N \to \infty.$
\end{assumption}

\begin{proposition}[Proposition 3.2 in \cite{sun2021improved}]\label{pro.11}
Suppose Assumption \ref{assump.5} satisfied and that \(\tau_N \to \infty\) and \(r_N^{-1/2} \tau_N \to 0\) as \(n \to \infty\).   Then the estimated functionals \(\widehat{\mathcal{S}}^{\prime}_g\) and \(\widehat{\mathcal{I}}^{\prime}_g\) satisfy the conditions imposed on \(\widehat{\mathcal{F}}^{\prime}_g\) in Assumption \ref{assump.6}.
\end{proposition}

\begin{proposition}\label{pro.12}
Under Assumptions  \ref{assump.1},\ref{assump.2},\ref{assump.3},\ref{assump.4},\ref{assump.5} and \ref{assump.6}, we have $ \widehat{\mathcal{F}}_\mathrm{T} \left( r_N^{1/2} (\widehat{\mathrm{T} }^B - \widehat{\mathrm{T} }) \right) 
\overset{\rm P}{\underset{W}{\rightsquigarrow}} 
\mathcal{F}'_\mathrm{T} (\bar{\mathcal{J}}) $
and $\widehat{\mathcal{F}}_\mathbb{T} \left( r_N^{1/2} (\widehat{\mathbb{T}  }^B - \widehat{\mathbb{T}  }) \right) 
\overset{\rm P}{\underset{W}{\rightsquigarrow}} 
\mathcal{F}'_\mathbb{T}  (\bar{\mathcal{K}})$ in $\mathbb{R}. $

\end{proposition}

Let \(\hat{c}^{\mathrm{T}}_{1-\alpha}\) and \(\hat{c}^{\mathbb{T}}_{1-\alpha}\) denote the \((1-\alpha)\)-quantiles of the bootstrap distributions of 
\(\widehat{\mathcal{F}}_{\widehat{\mathrm{T}}}\left(r_N^{1/2} (\widehat{\mathrm{T}}^B - \widehat{\mathrm{T}})\right)\)
and \( \widehat{\mathcal{F}}_{\widehat{\mathbb{T}}}\left(r_N^{1/2} (\widehat{\mathbb{T}}^B - \widehat{\mathbb{T}})\right)\), respectively.
In practice, \(\hat{c}_{1-\alpha}\) is approximated by computing the \([B(1-\alpha)]\)-th largest value among \(B\) independently generated bootstrap statistics, where \(N\) is chosen as large as computationally feasible. The decision rule for our test is as follows:
\[
\text{Reject } H^1_0 \, \text{ or }\,  H^2_0 \,\text{ if }\, r_N^{1/2} \mathcal{F}(\widehat{\mathrm{T}}) > \hat{c}^{\mathrm{T}}_{1-\alpha} \,\text{ or }\, r_N^{1/2} \mathcal{F}(\widehat{\mathbb{T}}) > \hat{c}^{\mathbb{T}}_{1-\alpha}.
\]

Using Proposition \ref{pro.4} and Theorem 3.10.11 in \cite{van1996weak}, the following result characterizes the asymptotic rejection probabilities of our test.

\begin{proposition}\label{pro.13}
Suppose Assumptions \ref{assump.1}, \ref{assump.2}, \ref{assump.3}, \ref{assump.4}, \ref{assump.5} and \ref{assump.6} hold:
\begin{itemize}
	\item[(i)] Under \(H^1_0\) or \(H^2_0\):
	\begin{itemize}
		\item If the contact set has an interior point (i.e.\ \(B_0(g) \) contains some \(p\in(0,1)\), then the CDF of \(\mathcal{F}^{\prime}_{\mathrm{T}}(\mathcal{J})\) and \(\mathcal{F}^{\prime}_{\mathbb{T}}(\mathcal{K})\)  is continuous and strictly increasing at its \(1-\alpha\) quantile, and we have
		\[
		\lim_{N \to \infty} \mathbb{P}\!\left(r_N^{1/2} \mathcal{F}(\widehat{\mathrm{S}}) > \hat{c}^{\mathrm{S}}_{1-\alpha}\right) = \alpha,
		\quad \text{where } \widehat{\mathrm{S}} \in \{\widehat{\mathrm{T}} - \mathrm{T}, \widehat{\mathbb{T}}- \mathbb{T}\}.
		\]
		\item If the contact set has no interior point, then the limit distribution degenerates at zero. 
	\end{itemize}
	\item[(ii)] Under \(H^1_1\) or \(H^2_1\):
	\[
	\lim_{N \to \infty} \mathbb{P}\!\left(r_N^{1/2} \mathcal{F}(\widehat{\mathrm{S}}) > \hat{c}^{\mathrm{S}}_{1-\alpha}\right) = 1,
	\quad \text{where } \widehat{\mathrm{S}} \in  \{\widehat{\mathrm{T}} - \mathrm{T}, \widehat{\mathbb{T}}- \mathbb{T} \}.
	\]
\end{itemize}
\end{proposition}
%
%
%

\begin{remark}
{\rm	When the contact set has no interior points, the test remains conservative. To address this, the critical value can be adjusted, e.g., using \( \max\{\hat c^{\mathrm{S}}_{1-\alpha}, \eta\} \) for small \( \eta > 0 \), ensuring asymptotic conservativeness. Proposition~\ref{pro.13} explains that in strict dominance cases, the test's empirical rejection rate may be zero because the test statistic's limiting distribution degenerates at zero, causing the probability of exceeding any non-degenerate critical value to approach zero.}
\end{remark}


\section{Experiments}\label{sec.6}
In this section, we present extensive numerical experiments, covering both synthetic and real-world datasets. A suite of Monte Carlo simulations is conducted to examine the finite-sample performance of the proposed tests. In all simulations, the entropic regularization parameter is fixed at $\varepsilon = 0.2$. Additional simulation results are provided in Appendix~\ref{appendix.d}. These include three‑dimensional experiments, the complete numerical studies for Examples~\ref{example.1} and~\ref{example.2}, and supplementary results for the experiments reported in this section.


\subsection{Experiments on the consistency of the critical value}\label{sec.6.1}

The first simulation study examines the consistency of the bootstrap-based critical values. Let $\mathbf{X}$ and $\mathbf{Y}$ be centered bivariate normal vectors, with $\mathbf{X}\sim\mathcal{N}((0,0)^\top, \bigl[\!\begin{smallmatrix}4&1\\1&\beta\end{smallmatrix}\!\bigr])$ and $\mathbf{Y}\sim\mathcal{N}((0,0)^\top, \bigl[\!\begin{smallmatrix}\beta&1\\1&4\end{smallmatrix}\!\bigr])$, where $\mathbf{Y}$ is obtained by a $90^\circ$ counterclockwise rotation of $\mathbf{X}$. The parameter $\beta$ takes values in $\{2,3,\ldots,7\}$. In this configuration, both null hypotheses $H_0^1$ and $H_0^2$ are satisfied.

In each iteration, we draw $N_1 = N_2 = 1200$ independent observations from $\mathbf{X}$ and $\mathbf{Y}$ respectively. Following the bootstrap procedure of \cite{sun2021improved} with 1000 bootstrap samples, we compute the critical values at nominal levels $\alpha = 0.05, 0.1, 0.2$. The empirical rejection rates are obtained from 1000 Monte Carlo replications. We compare three choices for the contact-set tuning parameter: $\tau_N = 1, 2, \infty$, with $\tau_N = \infty$ corresponding to the test of \cite{barrett2014consistent}. Tables~\ref{tab:order_1} and \ref{tab:order_2} report the first- and second-order results, respectively. The results indicate that for $\tau_N$ not too small, the bootstrap critical values exhibit satisfactory consistency, aligning with the findings of \cite{fang2019inference}.

We also present test results based on the original (unregularized) center-outward quantile method in Appendix~\ref{appendix.d1}. Since this method relies on classical optimal transport, it suffers from slower convergence rates. This not only leads to poorer finite-sample performance but also compromises the theoretical validity of bootstrap inference. Consequently, the original method exhibits unsatisfactory performance in several scenarios. 


\begin{table}[h]
\centering
\scalebox{0.70}{
	\begin{tabular}{c|c|cccccc|cccccc|cccccc}
		\toprule
		\cmidrule{3-20}
		& $\alpha$ & \multicolumn{6}{c|}{0.05} & \multicolumn{6}{c|}{0.1} & \multicolumn{6}{c}{0.2} \\
		\cmidrule{3-20}
		& $\beta$ 
		& 2 & 3 & 4 & 5 & 6 & 7 
		& 2 & 3 & 4 & 5 & 6 & 7
		& 2 & 3 & 4 & 5 & 6 & 7 \\
		\midrule
		\multirow{3}{*}{$\mathcal{S}$}
		& $\tau=$1       
		& 0.10 & 0.10 & 0.08 & 0.07 & 0.10 & 0.07
		& 0.16 & 0.14 & 0.13 & 0.11 & 0.14 & 0.11
		& 0.29 & 0.24 & 0.23 & 0.16 & 0.19 & 0.20 \\
		& $\tau=$2       
		& 0.06 & 0.06 & 0.06 & 0.03 & 0.08 & 0.05
		& 0.11 & 0.08 & 0.12 & 0.07 & 0.13 & 0.10
		& 0.22 & 0.21 & 0.22 & 0.15 & 0.18 & 0.14 \\
		& $\tau=$$\infty$ 
		& 0.05 & 0.05 & 0.06 & 0.03 & 0.07 & 0.05
		& 0.11 & 0.07 & 0.12 & 0.07 & 0.09 & 0.10
		& 0.22 & 0.18 & 0.22 & 0.14 & 0.18 & 0.14 \\
		\midrule
		\multirow{3}{*}{$\mathcal{I}$}
		& $\tau=$1        
		& 0.17 & 0.16 & 0.12 & 0.14 & 0.22 & 0.13
		& 0.22 & 0.20 & 0.16 & 0.16 & 0.26 & 0.18
		& 0.29 & 0.25 & 0.25 & 0.22 & 0.33 & 0.25 \\
		& $\tau=$2        
		& 0.07 & 0.06 & 0.05 & 0.06 & 0.10 & 0.07
		& 0.14 & 0.12 & 0.09 & 0.10 & 0.15 & 0.09
		& 0.25 & 0.22 & 0.22 & 0.19 & 0.22 & 0.19 \\
		& $\tau=$$\infty$ 
		& 0.04 & 0.04 & 0.05 & 0.03 & 0.07 & 0.05
		& 0.12 & 0.10 & 0.09 & 0.09 & 0.12 & 0.08
		& 0.24 & 0.22 & 0.21 & 0.18 & 0.21 & 0.19 \\
		\bottomrule
	\end{tabular}
}
\caption{Rejection rates under first-order multivariate stochastic dominance at various significance levels and tuning parameters}
\label{tab:order_1}
\end{table}

\subsection{Simulations under skew-$t$ and mixture normal distributions}\label{sec.t&mixture}
To evaluate the finite‑sample performance of the proposed multivariate first‑ and second‑order stochastic dominance tests, we conduct simulations under two representative distribution families: multivariate skew‑$t$ (featuring heavy tails and asymmetry) and Gaussian mixtures (exhibiting multimodality and non‑convex support). In each family, a scale parameter directly controls whether stochastic dominance holds. By examining the empirical rejection rates of $H^1_0$ and $H^2_0$ under the null and alternative regimes, we assess both the tests’ ability to maintain the nominal level when the null is true and their power to detect violations when it is false. The specific setups and results for each family are presented next. The reported average rejection rates are computed from 10000 Monte Carlo replications, where each replication uses 1000 bootstrap resamples.

\noindent\textbf{Skew-$t$ distributions.}
We generate centered multivariate skew‑$t$ distributions following \citep{azzalini2003distributions}. Let $\mathbf{S}_1$ be a skew‑$t$ vector with skewness $\boldsymbol{\alpha}=[2,1]^\top$, location $\boldsymbol{\xi}=[0,0]^\top$, scale matrix $\Sigb=\boldsymbol{I}_2$, and degrees of freedom $\nu=5$. A second distribution $\mathbf{S}_2$ is obtained by scaling the covariance to $c^2\Sigb$, with $c\in\{0.8,0.9,1.0,1.1,1.2\}$, while keeping $\boldsymbol{\alpha}$, $\boldsymbol{\xi}$ and $\nu$ unchanged. This yields a homothetic family of skew‑$t$ laws.

To obtain centered samples, we first draw a large Monte Carlo batch (50,000) to estimate the mean, then subtract this estimate from the final sample of size 2,400. This ensures the empirical distribution is approximately centered, as required by the center‑outward quantile construction.

First‑order and second‑order dominance of $\mathbf{S}_1$ over $\mathbf{S}_2$ holds precisely when $c \le 1$. Table~\ref{tab:rejection rates of skewt} reports the average rejection rates of $H^{1}_{0}: \mathbf{S}_1 \succeq^{1} \mathbf{S}_2$ and $H^{2}_{0}: \mathbf{S}_1 \succeq^{2} \mathbf{S}_2$ at the nominal level $\alpha = 0.1$. For $c<1$, the rejection rates are low, confirming that the test maintains size under the null. For $c>1$, the rejection probability rises, reflecting increasing power as the alternative strengthens. At $c=1$, the empirical rejection rate does not converge exactly to $0.1$; this mild discrepancy is likely attributable to the heavier tails and asymmetry of the skew‑$t$ distribution, which may require larger sample sizes for the asymptotic approximations to fully stabilize.

\begin{table}[htbp]
\centering
\begin{subtable}[t]{0.49\textwidth}
	\centering
	\begin{tabular}{c c c c c}
		\toprule
		$c$ & \multicolumn{2}{c}{$\mathcal{S}$} & \multicolumn{2}{c}{$\mathcal{I}$} \\
		\cmidrule(lr){2-3} \cmidrule(lr){4-5}
		& {$\tau_N = 2$} & {$\tau_N = \infty$} & {$\tau_N = 2$} & {$\tau_N = \infty$} \\
		\midrule
		0.8 & 0.017 & 0.042 & 0.008 & 0.133 \\
		0.9 & 0.022 & 0.117 & 0.019 & 0.192 \\
		1.0 & 0.116 & 0.222 & 0.144 & 0.222 \\
		1.1 & 0.239 & 0.400 & 0.267 & 0.406 \\
		1.2 & 0.239 & 0.450 & 0.267 & 0.406 \\
		\bottomrule
	\end{tabular}
	\caption{$H^1_0$}
\end{subtable}
\hfill
\begin{subtable}[t]{0.49\textwidth}
	\centering
	\begin{tabular}{c c c c c}
		\toprule
		$c$ & \multicolumn{2}{c}{$\mathcal{S}$} & \multicolumn{2}{c}{$\mathcal{I}$} \\
		\cmidrule(lr){2-3} \cmidrule(lr){4-5}
		& {$\tau_N = 2$} & {$\tau_N = \infty$} & {$\tau_N = 2$} & {$\tau_N = \infty$} \\
		\midrule
		0.8 & 0.008 & 0.283 & 0.008 & 0.283 \\
		0.9 & 0.000 & 0.033 & 0.000 & 0.033 \\
		1.0 & 0.083 & 0.167 & 0.083 & 0.158 \\
		1.1 & 0.458 & 0.583 & 0.517 & 0.583 \\
		1.2 & 0.350 & 0.683 & 0.350 & 0.683 \\
		\bottomrule
	\end{tabular}
	\caption{$H^2_0$}
\end{subtable}
\caption{Rejection rates of $H^1_0$ and $H^2_0$  at significance level $\alpha = 0.1$}
\label{tab:rejection rates of skewt}
\end{table}

\noindent \textbf{Mixture normal distribution.}
The first Gaussian mixture distribution $\mathbf{M}_1$ comprises two components with weights $p_1=0.4$ and $p_2=0.6$. Both components share the mean $\boldsymbol{\mu}_1=\boldsymbol{\mu}_2=\bigl[\begin{smallmatrix}0\\0\end{smallmatrix}\bigr]$, while the covariance matrices are $\boldsymbol{\Sigma}_1=\bigl[\begin{smallmatrix}1&0\\0&1\end{smallmatrix}\bigr]$ and $\boldsymbol{\Sigma}_2=\bigl[\begin{smallmatrix}2&0.5\\0.5&1\end{smallmatrix}\bigr]$. Hence $\mathbf{M}_1=\sum_{i=1}^2 p_i\,\mathcal{N}(\boldsymbol{\mu}_i,\boldsymbol{\Sigma}_i)$. A total of $1200$ samples are generated by first drawing component labels according to $\{p_i\}$ and then sampling from the corresponding Gaussian components. 

The second mixture $\mathbf{M}_2$ uses the same weights and means but modifies the first covariance matrix to $\boldsymbol{\Sigma}_1=\bigl[\begin{smallmatrix}\beta&0\\0&\beta\end{smallmatrix}\bigr]$, with $\beta\in\{1,2,3,4,5\}$, while keeping $\boldsymbol{\Sigma}_2=\bigl[\begin{smallmatrix}2&0.5\\0.5&1\end{smallmatrix}\bigr]$. Thus $\mathbf{M}_2=\sum_{i=1}^2 p_i\,\mathcal{N}(\boldsymbol{\mu}_i,\boldsymbol{\Sigma}_i)$, and its sample size is also $1200$.

The setup ensures that $\mathbf{M}_1$ dominates $\mathbf{M}_2$ in both first‑ and second‑order stochastic dominance only when $\beta = 1$. We report the empirical rejection rates of $H^1_0: \mathbf{M}_1 \succeq^{1} \mathbf{M}_2$ and $H^2_0: \mathbf{M}_1 \succeq^{2} \mathbf{M}_2$ at the nominal level $\alpha = 0.1$. Table~\ref{tab:rejection rates of mixnorm} displays the results for different values of $\beta$, for both statistics $\mathcal{S}$ and $\mathcal{I}$ and various choices of the tuning parameter $\tau_N$. When the null hypothesis holds ($\beta = 1$), the empirical rejection rates stay close to $0.1$; as $\beta$ increases, the rejection rates rise, reflecting the test’s power to detect violations of dominance. These findings align with the theoretical guarantee of Proposition~\ref{pro.13}.

\begin{table}[htbp]
\centering
\begin{subtable}[t]{0.48\textwidth}
	\centering
	\begin{tabular}{c c c c c}
		\toprule
		$\beta$ & \multicolumn{2}{c}{$\mathcal{S}$} & \multicolumn{2}{c}{$\mathcal{I}$} \\
		\cmidrule(lr){2-3} \cmidrule(lr){4-5}
		& {$\tau_N = 2$} & {$\tau_N = \infty$} & {$\tau_N = 2$} & {$\tau_N = \infty$} \\
		\midrule
		1 & 0.093 & 0.085 & 0.092 & 0.089 \\
		2 & 0.158 & 0.992 & 0.167 & 1.000 \\
		3 & 0.196 & 1.000 & 0.208 & 1.000 \\
		\bottomrule
	\end{tabular}
	\caption{$H^1_0$}
\end{subtable}
\hfill
\begin{subtable}[t]{0.48\textwidth}
	\centering
	\begin{tabular}{c c c c c}
		\toprule
		$\beta$ & \multicolumn{2}{c}{$\mathcal{S}$} & \multicolumn{2}{c}{$\mathcal{I}$} \\
		\cmidrule(lr){2-3} \cmidrule(lr){4-5}
		& {$\tau_N = 2$} & {$\tau_N = \infty$} & {$\tau_N = 2$} & {$\tau_N = \infty$} \\
		\midrule
		1 & 0.089 & 0.092 & 0.102 & 0.108 \\
		2 & 0.193 & 0.224 & 1.000 & 1.000 \\
		3 & 0.413 & 0.454 & 1.000 & 1.000 \\
		\bottomrule
	\end{tabular}
	\caption{$H^2_0$}
\end{subtable}
\caption{Rejection rates of $H^1_0$ and $H^2_0$  at significance level $\alpha = 0.1$}
\label{tab:rejection rates of mixnorm}
\end{table}

\subsection{Power Comparison}

To evaluate the power of the proposed test, we consider two centered bivariate normal vectors: $\mathbf{X}\sim\mathcal{N}((0,0)^\top, \bigl[\!\begin{smallmatrix}4&1\\1&\beta\end{smallmatrix}\!\bigr])$ and $\mathbf{Y}\sim\mathcal{N}((0,0)^\top, \bigl[\!\begin{smallmatrix}\beta&1\\1&\beta\end{smallmatrix}\!\bigr])$, with $\beta$ varying from $3$ to $5$ in steps of $0.5$. Here, $H_0^1$ and $H_0^2$ hold for $\beta \le 4$, whereas $H_1^1$ and $H_1^2$ hold for $\beta > 4$.

\begin{figure}[h]
\centering
\begin{subfigure}[b]{0.48\textwidth}
	\centering
	\includegraphics[width=\textwidth]{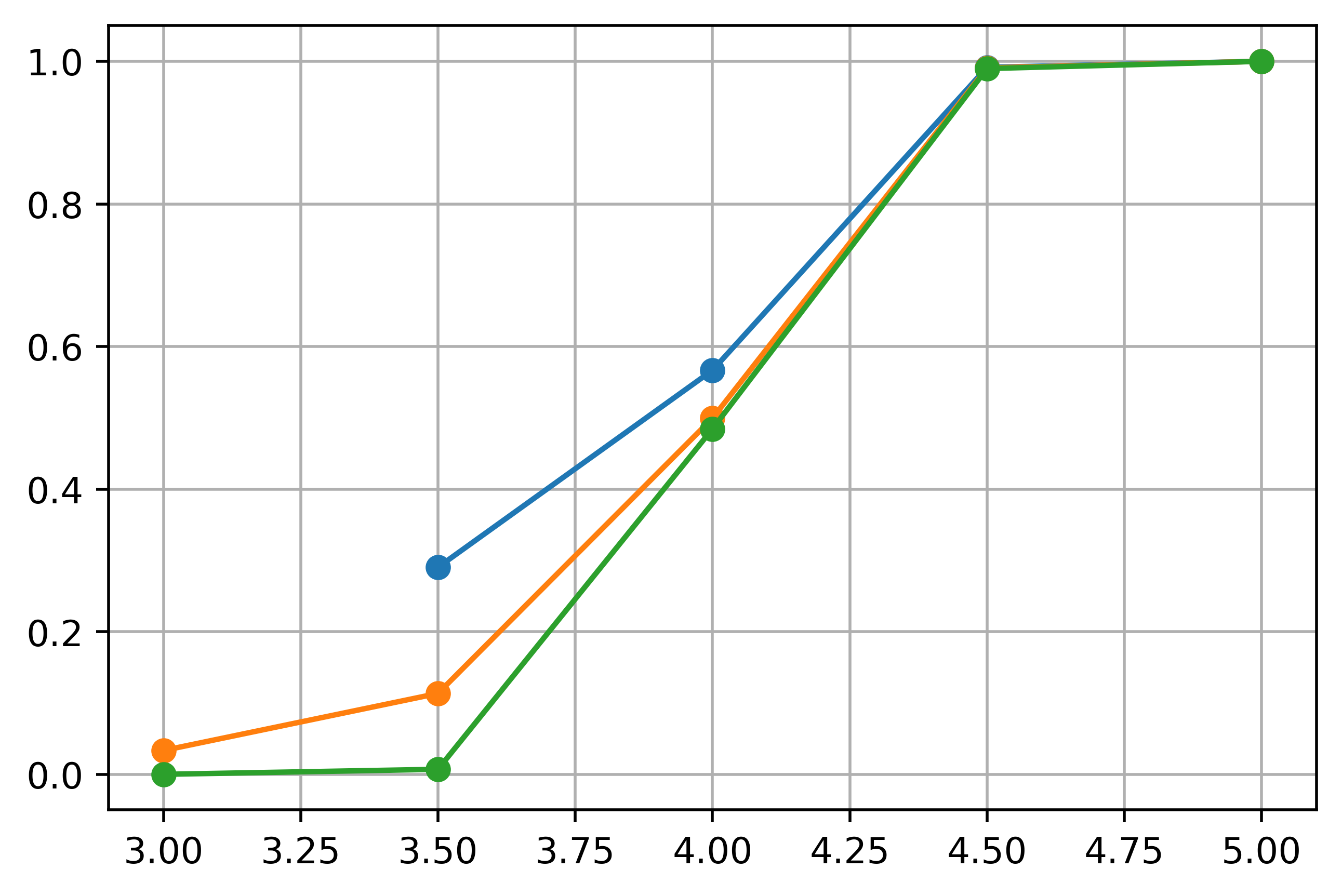}
\end{subfigure}
\begin{subfigure}[b]{0.48\textwidth}
	\centering
	\includegraphics[width=\textwidth]{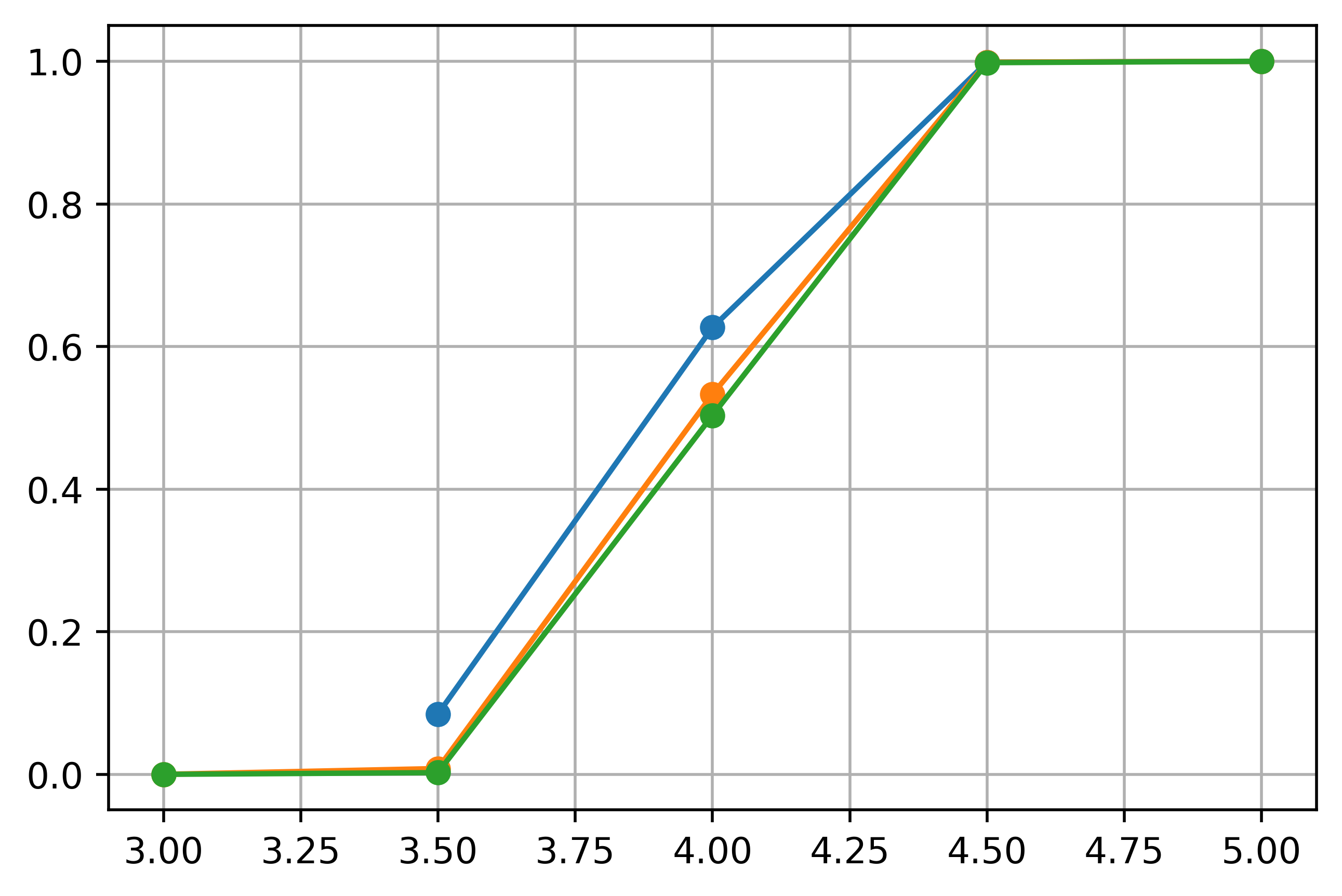}
\end{subfigure}
\\
\begin{subfigure}[b]{0.48\textwidth}
	\centering
	\includegraphics[width=\textwidth]{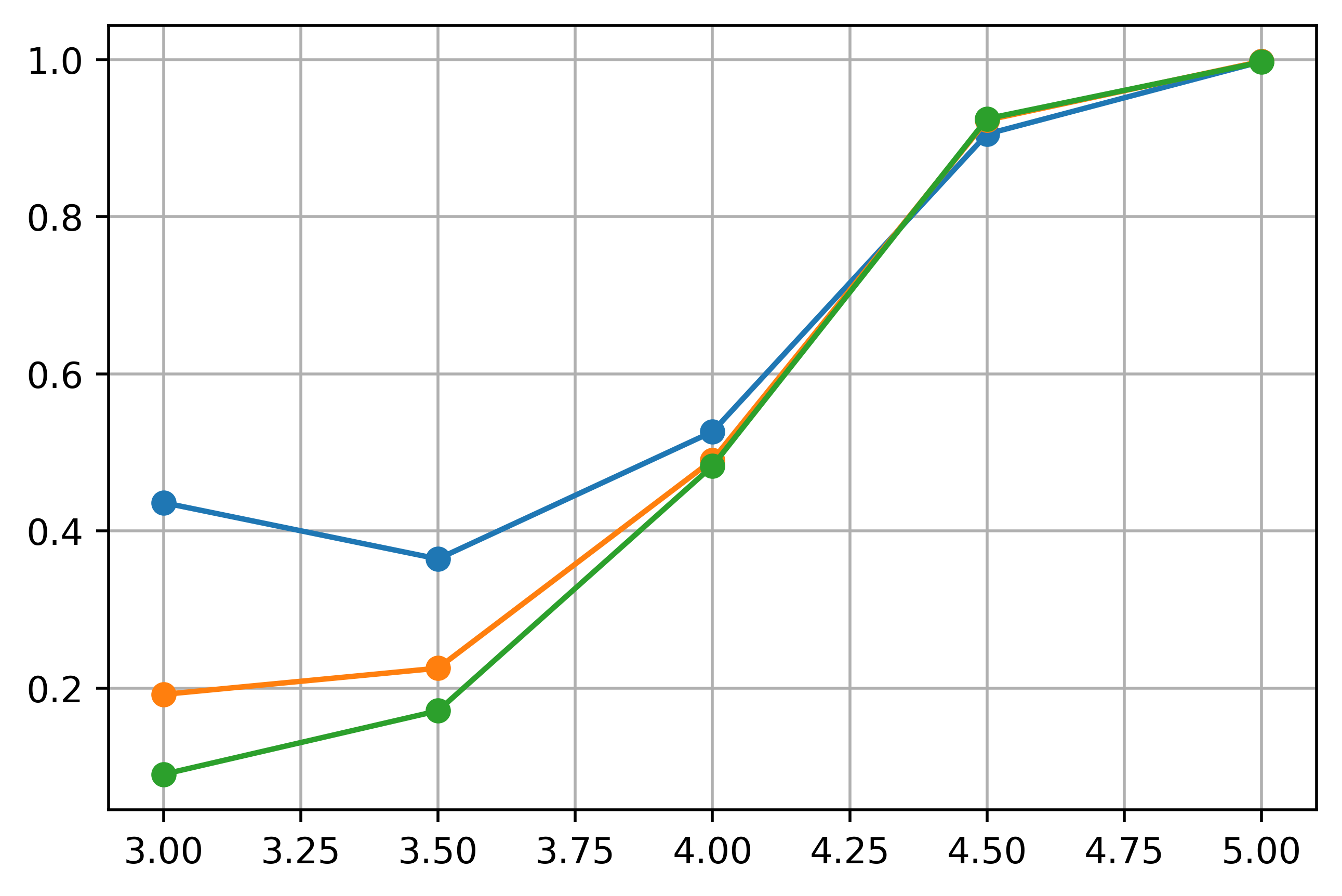}
\end{subfigure}
\begin{subfigure}[b]{0.48\textwidth}
	\centering
	\includegraphics[width=\textwidth]{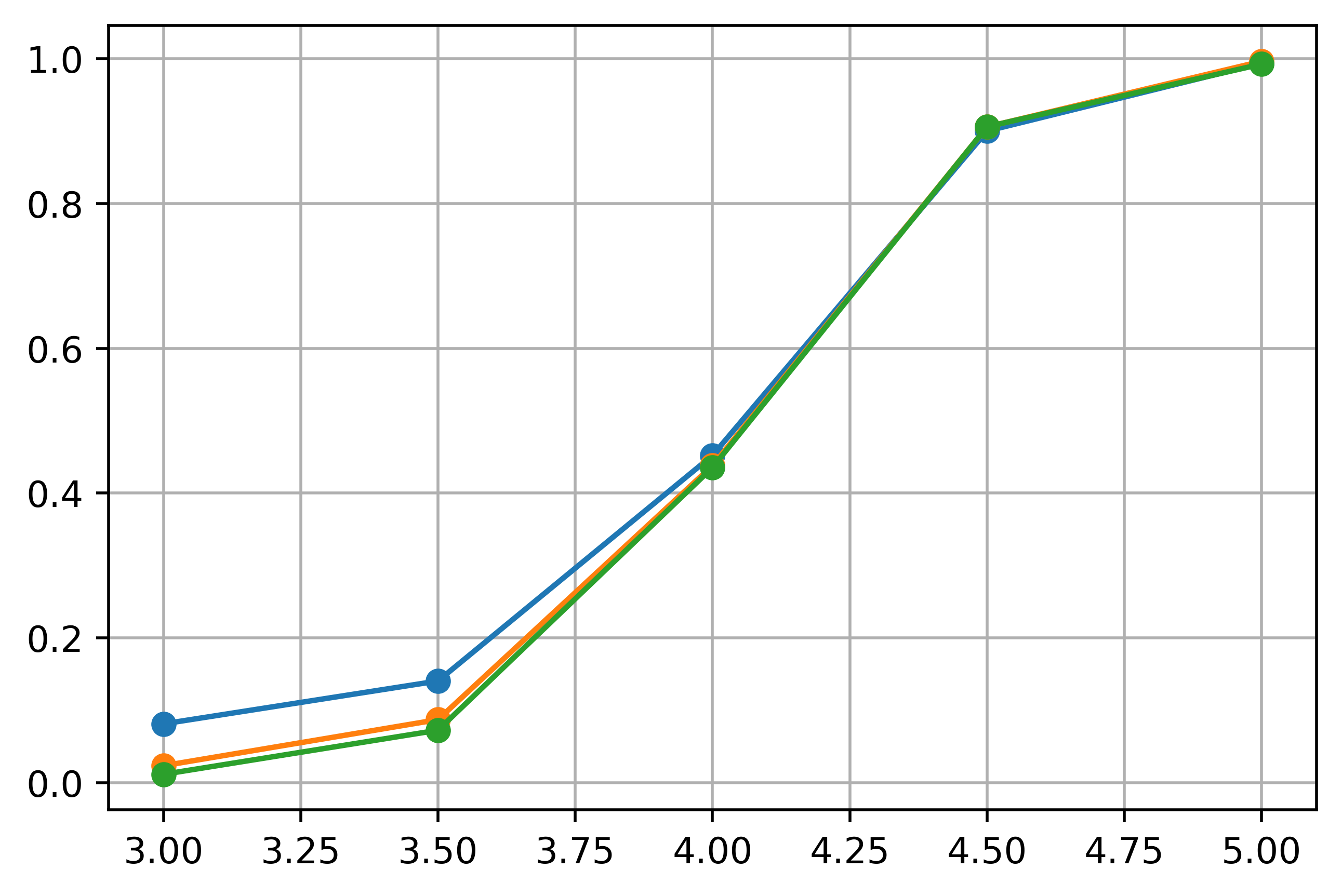}
\end{subfigure}

\caption{The first row shows the first‑order multivariate stochastic dominance results; the second row displays the second‑order results. The first column corresponds to the $\mathcal{S}$‑statistic, the second to the $\mathcal{I}$‑statistic. Within each panel, the blue, orange, and green lines represent the choices $\tau_N = 1$, $2$, and $\infty$, respectively.
}
\label{fig:power}
\end{figure}

In each simulation, we draw $N_1 = N_2 = 3000$ independent observations from each distribution. Power is estimated using 1000 bootstrap samples and averaged over 1000 Monte Carlo replications. Results are displayed in Figure~\ref{fig:power}. Reducing $\tau_N$ generally improves power; however, a very small $\tau_N$ (e.g., $\tau_N = 1$) can produce an empty contact set, rendering the test infeasible (see the missing entry for $\beta = 3$ in the first row of the figure). We therefore recommend a moderate choice such as $\tau_N = 2$, which yields stable performance for sample sizes up to 3000.

\subsection{Real Data Analysis}

The S\&P 500 and NASDAQ Composite indices represent distinct segments of the U.S. stock market: the S\&P 500 tracks the broad economy, whereas the NASDAQ is heavily weighted toward technology and high-growth sectors. While the two indices are positively correlated, the NASDAQ typically exhibits higher volatility due to its sector concentration, especially during periods of pronounced technology sector performance. The S\&P 500 and NASDAQ data used in this study are publicly available from Yahoo Finance (\url{https://finance.yahoo.com}).

\begin{figure}[h]
\centering
\includegraphics[width=0.95\textwidth]{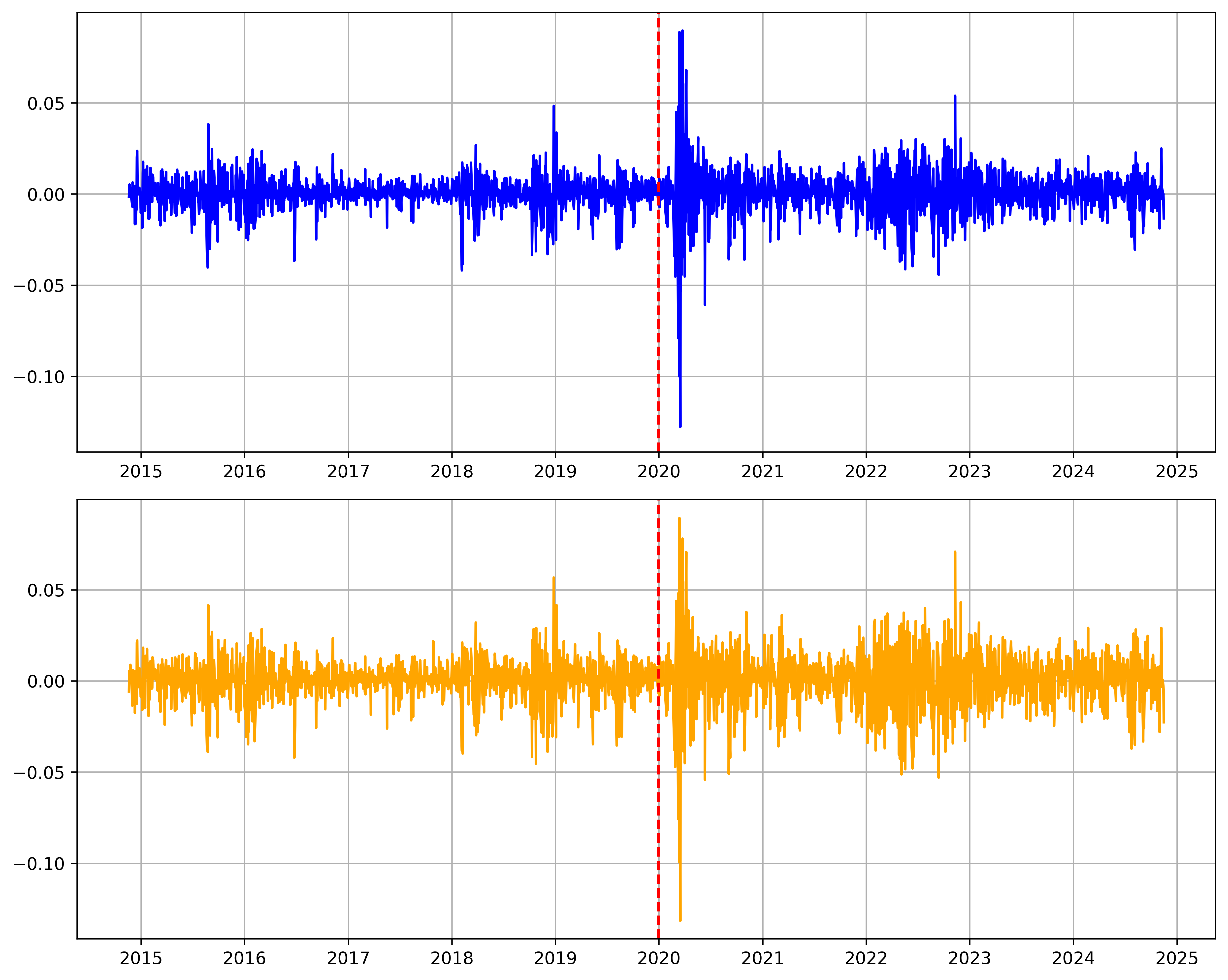}
\caption{Log return trends of the Nasdaq Composite Index (orange line) and the S\&P 500 Index (blue line) from 2014 to 2024, with the red dashed line indicating December 30, 2019.}
\label{fig:stocks}
\end{figure}

We collect daily closing prices of both indices from November 18, 2014, to November 18, 2024. Figure~\ref{fig:stocks} displays their log-return trajectories, which show clear comovement. To model their joint dynamics, we employ a vector autoregressive moving average (VARMA) specification, which captures lagged cross‑effects—for example, how technology‑driven moves in the NASDAQ propagate to the broader S\&P 500. The model is
$\Phi(\mathrm{L}) \mathbf{y}_t = \Theta(\mathrm{L}) \mathbf{e}_t,$
where $\mathbf{y}_t$ is the vector of log‑returns at time $t$, $\Phi(\mathrm{L}) = I - \Phi_1 \mathrm{L} - \cdots - \Phi_p \mathrm{L}^p$ and $\Theta(\mathrm{L}) = I + \Theta_1 \mathrm{L} + \cdots + \Theta_q \mathrm{L}^q$ are matrix polynomials in the lag operator $\mathrm{L}$, and $\mathbf{e}_t$ is a vector of innovations.

\begin{figure}[h]
\centering
\includegraphics[width=0.95\textwidth]{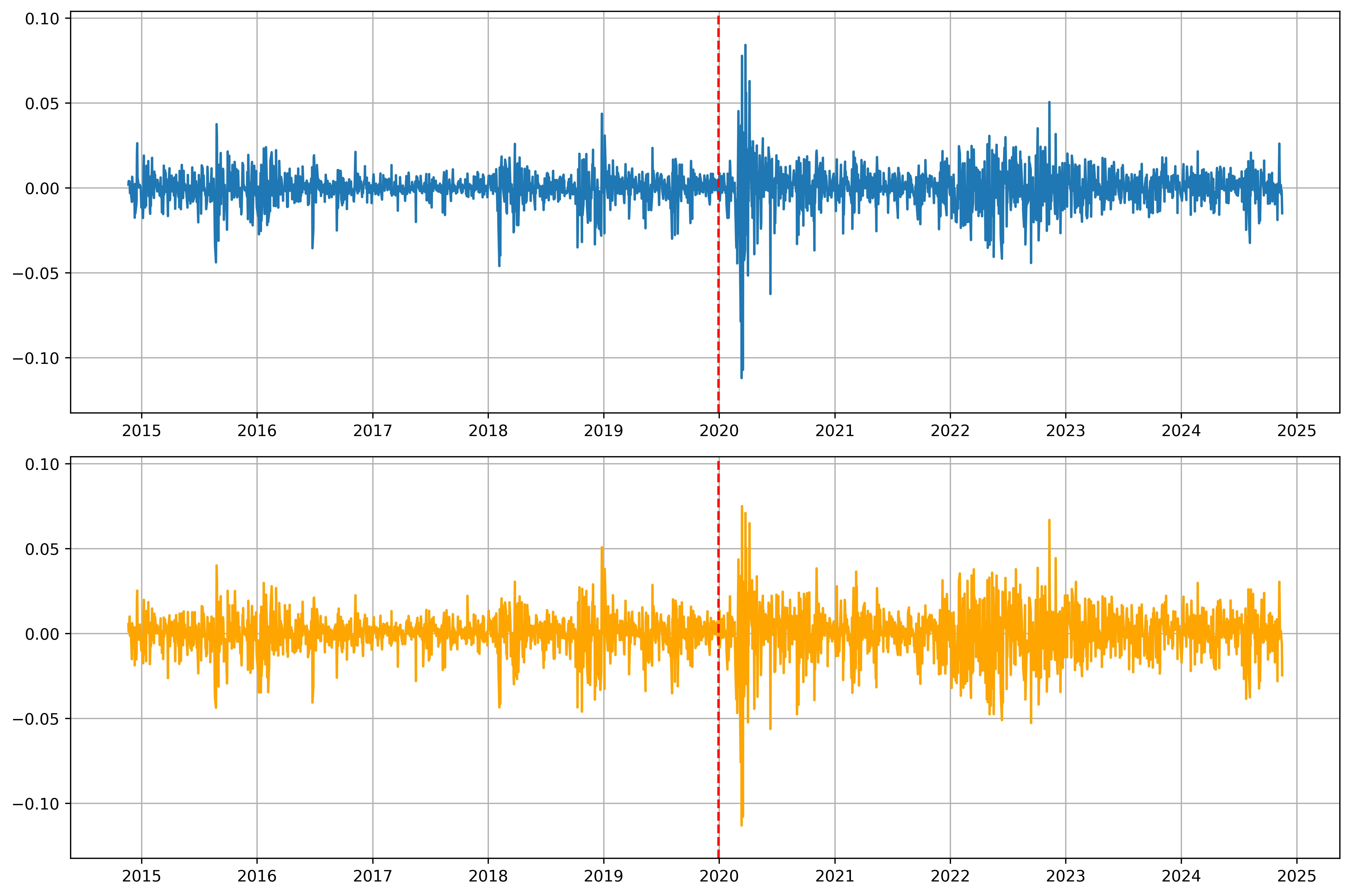}
\caption{Residuals of the Nasdaq Composite Index (orange line) and the S\&P 500 Index (blue line), with the red dashed line indicating December 30, 2019.}
\label{fig:resstocks}
\end{figure}

The Akaike Information Criterion selects $p=2$ and $q=0$, reducing the model to a VAR(2). We take December~31, 2019---the date China reported virus samples to the WHO---as a break point and compare the residual distributions over 1,200 trading days before and after this date (Figure~\ref{fig:resstocks}). Let $R_1$ and $R_2$ denote the residual distributions before and after the break, respectively. Table~\ref{tab:stocks} reports $p$-values based on 10000 bootstrap replications. Both the $\mathcal{S}$-test and $\mathcal{I}$-test fail to reject the hypothesis that $R_2$ first-order dominates $R_1$, while they reject the reverse hypothesis. The same conclusions hold for second-order dominance. Under our modeling assumptions, this indicates that both indices exhibited increased overall volatility after the pandemic.

\begin{table}[h]
\centering
\begin{tabular}{cc|cccc}
	\toprule
	& $\tau_N$   & $R_1  \succeq^1 R_2$ & $R_1  \succeq^2 R_2$ & $R_2  \succeq^1 R_1$ & $R_2  \succeq^2 R_1$ \\  
	\midrule
	\multirow{2}{*}{$\mathcal{S}$} 
	& 2 &  $\approx 0.006$ & $\approx 0$  & 1  & 1 \\  
	& $\infty$ & $\approx 0$ & $\approx 0$ & 1 & 1 \\  
	\midrule
	\multirow{2}{*}{$\mathcal{I}$} 
	& 2 & $\approx 0$  & $\approx 0$  & 1 & 1 \\  
	& $\infty$ & $\approx 0$  & $\approx 0$  & 1 & 1 \\  
	\bottomrule
\end{tabular}
\caption{$p$-values for different \(H_0\)}
\label{tab:stocks}
\end{table}

\section{Conclusion}

This paper proposes new concepts of first- and second-order multivariate stochastic dominance based on center-outward quantiles and contribution functions, overcoming key limitations of existing approaches. Through regularized approximations, we reduced computational costs, and we established rigorous asymptotic theory for the test statistics and resampling methods. Extensive numerical experiments illustrate good finite-performance of the test statistics.  Overall, this work provides a practical and theoretically sound framework for assessing dominance in multivariate data.

\section*{Acknowledgments}
This work was supported by the National Natural Science Foundation of China (grants 72571262 and 12401372).

\section*{Disclosure Statement}
The authors report there are no competing interests to declare.

\bigskip
\begin{center}

{\large\bf SUPPLEMENTARY MATERIAL}

\end{center}

\appendix
\date{}
\begin{abstract}
The file ``Supplementary Material for {\it Multidimensional Stochastic Dominance Test Based on Center-outward Quantiles}'' consists of five appendices. Appendix \ref{sec.appa} provides preliminary background on optimal transport theory, entropic regularization optimal transport, and Hadamard directional derivatives. Appendix \ref{sec.appb} contains proofs of the main theoretical results. Appendix \ref{app.b} discusses extensions to general distributions via transformations and background mixtures. Appendix \ref{appendix.c} details the estimation of center-outward quantiles. Appendix \ref{appendix.d} presents additional simulation studies.
\end{abstract}

\section{Preliminaries}\label{sec.appa}
\subsection{Optimal transport}\label{sec.OT}
Measure transport theory dates back to the seminal work of Monge \cite{monge1781memoire}, where he introduced a mathematical problem that, in modern terms, can be stated as follows. Let $\mathcal{X}$ and $\mathcal{Y}$ denote two Polish spaces with Borel $\sigma$- fields $\mathcal{B(X)}$ and $\mathcal{B(Y)}$, and let $\mathcal{P(X)}$ represent the set of all probability measures on $\mathcal{X}$. Given two probability measures $\mu \in \mathcal{P(X)}$ and $\nu \in \mathcal{P(Y)}$, along with a cost function $c: \mathcal{X} \times \mathcal{Y} \rightarrow [0, +\infty]$, Monge’s problem is to solve the following minimization problem:
\begin{equation} \label{equb1}
	\inf \left\lbrace \int_{\mathcal{X}} c(\mathbf{x}, T(\mathbf{x})) \, {\rm d} \mu : T\# \mu = \nu \right\rbrace,
\end{equation}
where, in measure transport terminology, $T\# \mu = \nu$ indicates that $T$ pushes $\mu$ forward to $\nu$, i.e., $(T\#\mu)(B) = \mu(T^{-1}(B))$ for any $B \in \mathcal{B(Y)}$.

The solution $T_0$ to Monge's problem (MP) \eqref{equb1} is called the \textit{optimal transport map}. Intuitively, $T_0$ describes how the mass of $\mu$ is transported to match the mass of $\nu$ with minimal cost.
However, Monge’s formulation has two main limitations. First, for some pairs of measures, no solution to \eqref{equb1} exists. For instance, if $\mu$ is a Dirac measure while $\nu$ is not, there is no transport map that can transform $\mu$ into $\nu$. Second, the set of all measurable transport maps $T_0$ that satisfy $T_0\# \mu = \nu$ is non-convex, making problem \eqref{equb1} computationally challenging.

Monge's problem was later reformulated by Kantorovich \cite{kantorovich}, who proposed a more flexible and computationally tractable version. Kantorovich’s approach allows for the mass to be split and recombined. Let $\Pi(\mu, \nu)$ denote the set of all joint probability measures on $\mathcal{X} \times \mathcal{Y}$ with marginals $\mu$ and $\nu$. Kantorovich’s problem seeks a joint distribution $\pi \in \Pi(\mu, \nu)$ that minimizes the expected transport cost:
\begin{equation} \label{equb2}
	\inf \left\lbrace \int_{\mathcal{X} \times \mathcal{Y}} c(\mathbf{x, y}) \, {\rm d}\pi(\mathbf{x, y}) : \pi \in \Pi(\mu, \nu) \right\rbrace.
\end{equation}


{A solution $\pi_0$ to Kantorovich’s problem (KP) \eqref{equb2} is known as an \textit{optimal transport plan}. Unlike Monge’s problem, Kantorovich’s formulation allows for mass splitting and is more general. Furthermore, the set $\Pi(\mu, \nu)$ is convex, and under mild conditions on $c$, such as lower semicontinuity, a solution to \eqref{equb2} exists \cite{villani2009optimal}. The relationship between optimal transport plans and maps was established by \cite{brenier1987decomposition} in the case where the cost function is the squared Euclidean distance; here, if $\mu$ is absolutely continuous, the optimal transport plan $\pi_0$ can be expressed as $(\mathrm{Id}, T_0)_{\#} \mu$. Thus, in this case, the optimal transport map can be uniquely determined by the Kantorovich formulation. We can first solve the Kantorovich problem and define a barycentric projection map as
	\begin{equation}\label{equ3}
		\mathbf{T}(\mathbf{x}) := \int \mathbf{y} \, \mathrm{d}\pi_0(\mathbf{y} \mid \mathbf{x}),
	\end{equation}
	which coincides with $T_0$ from the Monge problem $\mu$-a.e.\ when the above regularity conditions are satisfied.}

The dual form of Kantorovich’s primal minimization problem \eqref{equb2} can be expressed as the maximization problem
\begin{equation} \label{equb3}
	\begin{array}{ll}
		\sup \left\lbrace \int_{\mathcal{Y}} \phi \, {\rm d}\nu + \int_{\mathcal{X}} \psi \, {\rm d}\mu : \phi \in C_b(\mathcal{Y}), \, \psi \in C_b(\mathcal{X}) \right\rbrace \\
		\text{s.t.} \quad \phi(\mathbf{y}) + \psi(\mathbf{x}) \leq c(\mathbf{x, y}), \quad \forall (\mathbf{x, y}),
	\end{array}
\end{equation}
where $C_b(\mathcal{X})$ denotes the set of bounded continuous functions on $\mathcal{X}$. According to Theorem 5.10 in \cite{villani2009optimal}, if $c$ is lower semicontinuous, there is no duality gap between the primal and dual problems, so that the solutions to KD and KP coincide. In this case, the solution can be written as
\begin{equation} \label{equ4}
	\phi(y) = \inf_{x \in \mathcal{X}} [c(\mathbf{x, y}) - \psi(\mathbf{x})] \quad \text{and} \quad \psi(\mathbf{x}) = \inf_{y \in \mathcal{Y}} [c(\mathbf{x, y}) - \phi(\mathbf{y})],
\end{equation}
where the functions $\phi$ and $\psi$ are known as \textit{$c$-concave} and \textit{$c$-convex} functions, respectively, with $\phi$ (resp. $\psi$) referred to as the \textit{$c$-transform} of $\psi$ (resp. $\phi$).

\subsection{Entropic optimal transport}\label{sec.EOT}

Given the need for statistical inference based on $T_0$, achieving faster convergence is essential. To this end, we introduce entropic optimal transport (EOT), which improves statistical convergence rates while enhancing computational efficiency.

Building on recent theoretical advances in entropic regularization of optimal transport \cite{peyre2019computational}, we propose the following reformulation of \eqref{equb2}:

\begin{equation}\label{equ.6}
	\inf_{\pi \in \Pi(\mu, \nu)} \left\{ \int_{\mathbb{R}^d \times \mathbb{R}^d} \frac{1}{2}\|\mathbf{x} - \mathbf{y}\|^2 d\pi(\mathbf{x}, \mathbf{y}) + \varepsilon D_{\mathrm{KL}}(\pi \| \mu \otimes \nu) \right\},
\end{equation}
where the Kullback-Leibler divergence is defined as
\begin{equation}
	D_{\mathrm{KL}}(\pi \| \mu \otimes \nu) := \int_{\mathbb{R}^d \times \mathbb{R}^d} \ln\left(\frac{d\pi}{d(\mu \otimes \nu)}\right) d\pi(\mathbf{x}, \mathbf{y}),
\end{equation}
and the regularization parameter $\varepsilon > 0$ governs the trade-off between transport cost minimization and entropic smoothing.

This problem also admits a dual formulation (see \cite{genevay2019entropy}), which is a relaxed version of \eqref{equb3}:
\begin{equation}\label{equ.7}
	\mathcal{S}_{\varepsilon}(\mu,\nu) = \sup_{\substack{f \in L^1(\mu) \\ g \in L^1(\nu)}} \left\{ \int f \, {d}\mu + \int g \, {d}\nu - \varepsilon \iint e^{\left(f(\mathbf{x}) + g(\mathbf{y}) - \frac{1}{2} \|\mathbf{x - y}\|^2\right)/\varepsilon}{d}\mu {d}\nu + \varepsilon \right\},
\end{equation}
where $L^1(\mu)$ and $L^1(\nu)$ are the spaces of integrable functions with respect to $\mu$ and $\nu$, respectively.
Both \eqref{equ.6} and \eqref{equ.7}  possess solutions if $\mu$ and $\nu$ have finite second moments; moreover, if
we denote by $\pi_{\varepsilon}$ the solution to \eqref{equ.6}, which we call the optimal entropic plan, and $(\psi_{\varepsilon},\phi_{\varepsilon})$ the
solution to \eqref{equ.7}, which we call the optimal entropic potentials, then we obtain the optimality
relation (see \cite{csiszar1975divergence}):

\begin{equation}\label{equ.8}
	{d}\pi_\varepsilon(\mathbf{x,y}):=\exp((\psi_\varepsilon(\mathbf{x})+\phi_\varepsilon(\mathbf{y})-\frac12\|\mathbf{x-y}\|^2)/\varepsilon){d}\mu{d}\nu.
\end{equation}

A consequence of this relation is that we may choose optimal entropic potentials satisfying
\begin{gather} \label{equ.9} 
	\int e^{\frac{1}{\varepsilon} \left(\psi_{\varepsilon}(\mathbf{x}) + \phi_{\varepsilon}(\mathbf{y}) - \frac{1}{2} \|\mathbf{x - y}\|^2 \right)} \, \text{d}\mu = 1 \quad \forall \mathbf{y} \in \mathbb{R}^d \\
	\int e^{\frac{1}{\varepsilon} \left(\psi_{\varepsilon}(\mathbf{x}) + \phi_{\varepsilon}(\mathbf{y}) - \frac{1}{2} \|\mathbf{x - y}\|^2 \right)} \, \text{d}\nu = 1 \quad \forall \mathbf{x} \in \mathbb{R}^d. \label{equ.10}
\end{gather}

Given the optimal entropic plan $\pi_\varepsilon$ between measures $\mu$ and $\nu$, in accordance with \cite{deb2021rates} and \cite{pooladian2021entropic}, we estimate the entropic optimal transport map via barycentric projection as:
\begin{equation}\label{equ.11}
	\mathbf{T}^\varepsilon(\mathbf{x}) := \mathbb{E}_{\pi_\varepsilon}[\mathbf{Y} \mid \mathbf{X} = \mathbf{x}] = \int \mathbf{y} \, \mathrm{d}\pi_\varepsilon^\mathbf{x}(\mathbf{y}).
\end{equation}

Combing \eqref{equ.8}, \eqref{equ.9}, \eqref{equ.10} and \eqref{equ.11},  we have 

\begin{equation}
	\mathbf{T}^\varepsilon(\mathbf{x}):=\frac{\int \mathbf{\mathbf{y}}e^{\frac1\varepsilon(\phi_\varepsilon(y)-\frac12\|\mathbf{x-y}\|^2)} \mathrm{d}\nu}{\int e^{\frac1\varepsilon(\phi_\varepsilon(\mathbf{y})-\frac12\|\mathbf{x-y}\|^2)} \mathrm{d}\nu} .
\end{equation}

$\mathbf{T}^\varepsilon$ is called the entropic map between $\mu$ and $\nu$, and we must noted that $ \mathbf{T}^{\varepsilon }_{\#} \mu \neq \nu$. This is because entropy regularization makes the joint distribution smoother,  \(\pi_\varepsilon\) does not only take values on the set \(\{\mathrm{spt}(\mu) \times \mathbf{T}^\varepsilon(\mathrm{spt}(\mu))\}\).  As \(\varepsilon\) approaches 0, \cite{carlier2017convergence} shows that the joint distribution \(\pi_\varepsilon\) converges to \(\pi\) in the sense of weak topology and \cite{nutz2022entropic} provide strong compactness in $L_1 $ for $\phi_{\varepsilon}$ and $\psi_{\varepsilon}$. The convergence properties of \(\mathbf{T}^{\varepsilon}\)  is provided  in \cite{goldfeld2024limit}.

\subsection{Hadamard directional derivative}\label{sec.appendixf}

\begin{definition}[Hadamard derivative] \label{def.hadamard}
	Let \( G: V \to W \) be a mapping between normed spaces.  
	We say that \( G \) is \emph{Hadamard differentiable} at \( \xi \in V \) if there exists a continuous linear operator  
	\[
	G'_{\xi} : V \to W
	\]
	such that for any sequences \( t_n \to 0 \) and \( \zeta_n \to \zeta \in V \),  
	\[
	\lim_{n \to \infty}
	\frac{G(\xi + t_n \zeta_n) - G(\xi)}{t_n}
	= G'_{\xi}(\zeta).
	\]
	The operator \( G'_{\xi} \) is called the Hadamard derivative of \( G \) at \( \xi \).
\end{definition}

Standard accounts of the functional delta method \cite{kosorok2008introduction} require that $G$ is Hadamard differentiable when considered as a map from $\mathcal{C}[0, 1]$ to $\mathbb{R}$. \cite{fang2019inference} weakens this requirement: the functional delta method remains applicable even when the function is only Hadamard directionally differentiable.

\begin{definition} [Hadamard directional derivative] \label{def.hd}
	Let \( G: V \to W \) be a mapping between normed spaces \( V \) and \( W \).  
	Given \( \xi \in V \) and a direction \( \zeta \in V \), we say that \( G \) is Hadamard directionally differentiable at \( \xi \) in the direction \( \zeta \) if there exists a limit \( G^{\prime}_{\xi}(\zeta) \in W \) such that for any sequence \( t_n \to 0^{+} \) and any sequence \( \zeta_n \to \zeta \),  
	\[
	\lim_{n \to \infty} 
	\frac{G(\xi + t_n \zeta_n) - G(\xi)}{t_n} 
	= G^{\prime}_{\xi}(\zeta).
	\]
	The value \( G^{\prime}_{\xi}(\zeta) \) is called the Hadamard directional derivative of \( G \) at \( \xi \) in the direction \( \zeta \).
\end{definition}


Hadamard directional differentiability is weaker than Fréchet differentiability yet slightly stronger than Gateaux differentiability. It plays a central role in statistical inference on infinite-dimensional spaces, optimal transport theory, and bootstrap approximation. A distinguishing feature is that the perturbation sequence \( \zeta_n \) may converge to \( \zeta \) in a general manner rather than being restricted to the fixed direction \( \zeta \), providing additional flexibility for asymptotic analysis. For further details, we refer the reader to \cite{romisch2004delta,van1998asymptotic,van1996weak}.

\section{Proofs}\label{sec.appb}

\noindent{\bf Proof of Proposition \ref{pro.1}.} Let  $\mathtt{C}_{\mathbf{X},b}  =	\mathbf{Q}_{\pm} (\mathcal{S}_b(p))$, where $$\mathcal{S}_b(p)= \left\{ \mathbf{x} \in \mathbb{R}^d \mid (p - b) < \| \mathbf{x} \| \leq (p + b) \right\}.$$  
We define 
\begin{equation}\label{apeq.1}
	\widetilde{\mathrm{M}}_{b,N}(p) = \frac{1}{\#(\mathtt{I}_p(b))} \sum_{i \in \mathtt{I}_p(b)} \rho\left( \mathbf{Q}_{\pm}(\mathfrak{g}_i)\right).
\end{equation}

Then,
\[
\widehat{\mathrm{M}}_{b,N}(p)-\mathrm{M}(p) = \Bigl\{\widehat{\mathrm{M}}_{b,N}(p)-\widetilde{\mathrm{M}}_{b,N}(p)\Bigr\} + \Bigl\{\widetilde{\mathrm{M}}_{b,N}(p)-\mathrm{M}(p)\Bigr\}.
\]
The first term represents the error due to estimating $\mathbf{Q}_{\pm}$, and by the Lipschitz property of $\rho$ and uniform consistency of $\widehat{\mathbf{Q}}_{\pm}^{(N)}$ (according to Proposition 2.4 in \cite{hallin2021distribution}), it follows that as $N \rightarrow \infty$, one obtains

\[
\underset{0<p \leq 1}{\sup}  \Bigl|\widehat{\mathrm{M}}_{b,N}(p)-\widetilde{\mathrm{M}}_{b,N}(p)\Bigr| \overset{a.s.}{ \rightarrow} 0.
\]
Thus, the main task is to show that 
\[
 \sup_{0<p \leq 1} \Bigl|\widetilde{\mathrm{M}}_{b,N}(p)-\mathrm{M}(p)\Bigr| \stackrel{p}{\longrightarrow} 0  \quad (N \to \infty).
\]

By definition of $\mathrm{M}$ and continuity of quantile contours (see Section 2 in \cite{hallin2021distribution}), $\mathrm{M}$ is continuous and monotonically increasing. Therefore, $\mathrm{M}(p)$ is uniformly continuous with respect to $p$. 



The set $\{\mathfrak{g}_i\}_{i=1}^N$ in our construction forms a deterministic regular grid 
on the closed unit ball $\mathbb{S}_d$. 
Let $\widetilde{\mathrm{M}}_{b,N}^{\mathrm{(grid)}}(p)$ denote the statistic in \eqref{apeq.1} 
evaluated at these grid points.  
For comparison, let $\{\mathfrak{u}_i\}_{i=1}^N$ be i.i.d.\ from the product distribution 
$\mathbf{U}_d \times U[0,1)$, where $\mathbf{U}_d$ is the uniform distribution on the unit sphere 
and $U[0,1)$ the uniform distribution on $[0,1)$ for the radial part, and denote the corresponding 
version of the statistic by $\widetilde{\mathrm{M}}_{b,N}^{\mathrm{(iid)}}(p)$. 
By the regularity of the grid and the uniform continuity of $\rho(\mathbf{Q}_\pm(\cdot))$, 
a standard Riemann–sum argument yields
\[
\sup_{0<p\le 1} 
\big| \widetilde{\mathrm{M}}_{b,N}^{\mathrm{(grid)}}(p) 
- \widetilde{\mathrm{M}}_{b,N}^{\mathrm{(iid)}}(p) \big| = o(1) \quad (N \to \infty).
\]
Thus, without loss of generality, we may take $\{\mathfrak{g}_i\}_{i=1}^N$ 
to be i.i.d.\ draws from $\mathbf{U}_d \times U[0,1)$.

Next, we define ${\mathrm{M}}_{\mathbf{X},b}(p) = \dfrac{ \mathbb{E} \left[ \sum_{i \in \mathtt{I}_p(b)} \rho\left( \mathbf{Q}_{\pm}(\mathfrak{g}_i)\right) \right]}{\mathbb{E}\left[  \#(\mathtt{I}_p(b)) \right]}$. So that we can express ${\mathrm{M}}_{\mathbf{X},h}(p)$ as follows:
\[
{\mathrm{M}}_{\mathbf{X},b}(p) = \dfrac{\int_{\mathtt{C}_{\mathbf{X},b}(p)} \rho(\mathbf{x}) f_X(\mathbf{x}) \, d\mathcal{H}^{d-1}(\mathbf{x})}{2b}.
\]
By the mean value theorem for integrals, $ {\mathrm{M}}_{\mathbf{X},b}(p) $ converge to $\mathrm{M}(p)$ uniformly.

Now we have to prove that
\[
 \sup_{0<p \leq 1} \Bigl|\widetilde{\mathrm{M}}_{b,N}(p)- {\mathrm{M}}_{\mathbf{X},b}(p)  \Bigr| \stackrel{a.s.}{\longrightarrow} 0  \quad (N \to \infty).
\]

Define the numerator and denominator as
\[
N_N(p)=\frac{1}{N}\sum_{i=1}^N \rho\Bigl(\mathbf{Q}_{\pm}(\mathfrak{g}_i)\Bigr) \, \mathbf{1}\{p-b<\|\mathfrak{g}_i\|\le p+b\},
\]
\[
D_N(p)=\frac{1}{N}\sum_{i=1}^N \mathbf{1}\{p-b<\|\mathfrak{g}_i\|\le p+b\}.
\]
Their expectations are
\[
N(p)= \mathbb{E}\Bigl[\rho\Bigl(\mathbf{Q}_{\pm}(\mathfrak{g})\Bigr) \mathbf{1}\{p-b<\|\mathfrak{g}\|\le p+b\}\Bigr],
\]
\[
D(p)= \mathbb{E}\Bigl[\mathbf{1}\{p-b<\|\mathfrak{g}\|\le p+b\}\Bigr].
\]
Thus, we have
\[
\widetilde{\mathrm{M}}_{b,N}(p)=\frac{N_N(p)}{D_N(p)} \quad \text{and} \quad {\mathrm{M}}_{\mathbf{X},b}(p)=\frac{N(p)}{D(p)}.
\]
We define the class
\[
\mathcal{F} = \{ f_p(g)=\mathbf{1}\{p-b<\|g\|\le p+b\}: 0<p \leq 1\},
\]
and this class is $P$-Glivenko-Cantelli. By the Theorem 19.4 in \cite{van1998asymptotic}, we  have 
\[
 \sup_{0<p \leq 1} \Bigl|D_N(p)- D(p)  \Bigr| \stackrel{a.s.}{\longrightarrow} 0  \quad (N \to \infty).
\]

Define the weighted function class
\[
\mathcal{G} = \Bigl\{ g_p(g)=\rho\Bigl(\mathbf{Q}_{\pm}(g)\Bigr) \, \mathbf{1}\{p-h<\|g\|\le p+h\}: 0<p \leq 1\Bigr\}.
\]
similar arguments yield
\[
\sup_{0<p \leq 1} \Bigl|N_N(p) - N(p)\Bigr| \stackrel{a.s.}{\longrightarrow} 0 \quad (N \to \infty).
\]
Now we conclude that
\[
\sup_{0<p \leq 1} \Bigl|\widetilde{\mathrm{M}}_{b,N}(p)- {\mathrm{M}}_{\mathbf{X},b}(p)  \Bigr| \stackrel{a.s.}{\longrightarrow} 0  \quad (N \to \infty).
\]

For  the second part, we decompose the error as follows:

\[
\vert \widehat{\mathbb{M}}_N(p) - \mathbb{M}(p) \vert = \vert \widehat{\mathbb{M}}_N(p) - \widetilde{\mathbb{M}}_N(p)  \vert+  \vert  \widetilde{\mathbb{M}}_N(p) - \mathbb{M}(p)  \vert ,
\]
where
\[
\widetilde{\mathbb{M}}_N(p) = \frac{1}{N} \sum_{i \in \mathtt{J}_p} \rho\left(\mathbf{Q}_{\pm}^{(N)}(\mathfrak{g}_i)\right) .
\]
Since \(\rho(\cdot)\) is Lipschitz continuous , we have:

\[
\left| \widehat{\mathbb{M}}_N(p) - \widetilde{\mathbb{M}}_N(p) \right| \le L \cdot \sup_{ i } \left| \widehat{\mathbf{Q}}_{\pm}^{(N)}(\mathfrak{g_i}) - \mathbf{Q}_{\pm}(\mathfrak{g_i}) \right|.
\]
By Proposition 2.4 in \cite{hallin2021distribution}, we have
\[
\underset{0<p \leq 1}{\sup}  \Bigl|\widehat{\mathbb{M}}_{N}(p)-\widetilde{\mathrm{M}}_{N}(p)\Bigr| \overset{a.s.}{ \rightarrow} 0  \quad (N \to \infty).
\]

By the definition of  $\mathbb{M}$ and the continuity of the center-outward quantile regions, \(\mathbb{M}\) is continuous with respect to $p$ and bounded.
Since \(\mathbf{Q}_{\pm}(\mathfrak{g}_i), i = 1, \cdots, N,\) is an i.i.d. sequence, it follows from the large number law that as \(N \to \infty\), \(\widetilde{\mathbb{M}}_{N}(p)\) converges to \(\mathrm{M}(p)\) for a fixed \(p\) almost surely.

Given \(\eta > 0\), we can choose a partition \(0 = p_0 < p_1 < \ldots < p_k = 1\) such that \(\vert \mathbb{M}(p_i) - \mathbb{M}(p_{i-1}) \vert < \eta\) for each \(i\). The uniform convergence of \(\widetilde{\mathbb{M}}_{N}(p)\) over the finite set \(\{ p_1, p_2, \ldots, p_k \}\) is guaranteed.
Since \(\eta\) is arbitrary, we can   conclude that
\[
\sup_{1< p \leq 1} \vert \widetilde{\mathbb{M}}_{N}(p)- \mathbb{M}(p)\vert \overset{a.s.}{ \rightarrow} 0 \quad (N \to \infty).
\]
Since \(\eta\) is arbitrary and \(\widehat{\mathrm{M}}_{h,N}(p)\) converges uniformly to \(\tilde{\mathrm{M}}_{h,N}(p)\), this completes the proof for the first part. 
\qed

\vspace{1cm}

\noindent{\bf Proof of Proposition \ref{pro.2}.} We decompose the expectation as
\[
\mathbb{E}\bigl|\widehat{\mathbb{M}}_{N}(p)-\mathbb{M}(p)\bigr|
\le 
\mathbb{E}\bigl|\widehat{\mathbb{M}}_{N}(p)-\tilde{\mathbb{M}}_{N}(p)\bigr|
+
\mathbb{E}\bigl|\tilde{\mathbb{M}}_{N}(p)-\mathbb{M}(p)\bigr|.
\]

For the first term, Jensen’s inequality gives
\[
\bigl(\mathbb{E}\,|\widehat{\mathbb{M}}_{N}(p)-\tilde{\mathbb{M}}_{N}(p)|\bigr)^2
\le 
\mathbb{E}\,|\widehat{\mathbb{M}}_{N}(p)-\tilde{\mathbb{M}}_{N}(p)|^2.
\] 
Moreover,
\begin{align*}
	\mathbb{E}\,|\widehat{\mathbb{M}}_{N}(p)-\tilde{\mathbb{M}}_{N}(p)|^2
	&=
	\frac{1}{N^2}
	\mathbb{E}\Bigl|
	\sum_{i=1}^N
	\bigl[\rho(\widehat{\mathbf{Q}}^{(N)}_{\pm}(\mathfrak{g}_i))
	-
	\rho(\mathbf{Q}(\mathfrak{g}_i))\bigr]
	\mathbf{I}(i\in\mathtt{J}_p)
	\Bigr|^2 \\
	&\le
	\frac{1}{N^2}
	\mathbb{E}\Bigl|
	\sum_{i=1}^N
	\bigl[\rho(\widehat{\mathbf{Q}}^{(N)}_{\pm}(\mathfrak{g}_i))
	-
	\rho(\mathbf{Q}(\mathfrak{g}_i))\bigr]
	\Bigr|^2 \\
	&\le
	\frac{1}{N}\sum_{i=1}^N
	\mathbb{E}\bigl(\rho(\widehat{\mathbf{Q}}^{(N)}_{\pm}(\mathfrak{g}_i))
	-
	\rho(\mathbf{Q}(\mathfrak{g}_i))\bigr)^2.
\end{align*}

Since \(\rho\) is Lipschitz with constant \(L_1\), we have
\begin{align*}
	\frac{1}{N}\sum_{i=1}^N
	\mathbb{E}\bigl(\rho(\widehat{\mathbf{Q}}^{(N)}_{\pm}(\mathfrak{g}_i))
	-
	\rho(\mathbf{Q}(\mathfrak{g}_i))\bigr)^2
	&\le
	\frac{L_1^2}{N}
	\sum_{i=1}^N
	\mathbb{E}\bigl(\widehat{\mathbf{Q}}^{(N)}_{\pm}(\mathfrak{g}_i)
	-
	\mathbf{Q}(\mathfrak{g}_i)\bigr)^2 \\
	&=
	L_1^2\,
	\mathbb{E}\bigl(\widehat{\mathbf{Q}}^{(N)}_{\pm}(\mathfrak{g}_1)
	-
	\mathbf{Q}(\mathfrak{g}_1)\bigr)^2,
\end{align*}
where the last equality uses the i.i.d.\ property of 
\(\widehat{\mathbf{Q}}^{(N)}_{\pm}(\mathfrak{g}_i)\) and \(\mathbf{Q}(\mathfrak{g}_i)\).

By Corollary 2.3 in \cite{deb2021rates}, we obtain
\[
\mathbb{E} \left| \widehat{\mathbb{M}}_{N}(p) - \tilde{\mathbb{M}}_{N}(p) \right|^2 \leq C_1 r_d^{N},
\]
where \[ r^{(N)}_{d}=
\begin{cases}
	N^{-1/2},& d=2,3,\\[2pt]
	N^{-1/2}\log(1+N),& d=4,\\[2pt]
	N^{-2/d},& d\ge 5.
\end{cases} \]

Furthermore, since all \(\mathbf{Q}_{\pm}(\mathfrak{g}_i)\) are i.i.d., the variance of the estimator \(\tilde{\mathbb{M}}_{N}(p)\) for the population parameter \(\mathbb{M}(p)\) decreases at a rate of \(1/\sqrt{N}\). Based on properties of sample variance and convergence rates, we have
\[
\mathbb{E} \left| \tilde{\mathbb{M}}_{N}(p) - \mathbb{M}(p) \right| \leq C_2 \frac{1}{\sqrt{N}},
\]
for some constant \(C_2 > 0\).
Combining the two parts, we complete the proof. \qed

\vspace{1cm}

%

\noindent{\bf Proof of Proposition \ref{pro.3}.} 
By Lemma~\ref{lem.4}, $\varphi^{\varepsilon}$ is strictly convex and belongs to $C^1$; consequently, $\mathbf{Q}_{\pm}^{\varepsilon}=\nabla\varphi^{\varepsilon}$ is continuous and strictly monotone:
\[
\big(\mathbf{Q}_{\pm}^{\varepsilon}(x)-\mathbf{Q}_{\pm}^{\varepsilon}(y)\big)\!\cdot\!(x-y)>0\quad(x\neq y).
\]
In particular, $\mathbf{Q}_{\pm}^{\varepsilon}$ is injective on $\mathbb{S}_d$.

Fix $p\in(0,1)$ and consider the restriction
\[
\mathbf{Q}_{\pm}^{\varepsilon}\big|_{\,p\,\mathbb{S}_d}:\ p\,\mathbb{S}_d\longrightarrow\mathbb{C}^{\varepsilon}(p),
\qquad
\mathbb{C}^{\varepsilon}(p):=\mathbf{Q}_{\pm}^{\varepsilon}\!\big(p\,\mathbb{S}_d\big).
\]
Hence $\mathbf{Q}_{\pm}^{\varepsilon}\big|_{\,p\,\mathbb{S}_d}$ is a bijection.
Since the domain is compact and the codomain is Hausdorff, this continuous bijection
is a homeomorphism onto its image $\mathbb{C}^{\varepsilon}(p)$.
Because $p\,\mathbb{S}_d$ is simply connected, so is $\mathbb{C}^{\varepsilon}(p)$.

For $0<p_1<p_2<1$,
$p_1\mathbb{S}_d\subsetneq  p_2\mathbb{S}_d$.
The injectivity of $\mathbf{Q}_{\pm}^{\varepsilon}$ preserves strict inclusion, yielding
\[
\mathbb{C}^{\varepsilon}(p_1)\subsetneq  \mathbb{C}^{\varepsilon}(p_2).
\]\qed

\vspace{1cm}

\noindent{\bf Proof of Proposition \ref{pro.4}.}
We begin by establishing the following lemmas:
\begin{lemma}[Proposition 2 in  \cite{pooladian2021entropic}]\label{lem.2}
	Let $ ( \psi_{\varepsilon }, \phi_{\varepsilon })$ be optimal entropic potentials satisfying \eqref{equ.9} and \eqref{equ.10}, then $ T_0^\varepsilon=\mathrm{Id}-\nabla \psi_\varepsilon$.
\end{lemma}

%

{
	\begin{lemma}\label{lem.3}
		Let $ (\psi, \phi)$ be optimal entropic potentials satisfying \eqref{equb3} and assume that $\mu$ is absolutely continuous, then $ T= \mathrm{Id}-\nabla \psi$.
	\end{lemma}
	
	\begin{proof}
		Assume that $(\mathbf{x}_0, \mathbf{y}_0)$ is an interior point of $\mathrm{spt}(\pi)$. 
		Since the source measure $\mu$ is absolutely continuous, the convex potential $\psi$ is differentiable $\mu$-almost everywhere. 
		Let $\psi$ be differentiable at $\mathbf{x}_0$. Then,
		\[
		\psi^c (\mathbf{y}_0) = c(\mathbf{x}_0, \mathbf{y}_0) - \psi(\mathbf{x}_0) = \underset{\mathbf{x} \in \mathrm{spt}(\mu)}{\inf} \, \bigl(c(\mathbf{x}, \mathbf{y}_0) - \psi(\mathbf{x})\bigr).
		\]
		At $(\mathbf{x}_0, \mathbf{y}_0)$, the gradient of $\underset{\mathbf{x} \in \mathrm{spt}(\mu)}{\inf} \, c(\mathbf{x}, \mathbf{y}_0) - \psi(\mathbf{x})$ at $\mathbf{x}_0$ vanishes, i.e.,
		\[ 
		\nabla_\mathbf{x} c(\mathbf{x}_0, \mathbf{y}_0) = \nabla \psi(\mathbf{x}_0).
		\]
		Since $c(\mathbf{x}, \mathbf{y}) = \tfrac{1}{2} \| \mathbf{x} - \mathbf{y} \|^2$, we have 
		\[ 
		\nabla_\mathbf{x} \tfrac{1}{2} \| \mathbf{x}_0 - \mathbf{y}_0 \|^2 = \nabla \psi(\mathbf{x}_0).
		\]
		Hence, $\nabla \psi(\mathbf{x}_0) = \mathbf{x}_0 - \mathbf{y}_0$, and we conclude that
		\[
		T(x) = \mathrm{Id}(x) - \nabla \psi(x).
		\]
	\end{proof}
	
}

\begin{lemma}\label{lem.4}
	
	$ (\mathbf{Q}_{\pm}^{\varepsilon}(x) - \mathbf{Q}_{\pm}^{\varepsilon}(y))^{\top}(x - y) > 0.$
\end{lemma}

\begin{proof}
	Let $\varphi_\varepsilon$  be a function such that $\mathbf{Q}_{\pm}^{\varepsilon}(x) =\nabla \varphi_\varepsilon(x)$. 
	According to Lemma  1 in \cite{chewi2023entropic}, we know that 
	$$\nabla^2\varphi_\varepsilon(x)=\varepsilon^{-1}\operatorname{Cov}_{Y\thicksim\pi_\varepsilon^x}(Y).$$
	And since Proposition 10 in \cite{rigollet2022sample}, we know that \( \nabla^2\varphi_\varepsilon(x) \) is always greater than 0, which implies that \(\varphi_\varepsilon \) is a strongly convex function.  Therefore, \( \mathbf{Q}_{\pm}^{\varepsilon} \) is strongly monotone.

\end{proof}

For part (i), combining Lemmas \ref{lem.2}, \ref{lem.3}, and \ref{lem.4}, we conclude that for each \( \varepsilon \), the function \( \varphi_{\varepsilon} \) is convex. Furthermore, by Brézis' theorem and the fact that \( \mathbf{Q}_{\pm}(x) = \nabla \varphi(x) \), we infer that \( \varphi \) itself is also convex.

According to Theorem 1.1 in \cite{nutz2022entropic}, the entropic potential \( \varphi_{\varepsilon_k} \) converges to \( \varphi\) in \( L^1 \) as \( k \rightarrow \infty \), which, by the compactness of support, further implies pointwise convergence. Finally, applying Theorem 25.7 in \cite{borwein2006convex}, we establish uniform convergence.

For part (ii), suppose there exists \(\mathbf{u}_0 \in \mathbb{S}_d\) and \(\epsilon_0 > 0\) such that 
\[ 
\underset{k \to \infty}{\lim\sup} \mathbb{P} \left( \mathbf{Q}^{\varepsilon_k}(\mathbf{u}_0) \notin \mathbf{Q}_{\pm}(\mathbf{u}_0) + \epsilon_0 \mathbb{S}_d \right) = \delta > 0.
\]
This assumption implies the existence of a subsequence \(k_j\) such that
\begin{equation}\label{app.1}
	\lim\limits_{j \to \infty} \mathbb{P} \left( \mathbf{Q}^{\varepsilon_{k_j}}(\mathbf{u}_0) \notin \mathbf{Q}_{\pm}(\mathbf{u}_0) + \epsilon_0 \mathbb{S}_d \right) = \delta > 0.
\end{equation}

By the construction of entropic potentials (see \cite{nutz2021introduction}), each \(\psi_{\varepsilon_i}\) is continuous. Furthermore, the Proof of Lemma \ref{lem.7} establishes that each \(\psi_{\varepsilon_k}\) is convex. 

Since \(\psi_{\varepsilon_k}\) converges uniformly to \(\psi\), we conclude the epigraphical convergence of \(\psi_{\varepsilon_k}\) to \(\psi\) (see \cite{borwein1996epigraphical} for more details about epigraphical convergence).

Then, by Theorem 8.3 in \cite{bagh1996convergence}, there exists \(j_0 \in \mathbb{N}\) such that, for all \(j > j_0\),
\begin{equation}\label{app.openset}
	\mathbf{Q}^{\varepsilon_{k_j}}(\mathbf{u}_0) \in \mathbf{Q}(\mathbf{u}_0) + \epsilon_0 \mathbb{S}_d.
\end{equation}

We can therefore deduce that  
\[
\mathbb{P}\left(\bigcap_{j_0 \in \mathbb{N}} \bigcup_{j \geq j_0} \mathbf{Q}^{\varepsilon_{k_j}}(\mathbf{u}_0|\mathbf{X}) \not\in \mathbf{Q}(\mathbf{u}_0|\mathbf{X}) + \epsilon_0 \mathbb{S}_d \right) = 0,
\]
which implies that 
\[ 
\underset{k \to \infty}{\lim\sup}  \mathbb{P} \left( \mathbf{Q}^{\varepsilon_{k_j}}(\mathbf{u}_0) \notin \mathbf{Q}_{\pm}(\mathbf{u}_0) + \epsilon_0 \mathbb{S}_d \right) = \delta \leq 0.
\]
This contradicts \eqref{app.1}. We have thus proved that
\[
\mathbb{P} \left( \mathbf{Q}^{\varepsilon_k}(\mathbf{u}) \notin \mathbf{Q}_{\pm}(\mathbf{u}) + \epsilon \mathbb{S}_d \right) \to 0 \quad \text{as } k \rightarrow \infty.
\]

For part (iii), we proceed by similar reasoning. We will show the first of these cases explicitly. Suppose there exists \(p_0 \in (0,1)\), \(\epsilon_0 > 0\), and a subsequence \(k_j\) such that 
\begin{equation}\label{app.2}
	\lim\limits_{j \to \infty}\mathbb{P} \left( \mathbb{C}^{\varepsilon_{k_j}}(p) \not\subset \mathbb{C}(p) + \epsilon \mathbb{S}_d \right) = \delta > 0.
\end{equation}
Theorem 8.3 in \cite{bagh1996convergence} implies that for every \(\bar{\mathbf{u}} \in \mathbb{S}_d\), there exists \(I_{\bar{\mathbf{u}}} \in \mathbb{N}\) and \(\lambda_{\bar{\mathbf{u}}} > 0\) such that
\[
\mathbf{Q}^{\varepsilon_{k_j}}(\mathbf{u} ) \in \mathbf{Q}(\bar{\mathbf{u}}) + \epsilon_0 \mathbb{S}_d \quad \text{for all } \mathbf{u} \in \bar{\mathbf{u}} + \lambda_{\bar{\mathbf{u}}} \mathbb{S}_d \text{ and all } j> J_{\bar{\mathbf{u}}}.
\]
Since \(p\mathbb{S}_d \) is compact, the finite cover theorem provides a finite covering 
\[
p \mathbb{S}_{d}\subset\bigcup_{j=1}^{J_{\epsilon}} \mathbf{\bar{u}}_{j}+\delta_{\mathbf{\bar{u}}_{j}}\mathbb{S}_{d}.
\]

Set \( J_0 = \max(J_{\bar{u}_1}, \dots, J_{\bar{u}_{K_\epsilon}}) \); then for all \( j> J_0 \), we have
\[
\mathbb{C}^{\varepsilon_{k_i}}(q_0 \mid \mathbf{x}) \subset \mathbb{C}(q_0 \mid \mathbf{x}) + \epsilon_0 S_d,
\]
which holds with probability one, thus contradicting \eqref{app.2}. \qed

\vspace{1cm}

\noindent{\bf Proof of Proposition \ref{pro.7}.}
We aim to show that both ${\mathrm{M}}^{\varepsilon_k}$ and ${\mathbb{M}}^{\varepsilon_k}$ converge uniformly to ${\mathrm{M}}$ and ${\mathbb{M}}$ as $k \to \infty$.

We first prove
\[
\lim_{k \to \infty} \sup_{0 < p \leq 1} \big|{\mathrm{M}}^{\varepsilon_k}(p) - {\mathrm{M}}(p)\big| = 0.
\]
Recall that
\[
{\mathrm{M}}^{\varepsilon_k}(p) = \int_{\mathtt{C}^{\varepsilon_k}(p)} \rho(x) f(x)\, dx,
\]
where $\mathtt{C}^{\varepsilon_k}(p)$ is the entropic quantile region generated by $\mathbf{Q}_{\pm}^{\varepsilon_k}$. By Proposition \ref{pro.4}, $\mathbf{Q}_{\pm}^{\varepsilon_k} \to \mathbf{Q}_{\pm}$ uniformly as $\varepsilon_k \to 0$. Hence, for each fixed $p$, ${\mathrm{M}}^{\varepsilon_k}(p) \to {\mathrm{M}}(p)$ pointwise.

Since ${\mathrm{M}}$ is continuous on $[0,1]$, the pointwise convergence together with monotonicity of ${\mathrm{M}}^{\varepsilon_k}(p)$ implies, by Dini’s theorem, that convergence is uniform:
\[
\sup_{0 < p \leq 1} \big|{\mathrm{M}}^{\varepsilon_k}(p) - {\mathrm{M}}(p)\big| \to 0.
\]

The same reasoning applies to ${\mathbb{M}}^{\varepsilon_k}(p)$, which is continuous, uniformly bounded, and converges pointwise to ${\mathbb{M}}(p)$. Thus,
\[
\sup_{0 < p \leq 1} \big|{\mathbb{M}}^{\varepsilon_k}(p) - {\mathbb{M}}(p)\big| \to 0.
\]\qed

\vspace{1cm}

\noindent{\bf Proof of Proposition \ref{pro.entrouni}.} This proof, after obtaining  Lemma \ref{lem.entropotential} and Lemma \ref{lem.entromap}, is similar to Proposition \ref{pro.1}. \qed

\vspace{1cm}
Let \(\{\xb_i\}_{i=1}^{N_1}\) and \(\{\yb_j\}_{j=1}^{N_2}\) be i.i.d. samples from measures \(\mu\) and \(\nu\), respectively. The empirical measures are defined as
\[
\widehat{\mu}_{N_1} = \frac{1}{N_1} \sum_{i=1}^{N_1} \delta_{\xb_i}, \quad \widehat{\nu}_{N_2} = \frac{1}{N_2} \sum_{j=1}^{N_2} \delta_{\yb_j}.
\]
$\widehat{\phi} _{\varepsilon, N_1,N_2}(\mathbf{y}_j)$ and $\widehat{\psi} _{\varepsilon, N_1,N_2}(\mathbf{x}_i)$ are solved by the discretized optimal entropy transfer problem and satisfy the discretized constraint conditions:

\begin{gather}
	\frac{1}{N_1} \sum_{i=1}^{N_1} \exp\left( \frac{\widehat{\psi} _{\varepsilon, N_1,N_2}(\mathbf{x}_i) + \widehat{\phi} _{\varepsilon, N_1,N_2}(\mathbf{y}_j) - c(\mathbf{x}_i, \mathbf{y}_j)}{\varepsilon} \right)  = 1, \ \forall j = 1, \cdots, N_2, \\
	\frac{1}{N_2} \sum_{j=1}^{N_2} \exp\left( \frac{\widehat{\psi} _{\varepsilon, N_1,N_2}(\mathbf{x}_i) + \widehat{\phi} _{\varepsilon, N_1,N_2}(\mathbf{y}_j) - c(\mathbf{x}_i, \mathbf{y}_j)}{\varepsilon} \right) = 1, \ \forall i = 1, \cdots, N_1.
\end{gather}
Then we can obtain the empirical entropic map as 
$$
\widehat{\mathbf{T}}^{\varepsilon}_{N_1,N_2}(\mathbf{x}) = \frac{\sum_{j=1}^{N_2} \mathbf{y}_j \exp\left( \frac{\widehat{\phi} _{\varepsilon, N_1,N_2}(\mathbf{y}_j) - c(\mathbf{x}, \mathbf{y}_j)}{\varepsilon} \right)}{\sum_{j=1}^{N_2} \exp\left( \frac{\widehat{\phi} _{\varepsilon, N_1,N_2}(\mathbf{y}_j) - c(\mathbf{x}, \mathbf{y}_j)}{\varepsilon} \right)}.
$$

In the $L^{\infty}$ norm, define the norm of $(\phi_{\varepsilon},\psi_{\varepsilon})$ as:
$$
\|(\phi_{\varepsilon},\psi_{\varepsilon})\|_{L^{\infty}} = \|\phi_{\varepsilon}\|_{L^{\infty}} + \|\psi_{\varepsilon}\|_{L^{\infty}},
$$
where
\[
\|\phi_{\varepsilon}\|_{L^{\infty}} = \sup_{\mathbf{x}\in\mathcal{X}}|\phi_{\varepsilon}(\mathbf{x})|, \quad \|\psi_{\varepsilon}\|_{L^{\infty}} = \sup_{\mathbf{y}\in\mathcal{Y}}|\psi_{\varepsilon}(\mathbf{y})|.
\]
We write $\widehat{\phi} _{\varepsilon}$ and $\widehat{\psi} _{\varepsilon}$ for $\widehat{\phi} _{\varepsilon, N_1,N_2}$ and $\widehat{\psi} _{\varepsilon, N_1,N_2}$, respectively, when no confusion arises.

\noindent{\bf Proof of Lemma \ref{lem.entromap}.}
Since \(\mathcal{X}\) and \(\mathcal{Y}\) are compact and \(c\) is continuous, there exists \(M_c < \infty\) such that
\[
\sup_{\mathbf{x} \in \mathcal{X}, \mathbf{y} \in \mathcal{Y}} |c(\mathbf{x}, \mathbf{y})| \leq M_c.
\]
Assume \(\mathcal{Y} \subseteq \mathbb{R}^d\), with \(\|\cdot\|\) denoting the Euclidean norm. The continuity of the densities of \(\mu\) and \(\nu\), together with that of \(c\), ensures that both \(\phi_ \varepsilon\) and \(\widehat{\phi}_{ \varepsilon, N_1,N_2}\) are continuous. Since \(\mathcal{Y}\) is compact, there exists \(M_\phi < \infty\) such that
$$\sup_{\mathbf{y} \in \mathcal{Y}} |\phi_ \varepsilon(\mathbf{y})| \leq M_\phi \ \ \text{and} \ \ \sup_{\mathbf{y} \in \mathcal{Y}} |\widehat{\phi}_{ \varepsilon, N_1,N_2}(\mathbf{y})| \leq M_\phi.$$

Define the auxiliary functions:
\[
g_{\mathbf{x}}(\mathbf{y}) = e^{\frac{1}{ \varepsilon} ( \phi_ \varepsilon(\mathbf{y}) - c(\mathbf{x}, \mathbf{y}) )}, \quad 
f_{\mathbf{x}}(\mathbf{y}) = \mathbf{y} \, g_{\mathbf{x}}(\mathbf{y}),
\]
\[
\widehat{g}_{\mathbf{x}}(\mathbf{y}) = e^{\frac{1}{ \varepsilon} ( \widehat{\phi}_{ \varepsilon, N_1,N_2}(\mathbf{y}) - c(\mathbf{x}, \mathbf{y}) )}, \quad 
\widehat{f}_{\mathbf{x}}(\mathbf{y}) = \mathbf{y} \, \widehat{g}_{\mathbf{x}}(\mathbf{y}).
\]
Then the regularized transport map and its empirical approximation are given by
\[
\mathbf{T}^ \varepsilon(\mathbf{x}) = \frac{A_0(\mathbf{x})}{B_0(\mathbf{x})}, \quad
\widehat{\mathbf{T}}^{ \varepsilon}_{N_1,N_2}(\mathbf{x}) = \frac{A(\mathbf{x})}{B(\mathbf{x})},
\]
where
\[
A_0(\mathbf{x}) = \int f_{\mathbf{x}}(\mathbf{y}) \mathrm{d}\nu(\mathbf{y}), \quad 
B_0(\mathbf{x}) = \int g_{\mathbf{x}}(\mathbf{y}) \mathrm{d}\nu(\mathbf{y}),
\]
\[
A(\mathbf{x}) = \frac{1}{N_2} \sum_{j=1}^{N_2} \widehat{f}_{\mathbf{x}}(\mathbf{y}_j), \quad 
B(\mathbf{x}) = \frac{1}{N_2} \sum_{j=1}^{N_2} \widehat{g}_{\mathbf{x}}(\mathbf{y}_j).
\]

We aim to show that
\[
\sup_{\mathbf{x} \in \mathcal{X}} \left\| \widehat{\mathbf{T}}^{ \varepsilon}_{N_1,N_2}(\mathbf{x}) - \mathbf{T}^ \varepsilon(\mathbf{x}) \right\| \to 0.
\]
Using the standard error bound for quotients,
\[
\left\| \frac{A}{B} - \frac{A_0}{B_0} \right\| 
\leq \frac{ \|A - A_0\| B_0 + \|A_0\| |B - B_0| }{B B_0}.
\]

Since \(|c(\mathbf{x}, \mathbf{y})| \leq M_c\) and \(|\phi_ \varepsilon(\mathbf{y})|\), \(|\widehat{\phi}_{ \varepsilon,N_1,N_2}(\mathbf{y})| \leq M_\phi\), we obtain the bounds
\[
e^{-\frac{M_\phi + M_c}{ \varepsilon}} \leq g_{\mathbf{x}}(\mathbf{y}),  \widehat{g}_{\mathbf{x}}(\mathbf{y}) \leq e^{\frac{M_\phi + M_c}{ \varepsilon}}.
\]
Let \(b := e^{-\frac{M_\phi + M_c}{ \varepsilon}}\), then
\[
B_0(\mathbf{x}), B(\mathbf{x}) \geq b > 0.
\]

Compactness of \(\mathcal{Y}\) implies \(\|\mathbf{y}\| \leq R'\) for some \(R' > 0\), so
\[
\|f_{\mathbf{x}}(\mathbf{y})\|, \|\widehat{f}_{\mathbf{x}}(\mathbf{y})\| \leq R' e^{\frac{M_\phi + M_c}{ \varepsilon}}, \quad 
\|A_0(\mathbf{x})\| \leq R' e^{\frac{M_\phi + M_c}{ \varepsilon}}.
\]

We decompose the errors as
\[
A - A_0 = (A - \mathbb{E}A) + (\mathbb{E}A - A_0), \quad 
B - B_0 = (B - \mathbb{E}B) + (\mathbb{E}B - B_0),
\]
where expectations are taken with respect to the sampling of \(\mathbf{y}_j \sim \nu\).

By Lemma \ref{lem.entropotential}, uniform convergence of \(\widehat{\phi}_{ \varepsilon,N_1,N_2} \to \phi_ \varepsilon\) implies there exists \(\delta_{N_1,N_2} \to 0\) such that
\[
\sup_{\mathbf{y} \in \mathcal{Y}} |\widehat{\phi}_{ \varepsilon,N_1,N_2}(\mathbf{y}) - \phi_ \varepsilon(\mathbf{y})| \leq \delta_{N_1,N_2}.
\]
Therefore,
\[
\widehat{g}_{\mathbf{x}}(\mathbf{y}) = g_{\mathbf{x}}(\mathbf{y}) \cdot e^{\frac{1}{ \varepsilon} (\widehat{\phi}_{ \varepsilon,N_1,N_2}(\mathbf{y}) - \phi_ \varepsilon(\mathbf{y}))},
\]
and  there is a  $C_ \varepsilon$ such that 
\[
|\widehat{g}_{\mathbf{x}}(\mathbf{y}) - g_{\mathbf{x}}(\mathbf{y})| \leq C_ \varepsilon \frac{\delta_{N_1,N_2}}{ \varepsilon} e^{\frac{M_\phi + M_c}{ \varepsilon}},
\]
\[
\|\widehat{f}_{\mathbf{x}}(\mathbf{y}) - f_{\mathbf{x}}(\mathbf{y})\| \leq R' C_ \varepsilon \frac{\delta_{N_1,N_2}}{ \varepsilon} e^{\frac{M_\phi + M_c}{ \varepsilon}}.
\]
Integrating gives
\[
\sup_{\mathbf{x} \in \mathcal{X}} \left\| \int \widehat{f}_{\mathbf{x}} \mathrm{d}\nu - \int f_{\mathbf{x}} \mathrm{d}\nu \right\| \leq R' C_ \varepsilon \frac{\delta_{N_1,N_2}}{ \varepsilon} e^{\frac{M_\phi + M_c}{ \varepsilon}} \to 0,
\]
\[
\sup_{\mathbf{x} \in \mathcal{X}} \left| \int \widehat{g}_{\mathbf{x}} \mathrm{d}\nu - \int g_{\mathbf{x}} \mathrm{d}\nu \right| \leq C_ \varepsilon \frac{\delta_{N_1,N_2}}{ \varepsilon} e^{\frac{M_\phi + M_c}{ \varepsilon}} \to 0.
\]

For the empirical terms, define function classes
\[
\mathcal{F} = \{\widehat{f}_{\mathbf{x}} : \mathbf{x} \in \mathcal{X}\}, \quad 
\mathcal{G} = \{\widehat{g}_{\mathbf{x}} : \mathbf{x} \in \mathcal{X}\}.
\]
These classes are uniformly bounded and equicontinuous due to compactness and continuity, hence have finite covering numbers. By empirical process theory,
\[
\sup_{\mathbf{x} \in \mathcal{X}} \left\| \frac{1}{N_2} \sum_{j=1}^{N_2} \widehat{f}_{\mathbf{x}}(\mathbf{y}_j) - \int \widehat{f}_{\mathbf{x}} \mathrm{d}\nu \right\| \to 0, \quad 
\sup_{\mathbf{x} \in \mathcal{X}} \left| \frac{1}{N_2} \sum_{j=1}^{N_2} \widehat{g}_{\mathbf{x}}(\mathbf{y}_j) - \int \widehat{g}_{\mathbf{x}} \mathrm{d}\nu \right| \to 0
\]
almost surely as \(N_2 \to \infty\). Hence,
\[
\sup_{\mathbf{x} \in \mathcal{X}} \|A - A_0\| \to 0, \quad \sup_{\mathbf{x} \in \mathcal{X}} |B - B_0| \to 0.
\]

Combining the bounds yields
\[
\sup_{\mathbf{x} \in \mathcal{X}} \left\| \frac{A}{B} - \frac{A_0}{B_0} \right\| \leq \frac{ e^{\frac{M_\phi + M_c}{ \varepsilon}} \left( \sup_{\mathbf{x} \in \mathcal{X}} \|A - A_0\| + R' \sup_{\mathbf{x} \in \mathcal{X}} |B - B_0| \right) }{b^2} \to 0.
\]
Therefore,
\[
\sup_{\mathbf{x} \in \mathcal{X}} \left\| \widehat{\mathbf{T}}^{ \varepsilon}_{N_1,N_2}(\mathbf{x}) - \mathbf{T}^ \varepsilon(\mathbf{x}) \right\| \to 0, \quad \text{as } N_1, N_2 \to \infty.
\]\qed

\begin{lemma} \label{lem.entropotential}
	Under the conditions of Lemma~\ref{lem.entromap}, we have
	\[
	\|\widehat{\phi}_{\varepsilon} - \phi_{\varepsilon}\|_{L^\infty} + \|\widehat{\psi}_{\varepsilon} - \psi_{\varepsilon}\|_{L^\infty} \to 0 \quad \text{as } N_1, N_2 \to \infty.
	\]
\end{lemma}

\begin{proof}
The Sinkhorn algorithm is a widely used method for solving the entropic optimal transport problem. We begin by introducing the Sinkhorn operator.

In the continuous setting, given a pair of initial potential functions $(\phi, \psi)$, the Sinkhorn operator $\mathcal{G}(\phi, \psi) = (\phi', \psi')$ is defined as:
\[
\phi'(x) = -\varepsilon \log \int e^{\frac{\psi(y) - c(x,y)}{\varepsilon}} \nu(y) \, dy, \quad 
\psi'(y) = -\varepsilon \log \int e^{\frac{\phi(x) - c(x,y)}{\varepsilon}} \mu(x) \, dx,
\]
where $c(x,y)$ is the cost function, $\varepsilon > 0$ is the entropic regularization parameter, and $\mu(x)$, $\nu(y)$ are the prescribed marginal distributions. Iteratively applying $\mathcal{G}$ yields a sequence that converges to a fixed point $(\phi_\varepsilon, \psi_\varepsilon)$, which solves the entropic optimal transport problem.

In the discrete setting, given initial potentials $(\widehat{\phi}, \widehat{\psi})$, the discrete Sinkhorn operator $\widehat{\mathcal{G}}(\widehat{\phi}, \widehat{\psi}) = (\widehat{\phi}', \widehat{\psi}')$ is defined by
\[
\widehat{\phi}'(\mathbf{x}_i) = -\varepsilon \log \left( \frac{1}{N_2} \sum_{j=1}^{N_2} e^{\frac{\widehat{\psi}(\mathbf{y}_j) - c(\mathbf{x}_i, \mathbf{y}_j)}{\varepsilon}} \right), \quad \forall i= 1, \cdots, N_1,
\]
\[
\widehat{\psi}'(\mathbf{y}_j) = -\varepsilon \log \left( \frac{1}{N_1} \sum_{i=1}^{N_1} e^{\frac{\widehat{\phi}(\mathbf{x}_i) - c(\mathbf{x}_i, \mathbf{y}_j)}{\varepsilon}} \right), \quad \forall j= 1, \cdots, N_2.
\]
Repeated application of $\widehat{\mathcal{G}}$ yields a sequence converging to $(\widehat{\phi}_\varepsilon, \widehat{\psi}_\varepsilon)$, the discrete counterpart to the continuous solution.

In both cases, the Sinkhorn algorithm proceeds by iteratively applying the respective operator—$\mathcal{G}$ in the continuous case and $\widehat{\mathcal{G}}$ in the discrete case—until convergence to a fixed point, which corresponds to the solution of the entropic optimal transport problem.

We note that 
\[ 
(\widehat{\phi}_{\varepsilon}, \widehat{\psi}_{\varepsilon}) = \mathcal{G} (\widehat{\phi}_{\varepsilon}, \widehat{\psi}_{\varepsilon}), \quad (\phi_{\varepsilon}, \psi_{\varepsilon}) = \mathcal{G}(\phi_{\varepsilon}, \psi_{\varepsilon}).
\]

By the triangle inequality, we have 
\begin{equation}\label{sinkhorn_inequality}
	\|\widehat{\mathcal{G} }(\widehat{\phi}_{\varepsilon}, \widehat{\psi}_{\varepsilon}) - \mathcal{G}(\phi_{\varepsilon}, \psi_{\varepsilon})\|_{L^\infty} \leq \|\widehat{\mathcal{G} }(\widehat{\phi}_{\varepsilon}, \widehat{\psi}_{\varepsilon}) - \widehat{\mathcal{G} } (\phi_{\varepsilon}, \psi_{\varepsilon})\|_{L^\infty} + \|\widehat{\mathcal{G} } (\phi_{\varepsilon}, \psi_{\varepsilon}) - \mathcal{G}(\phi_{\varepsilon}, \psi_{\varepsilon})\|_{L^\infty}.
\end{equation}

Since \(\widehat{\mu}_{N_1} \to \mu\) and \(\widehat{\nu}_{N_2} \to \nu\) converge weakly, and \(c\) is continuous, \(\mathcal{X}\) and \(\mathcal{Y}\) are compact, \(e^{-c(x,y)}\) is continuous and bounded, and we have

\[
\frac{1}{N_2} \sum_{j=1}^{N_2} e^{\frac{\widehat{\psi}_{\varepsilon}(y_j) - c(x,y_j)}{\varepsilon}} \to \int e^{\frac{\psi_{\varepsilon}(y) - c(x,y)}{\varepsilon}} \nu(y) \, dy,
\]
and
\[
\frac{1}{N_1} \sum_{i=1}^{N_1} e^{\frac{\widehat{\phi}_{\varepsilon}(x_i) - c(x_i,y)}{\varepsilon}} \to \int e^{\frac{\phi_{\varepsilon}(x) - c(x,y)}{\varepsilon}} \mu(x) \, dx.
\]
Therefore,
\[
\|\widehat{\mathcal{G} }(\phi_{\varepsilon}, \psi_{\varepsilon}) - \mathcal{G}(\phi_{\varepsilon}, \psi_{\varepsilon})\|_{L^\infty} \to 0.
\]

We now prove that
\[
\left\| \widehat{\mathcal{G}}(\widehat{\phi}_{\varepsilon}, \widehat{\psi}_{\varepsilon}) - \widehat{\mathcal{G}}(\phi_{\varepsilon}, \psi_{\varepsilon}) \right\|_{L^\infty} \to 0.
\]
To this end, we first introduce the Hilbert metric. A key property is that the dual variables in the matrix Sinkhorn algorithm exhibit strict contraction with respect to this metric during the iterations. This contraction property will be instrumental in establishing the desired convergence.

The Hilbert metric is typically defined on the space of positive functions. For two strictly positive functions \( f, g \colon \mathcal{X} \to \mathbb{R}_+ \), the Hilbert metric between them is given by
\[
d_H(f, g) = \sup_{x, y \in \mathcal{X}} \log\left(\frac{f(x)/g(x)}{f(y)/g(y)}\right).
\]
This metric quantifies the maximal logarithmic distortion between the pointwise ratios of \( f \) and \( g \) over the domain \( \mathcal{X} \), and is often used to analyze the contraction properties of positive operators.

Let \( u = e^{\widehat{\phi}/\varepsilon} \) and \( v = e^{\widehat{\psi}/\varepsilon} \), where \(\widehat{\phi}\) and \(\widehat{\psi}\) are dual potentials. According to Theorem 4.1 in \cite{peyre2019computational}, the Sinkhorn operator \(\widehat{\mathcal{G}}\), defined for the dual potentials \(\widehat{\phi}\) and \(\widehat{\psi}\), is also applicable to \( u \) and \( v \), and is contractive with respect to the Hilbert metric:
\[
d_H\big(\widehat{\mathcal{G}}(u, v)\big) \leq k \cdot d_H(u, v),
\]
for some constant \( 0 < k < 1 \).

More generally, for any two pairs of potential functions \( (\widehat{\phi}_1, \widehat{\psi}_1) \) and \( (\widehat{\phi}_2, \widehat{\psi}_2) \), define the corresponding exponentiated functions:
\[
u_1(x) = e^{\frac{\widehat{\phi}_1(x)}{\varepsilon}}, \quad u_2(x) = e^{\frac{\widehat{\phi}_2(x)}{\varepsilon}}, \quad
v_1(y) = e^{\frac{\widehat{\psi}_1(y)}{\varepsilon}}, \quad v_2(y) = e^{\frac{\widehat{\psi}_2(y)}{\varepsilon}}.
\]
Then the Sinkhorn operator satisfies the joint contraction property:
\[
d_H\big(\widehat{\mathcal{G}}(u_1, v_1), \widehat{\mathcal{G}}(u_2, v_2)\big) \leq k \cdot d_H\big((u_1, v_1), (u_2, v_2)\big),
\]
where the Hilbert distance between two pairs is defined as
\[
d_H\big((u_1, v_1), (u_2, v_2)\big) = d_H(u_1, u_2) + d_H(v_1, v_2).
\]
This contraction property plays a crucial role in establishing the convergence and stability of the Sinkhorn algorithm.

The Hilbert metric of \(u_1\) and \(u_2\)  can be written as
\[
d_H(u_1, u_2) = \sup_{x, x' \in \mathcal{X}} \left( \frac{\widehat{\phi}_1(x) - \widehat{\phi}_2(x)}{\varepsilon} - \frac{\widehat{\phi}_1(x') - \widehat{\phi}_2(x')}{\varepsilon} \right).
\]
Define \(\delta(x) = \widehat{\phi}_1(x) - \widehat{\phi}_2(x)\), then
\[
d_H(u_1, u_2) = \frac{1}{\varepsilon} \sup_{x, x' \in \mathcal{X}} (\delta(x) - \delta(x')) = \frac{1}{\varepsilon} \left( \sup_{x \in \mathcal{X}} \delta(x) - \inf_{x' \in \mathcal{X}} \delta(x') \right).
\]
Let \(M = \sup_{x \in \mathcal{X}} \delta(x)\) and \( m = \inf_{x \in \mathcal{X}} \delta(x)\), so
\[
d_H(u_1, u_2) = \frac{1}{\varepsilon} (M - m).
\]
The \(L^\infty\) norm is
\[
\|\widehat{\phi}_1 - \widehat{\phi}_2\|_{L^\infty} = \sup_{x \in \mathcal{X}} |\delta(x)| = \max(|M|, |m|).
\]

Since \(M - m \leq M + |m|\), 
we have
\[
d_H(u_1, u_2) = \frac{1}{\varepsilon} (M - m) \leq \frac{1}{\varepsilon} (M + |m|) \leq \frac{2}{\varepsilon} \max(|M|, |m|) = \frac{2}{\varepsilon} \|\widehat{\phi}_1 - \widehat{\phi}_2\|_{L^\infty}.
\]
Similarly,
\[
d_H(v_1, v_2) \leq \frac{2}{\varepsilon} \|\widehat{\psi}_1 - \widehat{\psi}_2\|_{L^\infty}.
\]
Thus,
\[
d_H((u_1, v_1), (u_2, v_2)) \leq \frac{2}{\varepsilon} \left( \|\widehat{\phi}_1 - \widehat{\phi}_2\|_{L^\infty} + \|\widehat{\psi}_1 - \widehat{\psi}_2\|_{L^\infty} \right).
\]

The reverse direction requires additional conditions. Consider \(d_H(u_1, u_2) = \frac{1}{\varepsilon} (M - m)\); it does not directly yield \(\sup |\delta|\), as \(M - m\) measures oscillation rather than the absolute maximum. To address this, normalization is introduced.	

In the Sinkhorn algorithm, the potential functions \((\widehat{\phi}, \widehat{\psi})\) admit a degree of freedom under constant shifts: if \((\widehat{\phi}, \widehat{\psi})\) is a solution, then so is \((\widehat{\phi} + c, \widehat{\psi} - c)\) for any constant \(c \in \mathbb{R}\). To eliminate this degree of freedom, we impose the normalization condition:
\[
\int \widehat{\phi}(x) \, d\widehat{\mu}_{N_1}(x) = 0.
\]
Thus, for \(\delta(x) = \widehat{\phi}_1(x) - \widehat{\phi}_2(x)\), we have
\[
\int \delta(x) \, d\widehat{\mu}_{N_1}(x) = 0.
\]

Since \(\delta(x)\) integrates to zero under \(\widehat{\mu}_{N_1}\), it must take both positive and negative values unless it is identically zero. Therefore,
\[
M = \sup_{x \in \mathcal{X}} \delta(x) \geq 0, \quad m = \inf_{x \in \mathcal{X}} \delta(x) \leq 0,
\]
and we have
\[
d_H(u_1, u_2) = \frac{1}{\varepsilon}(M - m) = \frac{1}{\varepsilon}(M + |m|).
\]
Since
\[
\|\widehat{\phi}_1 - \widehat{\phi}_2\|_{L^\infty} = \sup_{x \in \mathcal{X}} |\delta(x)| = \max(M, |m|),
\]
and \(M + |m| \geq \max(M, |m|)\), it follows that
\[
d_H(u_1, u_2) \geq \frac{1}{\varepsilon} \|\widehat{\phi}_1 - \widehat{\phi}_2\|_{L^\infty},
\]
or equivalently,
\[
\|\widehat{\phi}_1 - \widehat{\phi}_2\|_{L^\infty} \leq \varepsilon \cdot d_H(u_1, u_2).
\]
Similarly, we obtain
\[
\|\widehat{\psi}_1 - \widehat{\psi}_2\|_{L^\infty} \leq \varepsilon \cdot d_H(v_1, v_2).
\]
Combining both estimates gives
\[
\|(\widehat{\phi}_1, \widehat{\psi}_1) - (\widehat{\phi}_2, \widehat{\psi}_2)\|_{L^\infty} \leq \varepsilon \cdot d_H((u_1, v_1), (u_2, v_2)).
\]

\vspace{1em}

Combining the two-sided bounds, we conclude:
\[
\frac{1}{\varepsilon} \| (\widehat{\phi}_1, \widehat{\psi}_1) - (\widehat{\phi}_2, \widehat{\psi}_2) \|_{L^\infty} \leq d_H((u_1, v_1), (u_2, v_2)) \leq \frac{2}{\varepsilon} \| (\widehat{\phi}_1, \widehat{\psi}_1) - (\widehat{\phi}_2, \widehat{\psi}_2) \|_{L^\infty}.
\]

Let \(\widehat{\mathcal{G}}(u_1, v_1) = (u_1', v_1')\), \(\widehat{\mathcal{G}} (u_2, v_2) = (u_2', v_2')\). Then the Hilbert contraction property implies:
\[
d_H((u_1', v_1'), (u_2', v_2')) \leq k \cdot d_H((u_1, v_1), (u_2, v_2)),
\]
where \(k < 1\) is the contraction constant.

Let the corresponding potentials be \((\widehat{\phi}'_1, \widehat{\psi}'_1)\) and \((\widehat{\phi}'_2, \widehat{\psi}'_2)\). Then
\[
\| (\widehat{\phi}'_1, \widehat{\psi}'_1) - (\widehat{\phi}'_2, \widehat{\psi}'_2) \|_{L^\infty} \leq \varepsilon \cdot d_H((u_1', v_1'), (u_2', v_2')) \leq \varepsilon \cdot k \cdot d_H((u_1, v_1), (u_2, v_2)).
\]

By the previous upper bound on \(d_H\), we obtain:
\begin{align*}
	\| (\widehat{\phi}'_1, \widehat{\psi}'_1) - (\widehat{\phi}'_2, \widehat{\psi}'_2) \|_{L^\infty} 
	&\leq \varepsilon \cdot k \cdot \frac{2}{\varepsilon} \| (\widehat{\phi}_1, \widehat{\psi}_1) - (\widehat{\phi}_2, \widehat{\psi}_2) \|_{L^\infty} \\
	&= 2k \| (\widehat{\phi}_1, \widehat{\psi}_1) - (\widehat{\phi}_2, \widehat{\psi}_2) \|_{L^\infty}.
\end{align*}

Replace  \(\|\widehat{\mathcal{G}}(\widehat{\phi}, \widehat{\psi}) - \mathcal{G} (\phi, \psi)\|_{L^\infty}\)   in \eqref{sinkhorn_inequality}

\[
\|(\widehat{\phi}, \widehat{\psi}) - (\phi, \psi)\|_{L^\infty} \leq 2 k \|(\widehat{\phi}, \widehat{\psi}) - (\phi, \psi)\|_{L^\infty} + \|\widehat{\mathcal{G}} (\phi, \psi)- \mathcal{G}(\phi, \psi)\|_{L^\infty}.
\]

Thus,
\[
(1 - 2 k) \|(\widehat{\phi}, \widehat{\psi}) - (\phi, \psi)\|_{L^\infty} \leq \|\widehat{\mathcal{G}}  (\phi, \psi) - \mathcal{G}(\phi, \psi)\|_{L^\infty}.
\]
If \(2k < 1\), \(1 - 2k > 0\), we have
\[
\|(\widehat{\phi}, \widehat{\psi}) - (\phi, \psi)\|_{L^\infty} \leq \frac{1}{1 - 2k} \|\widehat{\mathcal{G}} (\phi, \psi) - \mathcal{G}(\phi, \psi)\|_{L^\infty}.
\]
From the previous result (since \(\widehat{\mathcal{G}}  \to \mathcal{G}\)), we conclude that
\[
\|(\widehat{\phi}, \widehat{\psi}) - (\phi, \psi)\|_{L^\infty} \xrightarrow{\text{a.s.}} 0.
\]
For \(2k \geq 1\), we can iterate the operator \(\mathcal{G}\). Specifically, letting \((u_1'', v_1'') = \widehat{\mathcal{G}} (u_1', v_1')\) and \((u_2'', v_2'') = \widehat{\mathcal{G}}(u_2', v_2')\), we similarly have
\[
\| (\widehat{\phi}''_1, \widehat{\psi}''_1) - (\widehat{\phi}''_2, \widehat{\psi}''_2) \|_{L^\infty} \leq 2k^2 \| (\widehat{\phi}_1, \widehat{\psi}_1) - (\widehat{\phi}_2, \widehat{\psi}_2) \|_{L^\infty}.
\]

More generally, after \(l\) iterations of the contraction operator, we obtain
\[
\| (\widehat{\phi}^{(l)}_1, \widehat{\psi}^{(l)}_1) - (\widehat{\phi}^{(l)}_2, \widehat{\psi}^{(l)}_2) \|_{L^\infty} \leq 2k^l \| (\widehat{\phi}_1, \widehat{\psi}_1) - (\widehat{\phi}_2, \widehat{\psi}_2) \|_{L^\infty}.
\]
Choose \( l \) large enough such that \( 2k^l < 1 \), and consider the \( l \)-fold composition \(\widehat{\mathcal{G}}^l\). Replacing \(\widehat{\mathcal{G}}\) in \eqref{sinkhorn_inequality} with its iterate \(\widehat{\mathcal{G}}^l\), we conclude
\[
\|(\widehat{\phi}, \widehat{\psi}) - (\phi, \psi)\|_{L^\infty} \xrightarrow{\text{a.s.}} 0.
\]
That is, repeated application of \(\widehat{\mathcal{G}}\) guarantees geometric convergence in the \(L^\infty\) norm.	
\end{proof}

\begin{remark}
{\rm	In \cite{peyre2019computational}, the strict contraction property of the Sinkhorn algorithm is established in the matrix setting. 
	In this formulation, the update of the scaling factor for the target space is given by
	\[
	v'(y_j) = \frac{\widehat{\nu}_{N_2}(y_j)}{\sum_{i=1}^{N_1} u(x_i) K(x_i, y_j)}.
	\]
	Since \(\widehat{\nu}_{N_2}(y_j) = 1/N_2\) for each sample point \(y_j\), it follows that
	\[
	v'(y_j) = \frac{1/N_2}{\sum_{i=1}^{N_1} u(x_i) K(x_i, y_j)}.
	\]
	Substituting the expressions \(u(x_i) = e^{\widehat{\psi}_{\varepsilon}(x_i)/\varepsilon}\) and \(K(x_i, y_j) = e^{-c(x_i, y_j)/\varepsilon}\), we obtain
	\[
	\sum_{i=1}^{N_1} u(x_i) K(x_i, y_j) = \sum_{i=1}^{N_1} e^{(\widehat{\psi}_{\varepsilon}(x_i) - c(x_i, y_j))/\varepsilon}.
	\]
	Thus, the update becomes
	\[
	v'(y_j) = \frac{1/N_2}{\sum_{i=1}^{N_1} e^{(\widehat{\psi}_{\varepsilon}(x_i) - c(x_i, y_j))/\varepsilon}}.
	\]
	Taking the logarithm and multiplying by \(\varepsilon\) yields
	\[
	\varepsilon \log v'(y_j) = \varepsilon \log\left( \frac{1}{N_2} \right) - \varepsilon \log\left( \sum_{i=1}^{N_1} e^{(\widehat{\psi}_{\varepsilon}(x_i) - c(x_i, y_j))/\varepsilon} \right).
	\]
	Recalling that \(\widehat{\phi}_{\varepsilon}(y_j) = \varepsilon \log v'(y_j)\), and observing that the additive constant \(\varepsilon \log(1/N_2)\) can be absorbed into the normalization, we conclude
	\[
	\widehat{\phi}_{\varepsilon}(y_j) = -\varepsilon \log\left( \sum_{i=1}^{N_1} e^{(\widehat{\psi}_{\varepsilon}(x_i) - c(x_i, y_j))/\varepsilon} \right).
	\]
	This update rule coincides with the dual formulation of the Sinkhorn algorithm in the discrete setting.
}	
	
\end{remark}

\noindent{\bf Proof of Proposition \ref{pro.8}.}
Define 
\[
\Bar{\mathbb{M}}^{\varepsilon_{N}}_{N}(p) = \int_{p\mathbb{S}_d} \rho(\mathbf{Q}^{\varepsilon_{N}}_N(\mathbf{s})) g(\mathbf{s}) \, d\mathbf{s}.
\]
The expression can be decomposed into two terms:
\begin{align*}
	\mathbb{E} \left| \widehat{\mathbb{M}}^{\varepsilon_{N}}_{N}(p) - \mathbb{M}(p) \right| & \leq \mathbb{E} \left| \widehat{\mathbb{M}}^{\varepsilon_{N}}_{N}(p) - \Bar{\mathbb{M}}^{\varepsilon_{N}}_{N}(p) \right| + \mathbb{E} \left| \Bar{\mathbb{M}}^{\varepsilon_{N}}_{N}(p) - \mathbb{M}(p) \right|.
\end{align*}	
For the term \( \mathbb{E} \left| \Bar{\mathbb{M}}^{\varepsilon_{N}}_{N}(p) - \mathbb{M}(p) \right| \), by Theorem 3 in \cite{pooladian2021entropic}, we have
\begin{align*}
	\mathbb{E} \left| \Bar{\mathbb{M}}^{\varepsilon_{N}}_{N}(p) - \mathbb{M}(p) \right| 
	&= \int_{p\mathbb{S}_d} \vert \rho(\mathbf{Q}^{\varepsilon_{N}}_N(\mathbf{s})) - \rho(\mathbf{Q}(\mathbf{s})) \vert g(\mathbf{s}) \, d\mathbf{s} \\
	&\leq \left( \int_{p\mathbb{S}_d} \vert \rho(\mathbf{Q}^{\varepsilon_{N}}_N(\mathbf{s})) - \rho(\mathbf{Q}(\mathbf{s})) \vert^2 g(\mathbf{s}) \, d\mathbf{s} \right)^{\frac{1}{2}} \\
	&\leq L_1 \left( \int_{p\mathbb{S}_d} \vert \mathbf{Q}^{\varepsilon_{N}}_N(\mathbf{s}) - \mathbf{Q}(\mathbf{s}) \vert^2 g(\mathbf{s}) \, d\mathbf{s} \right)^{\frac{1}{2}} \\
	&\leq \left( (1 + \mathrm{I}_0(\mathrm{P},\mathrm{U}_d)) N^{-\frac{\bar{\alpha}+1}{2(d+\bar{\alpha}+1)}} \log N \right)^{\frac{1}{2}},
\end{align*}
where \( \mathrm{I}_0(\mathrm{P}, \mathrm{U}_d) \) represents the integrated Fisher information along the Wasserstein geodesic between the source measure \( \mathrm{P} \) and the target measure \( \mathrm{U}_d \), which is finite under our assumptions. Since \(\rho\) is continuously differentiable with bounded support, it follows that \(\rho\) is \(L_1\)-Lipschitz continuous for some constant \(L_1\). 

For the term \( \mathbb{E} \left| \widehat{\mathbb{M}}^{\varepsilon_{N}}_{N}(p) - \Bar{\mathbb{M}}^{\varepsilon_{N}}_{N}(p) \right| \), by the multivariate central limit theorem and the delta method, there exists a constant \( C_2 \) such that
\[
\mathbb{E} \left| \widehat{\mathbb{M}}^{\varepsilon_{N}}_{N}(p) - \Bar{\mathbb{M}}^{\varepsilon_{N}}_{N}(p) \right| \leq C_2 \dfrac{1}{\sqrt{N}}.
\]
Combining these results, the proof is complete.\qed

\vspace{1cm}

\noindent{\bf Proof of Lemma \ref{lem.1}.}
This result follows directly as a corollary of the  Lemmas \ref{lem.5}  and \ref{lem.6} and the independence property. \qed
\begin{lemma}\label{lem.5}
	Let \(\rho\) be a differentiable function in \(\mathbb{R}^d\). Then, we have
	\[
	\sqrt{N} \left( \rho( \widehat{\mathbf{Q}}^{(\varepsilon,N)}_{\pm} ) - \rho( \mathbf{Q}_{\pm}^{\varepsilon}) \right) \rightsquigarrow - \nabla \rho(\mathbf{Q}_{\pm}^{\varepsilon}) \cdot \nabla G,
	\]
	where \(- \nabla \rho(\mathbf{Q}_{\pm}^{\varepsilon}) \cdot \nabla G\) is a zero-mean Gaussian random variable in \( \mathcal{C}^{s-1} (\mathcal{X}; \mathbb{R}) \)(for any $s \in \mathbb{N}$). 
\end{lemma}

\begin{proof}
	By Corollary 1 in \cite{goldfeld2024limit}, we know that
	\[	\sqrt{N} \left(  \widehat{\mathbf{Q}}^{(\varepsilon,N)}_{\pm} - \mathbf{Q}_{\pm}^{\varepsilon} \right) \rightsquigarrow -\nabla G
	\]
	in \( \mathcal{C}^{s-1}(\mathbb{S}_d; \mathcal{X}) \). The limit \(-\nabla G\) is a zero-mean Gaussian random variable in \( \mathcal{C}^{s-1}(\mathbb{S}_d; \mathcal{X}) \).
	
	Since \(\rho\) is differentiable, the Hadamard derivative of \(\rho(\mathbf{Q}_{\pm}^{\varepsilon})\) in the direction of \(h\) is given by \(\nabla \rho(\mathbf{Q}_{\pm}^{\varepsilon})\). Therefore, we have
	\[
	\sqrt{N} \left( \rho( \widehat{\mathbf{Q}}^{(\varepsilon,N)}_{\pm}) - \rho( \mathbf{Q}_{\pm}^{\varepsilon}) \right) \rightsquigarrow - \nabla \rho(\mathbf{Q}_{\pm}^{\varepsilon}) \cdot \nabla G,
	\]
	where \(- \nabla \rho(\mathbf{Q}_{\pm}^{\varepsilon}) \cdot \nabla G\) is a random variable in \( \mathcal{C}^{s-1} (\mathcal{X}; \mathbb{R}) \).
\end{proof}

\begin{lemma}\label{lem.6}
	We have
	\[
	\sqrt{N} \left( { \widehat{\mathrm{M}}}^{\varepsilon}_{N} -  {\mathrm{M}^{\varepsilon}} \right) \rightsquigarrow {\mathcal{J}}
	\]
	and
	\[
	\sqrt{N} \left({\widehat{\mathbb{M}}}_{N} -{ \mathbb{M}} \right) \rightsquigarrow {\mathcal{K}}.
	\]
	Here, \(\mathcal{J}\) and \(\mathcal{K}\) are zero-mean Gaussian random variables in \(\mathcal{C}[0,1]\).
\end{lemma}

\begin{proof}
	The conclusion follows directly from the linearity of conditional expectation and Lemma \ref{lem.5}.\
\end{proof}


\noindent{\bf Proof of Proposition \ref{pro.10}.}
 Following the notation in \cite{goldfeld2024limit}, we present the following lemma.  
 
\begin{lemma}\label{lem.7}
	Let \(s\) be a positive integer with \(s > d/2\). Then, for every \(\mu \in\mathcal{P}(X)\), we have \(\sqrt{N}(\hat{\mu}_{N}^{B}-\mu)\stackrel{d}{\to}\mathbb{G}^{\mu}\) in \(\ell^{\infty}(B^{s}) \),  where \(\mathbb{G}^{\mu}\) is a tight \(\mu\)-Brownian bridges in \(\ell^{\infty}(B^{s})\).
\end{lemma}

\begin{proof}
	The set \( B^s \) is \(\mu\)-Donsker by Theorem 2.7.1 in \cite{van1996weak}. By Theorem 3.7.1 in \cite{van1996weak}, we know that \(\sqrt{N}(\hat{\mu}_{N}^{B} - \mu)\) converges to \(\mathbb{G}^{\mu}\) under the bounded Lipschitz metric with outer probability one. Finally, applying Theorem 1.13.1 in \cite{van1996weak}, we obtain the desired weak convergence.
\end{proof}
Lemma 9 in \cite{goldfeld2024limit} provide the result that the empirical measures   convergence weakly to the tight Brownian bridges. \qed

\vspace{0.5cm}

\noindent{\bf Proof of Proposition \ref{pro.12}.} This result can be deduced from Theorem 3.10.11 in \cite{van1996weak}, along with Lemmas \ref{lem.5}, \ref{lem.6}, and \ref{lem.7}. \qed

\vspace{0.5cm}

\noindent{\bf Proof of Proposition \ref{pro.13}.}
The argument is an application of Proposition~3.3 in \cite{fang2019inference}. 
We verify that the four assumptions therein hold in our setting. 
In particular, their Assumption~1 is implied by our Assumption~\ref{assump.4}, 
while their Assumption~2 follows from Lemma~\ref{lem.1}. 
The first three parts of their Assumption~3 are guaranteed by Proposition~\ref{pro.10}. 
For the fourth part of Assumption~3, note the following.

Fix the sample $\{\mathbf{X}_i\}_{i=1}^N$. 
As a function of the bootstrap weights $\{W_i\}_{i=1}^N$, the mapping
\[
(W_1,\ldots,W_N)\ \longmapsto\ \mathrm{T}_N^B(\{W_i\}_{i=1}^N)
\]
is measurable with respect to the product $\sigma$-algebra on the weights. 
Moreover, for any bounded continuous function $f$, 
the composition $f(\mathrm{T}_N^B)$ remains measurable as a function of $\{W_i\}$. 
Hence, with respect to the outer measure over $\{\mathbf{X}_i\}$, 
this property holds almost surely, thereby verifying condition~(iv). 
Finally, their Assumption~4 is ensured by our Assumption~\ref{assump.6}.\qed

{
	\section{General Distributions}\label{app.b}

A limitation of the current definition is its reliance on center‑outward symmetry, whereas real‑world multivariate data often exhibits directional monotonicity—increases in any component, \emph{ceteris paribus}, correspond to improved outcomes (e.g., higher income or better health). To accommodate such monotonic structures, we generalize the framework via a transformation that recenters the data while preserving the ordinal interpretation of each dimension.
	
	Consider a componentwise transformation
	\[
	\mathbf{l}(\mathbf{X}) = \big(l_1(X_1),\, l_2(X_2),\, \ldots,\, l_d(X_d)\big),
	\]
	where each $l_i$ is continuous and strictly increasing, and $\mathbf{l}$ maps into $[0,\infty)^d$. Because each $l_i$ acts independently, the transformation preserves the relative ordering of points. After transformation, a larger distance from the origin corresponds to a “better” outcome, which aligns naturally with the center‑outward quantile interpretation.
	
	Many monotone functions can serve as $l_i$, typically parameterized as $l_i(t) = f(a_i t + b_i)$ with $a_i > 0$. Common choices include:
	\begin{itemize}
		\item Exponential: $l_i(t) = \exp(a_i t + b_i)$, suitable for data with exponential trends;
		\item Shifted ReLU: $l_i(t) = \max(0, a_i t + b_i)$, useful for sparse or truncated data;
		\item Softplus: $l_i(t) = a_i^{-1}\log\big(1+\exp(a_i t + b_i)\big)$, a smooth approximation to ReLU;
		\item Logistic sigmoid: $l_i(t) = \big[1+\exp(-(a_i t + b_i))\big]^{-1}$, which compresses extreme values;
		\item Arctangent: $l_i(t) = \pi^{-1}\arctan(a_i t + b_i) + 1$, offering robust normalization for heavy‑tailed data.
	\end{itemize}
	The parameters $a_i, b_i$ can be chosen based on domain knowledge or estimated from the data.
	
We can establish a dominance relationship under the transformation \( \mathbf{l} \). This relationship, however, is sensitive to the specific choice of \( \mathbf{l} \), as different transformations highlight different structural features of the data. Thus, the choice of \( \mathbf{l} \) should be guided by the data characteristics and the analytical objective.

Henceforth, we assume the random vector is supported in $[0,\infty)^d$. For the transformed framework to be coherent, the quantile region $\mathbb{C}(0) = \lim_{q\to 0} \mathbb{C}(q)$ must contain the origin. An arbitrary transformation $\mathbf{l}$ need not satisfy this; we therefore enforce it by symmetrization.

Let $\mathbf{F}(x_1,\dots,x_d)$ be the distribution function of $\mathbf{X}$ with support in $[0,\infty)^d$. Define the symmetrized version $\mathbf{X}^*$ through the distribution function
\[
\mathbf{F}^*(x_1,\dots,x_d)
=
\frac{1}{2^d}
\sum_{I_1,\dots,I_d \in \{-1,1\}}
\mathbf{F}(I_1 x_1,\dots,I_d x_d).
\]
Dominance comparisons are then performed on the symmetrized variables.


The MSD framework assumes nonvanishing distributions, but this requirement is seldom restrictive in practice, as the statistic remains computable in most cases. If one seeks full theoretical validity, or if the lack of nonvanishing leads to numerical issues, mixing the target distribution with an appropriate background distribution provides a straightforward remedy.

Let $\mathbf{X} \sim \mathrm{P}$ be defined on a bounded compact set $\mathcal{X} \subset \mathbb{R}^d$ with density $f$. Assume $f$ is locally bounded, i.e., for every $D>0$ there exists $\Lambda_{D,f}>0$ such that $f(\mathbf{x}) \le \Lambda_{D,f}$ for all $\|\mathbf{x}\| \le D$. Introduce a symmetric $d$-dimensional distribution $\mathbf{Y} \in \mathcal{P}_d$ and define the mixture
\begin{equation}\label{eq.6}
	F_{\mathbf{Z}^{\eta}}(x) = (1-\eta) F_{\mathbf{X}}(x) + \eta F_{\mathbf{Y}}(x),
\end{equation}
where $\eta \in [0,1]$ is the background intensity. Multivariate stochastic dominance can then be evaluated for $\mathbf{Z}^{\eta}$.

Denote the quantile functions of $\mathbf{X}$ and $\mathbf{Z}^{\eta}$ by $\mathbf{Q}_{\pm}$ and $\mathbf{Q}_{\pm}^{\eta}$, respectively. By the stability of optimal transport maps under convex combinations of measures (Theorem~1.2 of \cite{segers2022graphical}),
\[
\sup_{s \in \mathbb{S}_d} \big\lvert\mathbf{Q}_{\pm}^{\eta}(s) - \mathbf{Q}_{\pm}(s) \big\rvert \overset{\mathrm{a.s.}}{\longrightarrow} 0 \quad \text{as } \eta \to 0.
\]
Thus, when the background contribution is small ($\eta \to 0$), the quantile region of $\mathbf{X}$ is well approximated by that of $\mathbf{Z}^{\eta}$.	
	
\begin{remark}
{\rm Although the quantile function of \( \mathbf{Z}^{\eta} \) converges to that of \( \mathbf{X} \) as \( \eta \to 0 \), the choice of background distribution \( \mathbf{Y} \) affects the geometric features being emphasized. In practice, a uniform distribution on a bounded convex set is typically sufficient, while alternative symmetric backgrounds may be adopted when better aligned with the data structure. 
}
\end{remark}
	
%

\section{Estimating Center-outward Quantiles}\label{appendix.c}

To estimate center‑outward quantiles, we first construct a regular grid $\mathfrak{G}^{(n)}$ of $\mathbb{U}_d$ comprising $n$ points $\mathfrak{g}_{1}^{(n)}, \dots, \mathfrak{g}_{n}^{(n)}$, where $n = n_r n_s + n_0$ with $n_0 \in \{0,1\}$. The grid is obtained as the intersection of
\begin{itemize}
	\item \textbf{Rays:} an $n_s$‑tuple of unit vectors $\mathbf{u}_1,\dots,\mathbf{u}_{n_s} \in S_{d-1}$ whose empirical distribution converges weakly to the uniform distribution on $S_{d-1}$ as $n_s \to \infty$;
	\item \textbf{Hyperspheres:} centered at $\mathbf{0}$ with radii $j/(n_r+1)$ for $j=1,\dots,n_r$, plus the origin if $n_0 = 1$.
\end{itemize}
The associated discrete uniform measure on $\mathfrak{G}^{(n)}$ is
\[
\mathbb{U}_d^{(n)} := \frac{1}{n}\sum_{j=1}^n \delta_{\mathfrak{g}_j^{(n)}}, \qquad n \in \mathbb{N}.
\]
Taking $n \to \infty$ (so that $n_r, n_s \to \infty$) guarantees $\mathbb{U}_d^{(n)} \leadsto \mathbb{U}_d$.

Given a sample $\{\mathbf{X}_j\}_{j=1}^m$, the empirical center‑outward quantile map is defined as the optimal transport map that pushes $\mathbb{U}_d^{(n)}$ forward to the empirical measure $\mathbb{P}^{(m)}_{\mathbf{X}}$. In the Kantorovich formulation, this map is obtained by solving the linear program:

\begin{equation}\label{eq.Kantorovich}
		\begin{aligned}
			&\min_{\pi:=\{\pi_{i,j}\}}\sum_{i=1}^n\sum_{j=1}^m\frac{1}{2}\left|\mathbf{X}_j-\mathfrak{g}_i\right|^2\pi_{i,j}, \\
			\text{s.t.}\quad & \sum_{j=1}^m\pi_{i,j}=\frac{1}{n},\quad i\in\{1,2,\ldots,n\},  \\
			& \sum_{i=1}^n\pi_{i,j}=\frac{1}{m},\quad j\in\{1,2,\ldots,m\}, \\
			& \pi_{i,j}\geq0,\quad i\in\{1,2,\ldots,n\},\ j\in\{1,2,\ldots,m\}.
		\end{aligned}
\end{equation}

Without loss of generality, we may take $n = m$. Let $\pi^\star$ be a solution of \eqref{eq.Kantorovich}. By Theorem~2.12 of \cite{villani2003topics}, $\pi^\star = (\mathrm{Id} \times \mathbf{Q}^{m}_{\pm})_{\#} \mathbb{U}_d^{(n)}$, and its support $\mathrm{supp}(\pi^\star)$ is cyclically monotone. Thus we define
\[
\widehat{\mathbf{Q}}_{\pm}(\mathfrak{g}_i) := \mathbf{X}_j \quad \text{if } (\mathfrak{g}_i, \mathbf{X}_j) \in \mathrm{supp}(\pi^\star).
\]

In the general case $m \neq n$, the estimated conditional center‑outward quantile map $\widehat{\mathbf{Q}}_{\pm}$ is defined via barycentric projection:

\[
\widehat{\mathbf{Q}}_{\pm}(\mathfrak{g}_i) := \sum_{j=1}^m \mathbf{X}_j \, \pi_{i,j}^\star.
\]

The entropically regularized counterpart is defined analogously by substituting the optimal coupling $\pi_{i,j}^\star$ with the entropic optimal coupling $\pi_{i,j}^\varepsilon$:

\[
\widehat{\mathbf{Q}}_{\pm}^\varepsilon(\mathfrak{g}_i) := \sum_{j=1}^m \mathbf{X}_j \, \pi_{i,j}^\varepsilon .
\]

Here $\pi^{\varepsilon}$ solves the regularized optimal transport problem
\[
\pi^{\varepsilon} := \arg\min_{\pi} \Big[ \sum_{i=1}^n\sum_{j=1}^m c_{i,j} \pi_{i,j} + \varepsilon \sum_{i=1}^n\sum_{j=1}^m \pi_{i,j} (\log \pi_{i,j} - 1) \Big],
\]
where $c_{i,j} = \frac{1}{2} \lVert \mathbf{X}_j - \mathfrak{g}_i \rVert^2$. The solution takes the Sinkhorn form
\[
\pi_{i,j}^{\varepsilon} = a_i \exp\!\Big(-\frac{c_{i,j}}{\varepsilon}\Big) b_j,
\]
with positive scaling vectors $a \in \mathbb{R}^n_{>0}$, $b \in \mathbb{R}^m_{>0}$ determined by the marginal constraints via the Sinkhorn algorithm.

\section{Additional experiments}\label{appendix.d}

In Appendix~\ref{appendix.d1} we replicate the experimental setup of Section~\ref{sec.6.1}; in Appendix~\ref{appendix.d2} we adopt the setup of Section~\ref{sec.t&mixture}.

\subsection{Supplementary experiment of Section \ref{sec.6.1}}\label{appendix.d1}

\noindent\textbf{Test based on Center-outward quantile} As noted in Section~\ref{sec.4}, the convergence rate of the original (unregularized) center‑outward quantile is too slow to support valid bootstrap inference, thereby precluding the theoretical guarantees needed for consistent hypothesis testing. For comparison, Tables~\ref{tab:emdorder_1} and \ref{tab:emdorder_2} display the results obtained with the original method, alongside the entropic results in Tables~\ref{tab:order_1} and \ref{tab:order_2}. The original quantiles exhibit larger bias and worse control of the rejection rate, confirming that only the entropic regularization attains a convergence rate sufficient for the required statistical properties.
\begin{table}[htbp]
	\centering
	\begin{tabular}{ccc|cccccc}
		\toprule
		& \textbf{Significance level} & $\mathbf{\tau_N}$ & \multicolumn{6}{c}{$\beta$} \\
		\cmidrule{4-9}
		&  &  & 2 & 3 & 4 & 5 & 6 & 7 \\
		\midrule
		\multirow{9}{*}{$\mathcal{S}$} 
		& \multirow{3}{*}{0.05} & 1 & 0.08 & 0.14 & 0.09 & 0.14 & 0.10 & 0.09 \\
		&                       & 2 & 0.06 & 0.04 & 0.07 & 0.08 & 0.06 & 0.05 \\
		&                       & $\infty$ & 0.05 & 0.04 & 0.06 & 0.05 & 0.06 & 0.04 \\
		\cmidrule{2-9}
		& \multirow{3}{*}{0.1}  & 1 & 0.09 & 0.15 & 0.09 & 0.18 & 0.11 & 0.09 \\
		&                       & 2 & 0.11 & 0.11 & 0.09 & 0.12 & 0.08 & 0.06 \\
		&                       & $\infty$ & 0.11 & 0.11 & 0.09 & 0.09 & 0.08 & 0.07 \\
		\cmidrule{2-9}
		& \multirow{3}{*}{0.2}  & 1 & 0.13 & 0.19 & 0.16 & 0.21 & 0.16 & 0.09 \\
		&                       & 2 & 0.21 & 0.19 & 0.20 & 0.15 & 0.17 & 0.16 \\
		&                       & $\infty$ & 0.21 & 0.19 & 0.20 & 0.18 & 0.17 & 0.17 \\
		\midrule
		\multirow{9}{*}{$\mathcal{I}$} 
		& \multirow{3}{*}{0.05} & 1 & 0.09 & 0.17 & 0.11 & 0.16 & 0.13 & 0.13 \\
		&                       & 2 & 0.07 & 0.04 & 0.07 & 0.09 & 0.05 & 0.11 \\
		&                       & $\infty$ & 0.05 & 0.04 & 0.07 & 0.06 & 0.05 & 0.04 \\
		\cmidrule{2-9}
		& \multirow{3}{*}{0.1}  & 1 & 0.10 & 0.19 & 0.12 & 0.21 & 0.14 & 0.13 \\
		&                       & 2 & 0.12 & 0.08 & 0.11 & 0.11 & 0.06 & 0.12 \\
		&                       & $\infty$ & 0.12 & 0.08 & 0.11 & 0.10 & 0.08 & 0.09 \\
		\cmidrule{2-9}
		& \multirow{3}{*}{0.2}  & 1 & 0.13 & 0.20 & 0.14 & 0.24 & 0.19 & 0.17 \\
		&                       & 2 & 0.20 & 0.20 & 0.18 & 0.15 & 0.17 & 0.20 \\
		&                       & $\infty$ & 0.20 & 0.20 & 0.18 & 0.17 & 0.18 & 0.21 \\
		\bottomrule
	\end{tabular}
	\caption{Rejection rates under second-order multivariate stochastic dominance at various significance levels and tuning parameters}
	\label{tab:order_2}
\end{table}

\begin{table}[htbp]
		\centering
		\begin{tabular}{ccc|cccccc}
			\toprule
			& \textbf{Significance level} & $\mathbf{\tau_N}$ & \multicolumn{6}{c}{$\beta$} \\
			\cmidrule{4-9}
			&  &  & 2 & 3 & 4 & 5 & 6 & 7 \\
			\midrule
			\multirow{9}{*}{$\mathcal{S}$} 
			& \multirow{3}{*}{0.05} & 1 & 0.03 & 0.07 & 0.08 & 0.04 & 0.08 & 0.08 \\
			&                       & 2 & 0.02 & 0.06 & 0.05 & 0.03 & 0.04 & 0.04 \\
			&                       & $\infty$ & 0.02 & 0.05 & 0.05 & 0.03 & 0.04 & 0.03 \\
			\cmidrule{2-9}
			& \multirow{3}{*}{0.1}  & 1 & 0.07 & 0.09 & 0.11 & 0.10 & 0.09 & 0.12 \\
			&                       & 2 & 0.06 & 0.07 & 0.07 & 0.06 & 0.07 & 0.08 \\
			&                       & $\infty$ & 0.06 & 0.07 & 0.06 & 0.06 & 0.06 & 0.06 \\
			\cmidrule{2-9}
			& \multirow{3}{*}{0.2}  & 1 & 0.12 & 0.12 & 0.17 & 0.18 & 0.14 & 0.20 \\
			&                       & 2 & 0.12 & 0.12 & 0.13 & 0.15 & 0.12 & 0.16 \\
			&                       & $\infty$ & 0.12 & 0.12 & 0.12 & 0.15 & 0.12 & 0.15 \\
			\midrule
			\multirow{9}{*}{$\mathcal{I}$} 
			& \multirow{3}{*}{0.05} & 1 & 0.11 & 0.16 & 0.18 & 0.18 & 0.11 & 0.15 \\
			&                       & 2 & 0.03 & 0.06 & 0.09 & 0.10 & 0.05 & 0.07 \\
			&                       & $\infty$ & 0.03 & 0.05 & 0.04 & 0.08 & 0.05 & 0.04 \\
			\cmidrule{2-9}
			& \multirow{3}{*}{0.1}  & 1 & 0.14 & 0.19 & 0.24 & 0.22 & 0.14 & 0.22 \\
			&                       & 2 & 0.08 & 0.12 & 0.15 & 0.13 & 0.10 & 0.14 \\
			&                       & $\infty$ & 0.08 & 0.11 & 0.10 & 0.11 & 0.09 & 0.11 \\
			\cmidrule{2-9}
			& \multirow{3}{*}{0.2}  & 1 & 0.17 & 0.25 & 0.32 & 0.30 & 0.22 & 0.28 \\
			&                       & 2 & 0.15 & 0.21 & 0.22 & 0.23 & 0.15 & 0.20 \\
			&                       & $\infty$ & 0.14 & 0.21 & 0.21 & 0.21 & 0.15 & 0.18 \\
			\bottomrule
		\end{tabular}
		\caption{First-order multivariate stochastic dominance without entropic estimation}
		\label{tab:emdorder_1}
	\end{table}

\begin{table}[htbp]
		\centering
		\begin{tabular}{ccc|cccccc}
			\toprule
			& \textbf{Significance level} & $\mathbf{\tau_N}$ & \multicolumn{6}{c}{$\beta$} \\
			\cmidrule{4-9}
			&  &  & 2 & 3 & 4 & 5 & 6 & 7 \\
			\midrule
			\multirow{9}{*}{$\mathcal{S}$} 
			& \multirow{3}{*}{0.05} & 1 & 0.06 & 0.11 & 0.09 & 0.09 & 0.10 & 0.08 \\
			&                       & 2 & 0.04 & 0.06 & 0.05 & 0.04 & 0.06 & 0.08 \\
			&                       & $\infty$ & 0.05 & 0.04 & 0.05 & 0.03 & 0.06 & 0.08 \\
			\cmidrule{2-9}
			& \multirow{3}{*}{0.1}  & 1 & 0.07 & 0.12 & 0.11 & 0.10 & 0.12 & 0.09 \\
			&                       & 2 & 0.11 & 0.10 & 0.11 & 0.07 & 0.11 & 0.13 \\
			&                       & $\infty$ & 0.11 & 0.10 & 0.11 & 0.08 & 0.11 & 0.13 \\
			\cmidrule{2-9}
			& \multirow{3}{*}{0.2}  & 1 & 0.15 & 0.15 & 0.15 & 0.13 & 0.19 & 0.20 \\
			&                       & 2 & 0.21 & 0.20 & 0.18 & 0.13 & 0.20 & 0.25 \\
			&                       & $\infty$ & 0.21 & 0.20 & 0.18 & 0.14 & 0.20 & 0.25 \\
			\midrule
			\multirow{9}{*}{$\mathcal{I}$} 
			& \multirow{3}{*}{0.05} & 1 & 0.14 & 0.12 & 0.12 & 0.14 & 0.13 & 0.12 \\
			&                       & 2 & 0.07 & 0.08 & 0.05 & 0.07 & 0.06 & 0.08 \\
			&                       & $\infty$ & 0.06 & 0.06 & 0.04 & 0.03 & 0.06 & 0.08 \\
			\cmidrule{2-9}
			& \multirow{3}{*}{0.1}  & 1 & 0.15 & 0.13 & 0.14 & 0.16 & 0.15 & 0.15 \\
			&                       & 2 & 0.11 & 0.12 & 0.12 & 0.09 & 0.10 & 0.12 \\
			&                       & $\infty$ & 0.11 & 0.12 & 0.12 & 0.07 & 0.10 & 0.12 \\
			\cmidrule{2-9}
			& \multirow{3}{*}{0.2}  & 1 & 0.18 & 0.18 & 0.18 & 0.20 & 0.20 & 0.21 \\
			&                       & 2 & 0.20 & 0.22 & 0.20 & 0.15 & 0.21 & 0.24 \\
			&                       & $\infty$ & 0.21 & 0.22 & 0.20 & 0.15 & 0.21 & 0.24 \\
			\bottomrule
		\end{tabular}
		\caption{Second-order multivariate stochastic dominance without entropic estimation}
		\label{tab:emdorder_2}
	\end{table}

\noindent\textbf{Normal distribution with varying correlation} In this experiment, we examine the performance of the test for bivariate normal vectors with varying degrees of dependence. Let
\[
\mathbf{X} \sim \mathcal{N}\!\left(
\begin{pmatrix} 0 \\ 0 \end{pmatrix},
\begin{pmatrix}
	4 & 4\alpha \\[2pt]
	4\alpha & 4
\end{pmatrix}
\right),
\]
and let $\mathbf{Y}$ have the same distribution as $\mathbf{X}$. The remaining experimental settings follow Section~\ref{sec.6.1}. Tables~\ref{tab:cor_1} and \ref{tab:cor_2} present the results. Even under strong component dependence ($\alpha$ close to $\pm 1$), the empirical rejection rates remain close to the nominal levels, demonstrating the robustness of the method to high correlation between components.

\begin{table}[htbp]
	\centering
	\begin{tabular}{ccc|cccccc}
		\toprule
		& \textbf{Significance level} & $\mathbf{\tau_N}$ & \multicolumn{6}{c}{$\beta$} \\
		\cmidrule{4-9}
		&  &  & 0.1& 0.2 & 0.3 & 0.4 & 0.5 & 0.6 \\
		\midrule
		\multirow{6}{*}{$\mathcal{S}$} 
		& \multirow{3}{*}{0.05} & 1 & 0.09 & 0.06& 0.09 & 0.11& 0.08& 0.08 \\
		&                       				& 2 & 0.08 & 0.06 & 0.05 & 0.07 & 0.07& 0.06\\
		&                    & $\infty$ & 0.07& 0.05& 0.05 & 0.06 & 0.07 & 0.06\\
		\cmidrule{2-9}
		& \multirow{3}{*}{0.1}  & 1 & 0.12& 0.11 & 0.12& 0.14& 0.11 & 0.13 \\
		&                       			& 2 & 0.11 & 0.10& 0.11 & 0.10 & 0.09& 0.11 \\
		&                   	& $\infty$ & 0.09& 0.09 & 0.11 & 0.09 & 0.1& 0.11\\
		\cmidrule{2-9}
		
		\multirow{6}{*}{$\mathcal{I}$} 
		& \multirow{3}{*}{0.05} & 1 & 0.12 & 0.16& 0.09 & 0.20 & 0.13& 0.16 \\
		&                       & 2 & 0.07 & 0.06 & 0.05 & 0.12 & 0.05& 0.09\\
		&                       & $\infty$ & 0.06& 0.06& 0.05 & 0.06 & 0.05 & 0.08 \\
		\cmidrule{2-9}
		& \multirow{3}{*}{0.1}  & 1 & 0.15 & 0.21 & 0.13 & 0.24& 0.16& 0.19\\
		&                       & 2 & 0.10 & 0.14& 0.09& 0.13& 0.10& 0.11 \\
		&                       & $\infty$ & 0.09 & 0.13 & 0.08 & 0.12& 0.10 & 0.11 \\
		\bottomrule
	\end{tabular}
	\caption{Rejection rates under first-order multivariate stochastic dominance at various significance levels and tuning parameters}
	\label{tab:cor_1}
\end{table}

\begin{table}[htbp]
	\centering
	\begin{tabular}{ccc|cccccc}
		\toprule
		& \textbf{Significance level} & $\mathbf{\tau_N}$ & \multicolumn{6}{c}{$\beta$} \\
		\cmidrule{4-9}
		&  &  & 0.1& 0.2 & 0.3 & 0.4 & 0.5 & 0.6 \\
		\midrule
		\multirow{6}{*}{$\mathcal{S}$} 
		& \multirow{3}{*}{0.05} & 1 & 0.09 & 0.08& 0.08 & 0.07 & 0.11 & 0.12 \\
		&                       & 2 & 0.05 & 0.06 & 0.07 & 0.05 & 0.05 & 0.06\\
		&                       & $\infty$ & 0.05 & 0.04 & 0.05 & 0.05 & 0.06 & 0.04\\
		\cmidrule{2-9}
		& \multirow{3}{*}{0.1}  & 1 & 0.11& 0.11 & 0.09 & 0.13 & 0.09 & 0.13 \\
		&                       & 2 & 0.11 & 0.09& 0.09 & 0.12 & 0.12& 0.10 \\
		&                       & $\infty$ & 0.08& 0.11 & 0.09 & 0.09 & 0.11 & 0.10\\
		\cmidrule{2-9}
		
		\multirow{6}{*}{$\mathcal{I}$} 
		& \multirow{3}{*}{0.05} & 1 & 0.12 & 0.11& 0.11 & 0.09 & 0.16& 0.13 \\
		&                       & 2 & 0.05 & 0.06 & 0.04 & 0.09 & 0.08& 0.06\\
		&                       & $\infty$ & 0.04& 0.04 & 0.07 & 0.05 & 0.05 & 0.04 \\
		\cmidrule{2-9}
		& \multirow{3}{*}{0.1}  & 1 & 0.15 & 0.13 & 0.13 & 0.11 & 0.17& 0.13 \\
		&                       & 2 & 0.10 & 0.12 & 0.08 & 0.11 & 0.11& 0.10 \\
		&                       & $\infty$ & 0.09 & 0.11 & 0.08 & 0.10 & 0.09 & 0.10 \\
		\bottomrule
	\end{tabular}
	\caption{Rejection rates under second-order multivariate stochastic dominance at various significance levels and tuning parameters}
	\label{tab:cor_2}
\end{table}

\subsection{Examples of more distributions} \label{appendix.d2}

\textbf{Three-dimensional normal distribution. } We generate data to analyze the properties of the defined normal distributions. Let \(\mathbf{X}\) and \(\mathbf{Y}\) be two centered 3-dimensional normal distributions. Specifically, \(\mathbf{X}\) follows:

\[
\mathbf{X} \sim \mathcal{N}(\mathbf{\mu}_1, \Sigma_1),
\]

where
\[
\mu_1 = \begin{bmatrix}
	0 \\
	0 \\
	0
\end{bmatrix}, \quad
\Sigma_1 = \begin{bmatrix}
	4& 0 & 0 \\
	0 & 4 & 0 \\
	0 & 0 & 4
\end{bmatrix}.
\]

Similarly, \(\mathbf{Y}\) follows:

\[
\mathbf{Y} \sim \mathcal{N}(\mathbf{\mu}_2, \Sigma_2),
\]

where
\[
\mu_2 = \begin{bmatrix}
	0 \\
	0 \\
	0
\end{bmatrix}, \quad
\Sigma_2 =\begin{bmatrix}
	4 & 0 & 0 \\
	0 & \beta & 0 \\
	0 & 0 & \beta
\end{bmatrix}.
\]
%

For both distributions, we generate independent datasets of size $N_1 = N_2 = 1200$. The samples $\{X_i^1\}_{i=1}^{N_1}$ and $\{X_i^2\}_{i=1}^{N_2}$ are drawn from $\mathbf{X}$ and $\mathbf{Y}$, respectively. The parameter $\beta$ takes specific values, with $\mathbf{Y}$ obtained by a 3‑dimensional rotation of $\mathbf{X}$. The dominance pattern is as follows: $\mathbf{X}$ dominates $\mathbf{Y}$ for $\beta = 3$; the two are equivalent for $\beta = 4$; and $\mathbf{Y}$ dominates $\mathbf{X}$ for $\beta = 5$. Table~\ref{tab:rejection rates of 3norm} reports the average rejection rates of $H^1_0: \mathbf{X} \succeq^{1} \mathbf{Y}$ and $H^2_0: \mathbf{X} \succeq^{2} \mathbf{Y}$ at the nominal level $\alpha = 0.1$. When the null hypothesis holds ($\beta \le 4$), the empirical rejection rates remain below $0.1$; when it is violated ($\beta = 5$), the rejection rates exceed $0.1$. This pattern agrees with the theoretical result stated in Proposition~\ref{pro.13}.

\begin{table}[h!]
	\centering
	\begin{tabular}{c c c c c c}
		\toprule
		\multirow{2}{*}{} & \multirow{2}{*}{$\beta$} & \multicolumn{2}{c}{$\mathcal{S}$} & \multicolumn{2}{c}{$\mathcal{I}$} \\
		\cmidrule(lr){3-4} \cmidrule(lr){5-6}
		& & {$\tau_N = 2$} & {$\tau_N = \infty$} & {$\tau_N = 2$} & {$\tau_N = \infty$} \\
		\midrule
		\multirow{3}{*}{$H^1_0$} 
		& 3 & $\approx 0$ & $\approx 0$ & 0.005 & $\approx 0$ \\
		& 4 & 0.133 & 0.119 & 0.121 & 0.103 \\
		& 5 & 0.962 & 0.965 & 0.995 & 1.000 \\
		\midrule
		\multirow{3}{*}{$H^2_0$} 
		& 3 & $\approx 0$ & $\approx 0$ & 0.004 & $\approx 0$ \\
		& 4 & 0.116 & 0.136 & 0.103 & 0.110 \\
		& 5 & 0.603 & 0.592 & 1 & 0.992 \\
		\bottomrule
	\end{tabular}
	\caption{Rejection rates of $H^1_0$ and $H^2_0$  at significance level $\alpha = 0.1$}
	\label{tab:rejection rates of 3norm}
\end{table}

\noindent\textbf{Example \ref{example.1}  }
To compare the happiness distributions of Communities A and B, we conduct multivariate stochastic dominance tests. Community A follows a bivariate normal distribution; Community B has normal marginals linked by a Clayton copula ($\theta=2$), which induces lower‑tail dependence. We simulate $n=2000$ observations from each community and apply two testing procedures.
%

First, we implement a grid-based resampling test where empirical cumulative distribution functions are approximated on a two-dimensional quantile grid (see \cite{davidson2000statistical, barrett2003consistent}). 
	At each grid point, the difference between the two empirical CDFs is computed, and the test statistic is defined as the supremum (first-order) or the integrated difference (second-order) of these discrepancies. 
	A permutation procedure with 10000 replications provides finite-sample $p$-values. 
	Second, we apply the proposed center-outward quantile-based MSD test using statistics $\mathcal{S}$ and $\mathcal{I}$, together with the transformation described in Appendix~\ref{app.b}.
	
Results are reported in Tables~\ref{tab:9} and \ref{tab:10}. 
	As shown in Table~\ref{tab:9}, the grid-based test yields high $p$-values for both directions (e.g., 1 and 0.537 for SSD), failing to distinguish the dominance order. 
	In contrast, the proposed method in Table~\ref{tab:10} demonstrates higher sensitivity. 
	Specifically, the test based on statistic $\mathcal{I}$ strongly rejects the hypothesis $B \succeq^2 A$ ($p \approx 0$) while retaining $A \succeq^2 B$ ($p = 0.725$). 
	Although statistic $\mathcal{S}$ yields a slightly higher $p$-value for the non-dominant direction, it is still notably lower than that of the grid-based test ($0.125$ vs $0.537$), indicating a superior detection capability.

\begin{table}[htbp]
	\centering
	\begin{tabular}{lccc}
		\toprule
		\textbf{Order} & \textbf{$A \preceq B$} & \textbf{$B \preceq A$} \\
		\midrule
		FSD  & 0.896 & 0.694 \\
		SSD & 1 & 0.537 \\
		\bottomrule
	\end{tabular}
	\caption{$p$-values of bivariate stochastic dominance test }
	\label{tab:9}
\end{table}

\begin{table}[htbp]
	\centering
	\begin{tabular}{cc|cccc}
		\toprule
		& $\tau_N$   & $A \succeq^1 B$ & B $ \succeq^1 A$      & $A \succeq^2 B$ & $B  \succeq^2 A$ \\  
		\midrule
		\multirow{2}{*}{$\mathcal{S}$} 
		& 2 &  0.207  & 0.093  & 0.635 & 0.101 \\  
		& $\infty$ & 0.223 & 0.113 & 0.995 & 0.125 \\  
		\midrule
		\multirow{2}{*}{$\mathcal{I}$} 
		& 2 & 0.427  & 0.086 & $0.624$ & 0.001 \\  
		& $\infty$ & 0.455  & 0.075  & $0.725$& $\approx 0$ \\  
		\bottomrule
	\end{tabular}
	\caption{    $p$-values  of MSD test based on center-outward quantile}
	\label{tab:10}
\end{table}

\noindent\textbf{Example \ref{example.2}}
we consider a source distribution defined as a two-dimensional Gaussian distribution $\mathcal{N}\left(\mathbf{0}, \begin{bmatrix} a & 0 \\ 0 & b \end{bmatrix}\right)$. Its probability density function is given by:
\[
f(x, y) = \frac{1}{2\pi\sqrt{ab}} \exp\left(-\frac{x^2}{2a} - \frac{y^2}{2b}\right)
\]
The target distribution is chosen to be the uniform distribution over the unit disk $\{(u, v) : u^2 + v^2 \leq 1\}$, characterized by the probability density:
\[
g(u, v) = 
\begin{cases} 
	\displaystyle \frac{1}{\pi}, & u^2 + v^2 \leq 1, \\[6pt]
	0, & \text{otherwise}.
\end{cases}
\]
We can derive the following  optimal transport  mapping from the source to the target distribution:
\[
u(x, y) = \sqrt{1 - \exp\left(-\frac{1}{2}\left(\frac{x^2}{a} + \frac{y^2}{b}\right)\right)} \cdot \frac{x}{\sqrt{x^2 + y^2}},
\]
\[
v(x, y) = \sqrt{1 - \exp\left(-\frac{1}{2}\left(\frac{x^2}{a} + \frac{y^2}{b}\right)\right)} \cdot \frac{y}{\sqrt{x^2 + y^2}}.
\]

Therefore, for a two-dimensional normal distribution, its center-outward quantile contours are ellipses determined by the parameters $a$ and $b$.

For the concatenated vector $\mathbf{Y}$, we present the following assertion:
\begin{equation*}
	\mathbf{T}(\mathbf{y}) = \sqrt{1 - e^{-\frac{1}{2}\left( \frac{y_1^2}{a} + \frac{y_2^2}{b} \right)}} \cdot \frac{\left( \frac{y_1}{\sqrt{a}}, \frac{y_2}{\sqrt{b}} \right)}{\sqrt{\frac{y_1^2}{a} + \frac{y_2^2}{b}}},
\end{equation*}
where $b = 1$, and the parameter $a$ is defined as follows: $a = 2$ when $y_1 < 0$, and $a = 1$ otherwise.
The above assertion is motivated by the following reasoning. According to \cite{bauschke2016convexity}, we can conclude that $\mathbf{T}$ is the gradient of a globally convex function $\varphi(\mathbf{y})$; consequently, $\mathbf{T}$ also corresponds to the optimal transport map. The specific form of $\varphi(\mathbf{y})$ is given by
\begin{equation*}
	\varphi(\mathbf{y}) = 
	\begin{cases} 
		\displaystyle \int_{0}^{\sqrt{\frac{y_1^2}{2} + y_2^2}} \sqrt{1 - e^{-t^2 / 2}} \, dt, & y_1 < 0, \\
		\displaystyle \int_{0}^{\sqrt{y_1^2 + y_2^2}} \sqrt{1 - e^{-t^2 / 2}} \, dt, & y_1 \geq 0.
	\end{cases}
\end{equation*}

When $\rho$ is the Euclidean distance from the origin, stochastic dominance can be assessed via the average distance over the quantile ellipses. For $t=2.9$, the average distance of $\mathbf{X}$ is 1.534, which exceeds the corresponding value 1.542 for $\mathbf{Y}$.

We also perform first‑ and second‑order multivariate stochastic dominance tests of $\mathbf{X}$ over $\mathbf{Y}$ using $t = 2.8$, $2.9$, and $3$. The resulting $p$‑values for $H^1_0: \mathbf{X} \succeq^{1} \mathbf{Y}$ and $H^2_0: \mathbf{X} \succeq^{2} \mathbf{Y}$ are reported in Table~\ref{tab:11}. The outcomes align with Proposition~\ref{pro.13}. Notably, the small change from $t=2.8$ to $t=2.9$ shifts the case from satisfying the null hypothesis to violating it; the test correctly distinguishes between the two, demonstrating its sensitivity to fine differences.

\begin{table}[h!]
	\centering
	\begin{tabular}{c c c c c c}
		\toprule
		\multirow{2}{*}{} & \multirow{2}{*}{$\beta$} & \multicolumn{2}{c}{$\mathcal{S}$} & \multicolumn{2}{c}{$\mathcal{I}$} \\
		\cmidrule(lr){3-4} \cmidrule(lr){5-6}
		& & {$\tau_N = 2$} & {$\tau_N= \infty$} & {$\tau_N = 2$} & {$\tau_N = \infty$} \\
		\midrule
		\multirow{3}{*}{$H^1_0$} 
		& 2.8 & 0.047 &  0.012& 0.061 & 0.0753 \\
		& 2.9 & 0.794 & 0.080 & 0.406 & 0.411  \\
		& 3 & 0.597 & 0.604& 0.749 & 0.755 \\
		\midrule
		\multirow{3}{*}{$H^2_0$} 
		& 2.8 & 0.083 & .088& 0.175 & 0.197\\
		& 2.9 & 0.446 & 0.458 & 0.542& 0.576 \\
		& 3 & 0.966 & 0.998 & 1 & 1 \\
		\bottomrule
	\end{tabular}
	\caption{$p$-values  of different $H_0$ }
	\label{tab:11}
\end{table}

\bibliographystyle{chicago}
\bibliography{sd}

\end{document}